\documentclass[twocolumn]{aastex6}
\usepackage{times}
\usepackage{amsmath}
\usepackage{textcomp}
\usepackage{amssymb}

\newcommand{\msun}{\ensuremath{\rm M_{\odot}}}
\newcommand{\lum}{erg\,s$^{-1}$}
\newcommand{\fermi}{{\it Fermi}}
\newcommand{\nustar}{{\it NuSTAR}}
\newcommand{\xmm}{{\it XMM-Newton}}
\newcommand{\suzaku}{{\it Suzaku}}
\newcommand{\swift}{{\it Swift}}

\newcommand{\gm}{$\gamma$}

\shorttitle{\gm-NLSy1 galaxies}
\shortauthors{Paliya et al.}

\begin{document}

\title{General Physical Properties of Gamma-ray Emitting Narrow-Line Seyfert 1 Galaxies}
\author{Vaidehi S. Paliya$^{1}$, M. L. Parker$^{2}$, J. Jiang$^3$, A. C. Fabian$^{3}$, L. Brenneman$^4$, M. Ajello$^5$, D. Hartmann$^{5}$} 
\affil{$^1$Deutsches Elektronen Synchrotron DESY, Platanenallee 6, 15738 Zeuthen, Germany}
\affil{$^2$European Space Agency (ESA), European Space Astronomy Centre (ESAC), E-28691 Villanueva de la Ca\~{n}ada, Madrid, Spain}
\affil{$^3$Institute of Astronomy, Madingley Road, Cambridge CB3 0HA, UK}
\affil{$^4$Harvard-Smithsonian Center for Astrophysics, 60 Garden Street, Cambridge, MA, 02138, USA}
\affil{$^5$Department of Physics and Astronomy, Clemson University, Kinard Lab of Physics, Clemson, SC 29634-0978, USA}
\email{vaidehi.s.paliya@gmail.com}

\begin{abstract}

We present the results of a detailed multi-wavelength study of a sample of sixteen narrow-line Seyfert 1 (NLSy1) galaxies known in \gm-rays so far. This includes a focused X-ray spectral analysis, to compare them with a more general NLSy1 population, and a broadband spectral energy distribution (SED) modeling considering the leptonic radiative processes to understand their similarity/dissimilarity with blazars. The X-ray spectra of \gm-NLSy1 galaxies exhibit similarities both with typical NLSy1 sources below $\lesssim$2 keV and blazars at higher energies. We also find weak evidences of excess absorption in the soft X-ray spectra of 3 sources and a tentative detection of Fe line in 1H~0323+342. A comparison of the broadband parameters derived from the SED modeling reveals the similarity of \gm-NLSy1 objects with blazars, in particular more with flat spectrum radio quasars. The \gm-NLS1 galaxies host relatively low power jets with small bulk Lorentz factors with respect to blazars and this explains the paucity of such sources in the \gm-ray catalogs and identification of an even fewer parent population members. Many of the observed features of these intriguing objects suggest they host low-mass black holes. The accretion rate in Eddington units can be considered as the driving factor to explain the physical properties of \gm-NLSy1 sources, similar to that which is known for general blazar population.

\end{abstract}


\section{Introduction} \label{sec:intro}
The physical processes launching the relativistic jet from the core of active galactic nuclei (AGN) remain one of the most outstanding unexplained mystery in AGN physics \citep[e.g.,][]{2007ApJ...658..815S}. The origin of the radio-loudness \citep[which indicates the presence of the relativistic jet, but see][for recent updates]{2017NatAs...1E.194P} of an AGN is often associated to the properties of the central black hole \citep[i.e., mass and spin,][]{2000ApJ...543L.111L,2017ApJ...849....4S} and accretion rate \citep[e.g.,][]{2017MNRAS.466..921C}. It is also postulated that magnetic flux threading the accretion disk can be a driving factor for the jet formation \citep[][]{2013ApJ...764L..24S}. Observationally, these theoretical arguments can be tested by studying the X-ray coronal radiation and its interaction with the accretion disk at the base of the jet \citep[see, e.g.,][]{2016AN....337..375F,2017ApJ...841...80L}. In the AGN unification scheme, X-ray corona is expected to be present in all types of AGN, however, only a minor fraction of them are found to exhibit powerful jets \citep[][]{1989AJ.....98.1195K}. The most profound member of AGN family hosting powerful relativistic jets are blazars. These are sources with jets aligned close to the line of sight to the observer \citep[cf.][]{1995PASP..107..803U}. The blazar jet emission is strongly amplified due to Doppler boosting. This phenomenon makes the thermal radiation from the X-ray corona comparatively negligible and therefore rarely observed from blazars \citep[e.g.,][]{2015ApJ...812...14M}.

Based on the rest-frame equivalent width (EW) of the optical emission lines, blazars are classified as flat spectrum radio quasars (FSRQs, EW$>$5\AA) and BL Lac objects \citep[EW$<$5\AA,][]{1991ApJ...374..431S}. The broadband spectral energy distribution (SED) of a blazar exhibits a typical double hump structure with low energy emission originates via synchrotron process and peaks at sub-mm to X-ray frequencies. On the other hand, inverse Compton scattering of low energy photons, present inside the jet \citep[synchrotron self Compton or SSC,][]{1985ApJ...298..114M} and/or outside \citep[external Compton or EC,][]{1994ApJ...421..153S}, is proposed to explain the high energy radiation peaking at X-ray to TeV energies. In general, the broadband SED of a BL Lac source is well explained by synchrotron and SSC processes \citep[e.g.,][]{2000NewA....5..377A}, whereas, EC mechanism is proposed as the origin of the high-energy \gm-ray emission in FSRQs \citep[e.g.,][]{2000ApJ...545..107B}. Accordingly, FSRQs show a Compton dominated SED, whereas, the SEDs of BL Lac objects are typically synchrotron dominated \citep[cf.][]{2013ApJ...763..134F}.

At the low-luminosity end, Seyfert galaxies can be useful to probe the jet-corona-disk connection \citep[e.g.,][]{2015MNRAS.454.4440W}. In fact, properties of the X-ray emitting corona are mostly understood by studying Seyferts \citep[][]{2015MNRAS.451.4375F,2017MNRAS.467.2566F}. However, most of them are radio-quiet \citep[][]{2006AJ....132..531K}, thus making it tedious to study the connection of the corona with the jet. In this regard, a few radio-loud narrow-line Seyfert 1 (NLSy1) galaxies gained attention in the last two decades due to them exhibiting dual blazar-Seyfert characteristics \citep[][]{2008ApJ...685..801Y}. Historically, NLSy1 galaxies are identified as a sub-class of AGN solely by their optical spectral characteristics. In particular, they exhibit broad permitted lines with small widths (${\rm H_{\beta,~FWHM}<2000}$ km s$^{-1}$), weak [O~{\sc iii}] emission ([O~{\sc iii}]/H$_{\beta}$ flux ratio $<$3), and strong Fe complexes \citep[][]{1985ApJ...297..166O,1989ApJ...342..224G}. NLSy1s are typically found to be hosted in spiral galaxies with low mass black holes ($\sim 10^{6-8}$ \msun) residing at their centers \citep[][]{2004ApJ...606L..41G,2006AJ....132..321D}. Moreover, these objects are postulated to be powered by high-accretion process \citep[e.g.,][]{2000ApJ...542..161P}. Many NLSy1 galaxies are bright in X-rays, exhibit a prominent soft X-ray excess, and a reflection dominated hard X-ray emission \citep[][]{2009Natur.459..540F}.

The \fermi-Large Area Telescope \citep[\fermi-LAT,][]{2009ApJ...697.1071A} detection of significant \gm-ray emission from a few radio-loud NLSy1 (hereafter \gm-NLSy1) galaxies confirmed the presence of the closely aligned relativistic jet in them \citep[][]{2009ApJ...707L.142A}. This provided us a unique opportunity to study both non-thermal jet and thermal coronal properties, and their possible interplay, in the same source. In fact, X-ray spectral measurements of a few \gm-NLSy1 sources revealed the presence of soft X-ray excess (typically seen in radio-quiet NLSy1s) and a flat rising power-law spectrum (signature of the jet emission) above a few keV \citep[e.g.,][]{2014MNRAS.440..106B}. Various other physical properties of these objects are found to be similar to blazars. This includes the detection of superluminal motion \citep[see, e.g.,][for a recent review]{2018arXiv180505258L}, violent optical and infrared (IR) flux variability and optical polarization variations \citep[][]{2010ApJ...715L.113L,2013MNRAS.428.2450P,2013ApJ...762..124M,2013ApJ...775L..26I} and a characteristic double hump SED \citep[][]{2009ApJ...707L.142A,2013ApJ...768...52P,2015ApJ...798...43S}. Interestingly, among blazars, \gm-NLSy1 galaxies resemble more with FSRQs rather than BL Lac objects. This is due to the detection of rapid \gm-ray variability \citep[][]{2014ApJ...789..143P}, curved \gm-ray spectrum \citep[][]{2015AJ....149...41P}, and a Compton dominated SED \citep[][]{2015PASJ...67..124Y}. These features are typically observed in FSRQs \citep[][]{2010ApJ...710.1271A,2011A&A...530A..77F,2012A&A...541A.160G}.

The \gm-NLSy1s have been extensively studied since their discovery in the \gm-ray band \citep[see, e.g.,][]{2011MNRAS.413.1671F,2012ApJ...759L..31J,2013MNRAS.436..191D,2014ApJ...781...75W,2014ApJ...794...93M,2014PASJ...66..108I,2015ApJ...800L...8R,F15,2015AJ....150...23Y,2015MNRAS.453.4037O,2016A&A...585A..91K,2016ApJ...819..121P,2016ApJ...820...52P,2016ApJ...824..149W,2016RAA....16..170Z,2017MNRAS.469L..11D,2017A&A...603A.100L,2017ApJ...849...42Z,2018MNRAS.475..404K,2018ApJ...857L...6D,2018MNRAS.476...43L,2018MNRAS.479.2464G}. However, a detailed comparison of their physical properties with blazars/Seyferts could not be made due to small sample size. With the latest addition of seven new \gm-NLSy1 galaxies \citep[][]{2018ApJ...853L...2P}, the total number of the known sources is now doubled. In this work, we perform a broadband study of all \gm-NLSy1 sources with the primary motivation to compare their physical properties with blazars. By analyzing good quality X-ray measurements taken from \xmm, \suzaku, and \nustar, we also focus on their X-ray spectral characteristics and compare them with that known for general Seyfert population. The overall idea is to present a coherent picture of \gm-NLSy1s with respect to that known for more common blazars and Seyfert galaxies. We briefly describe the adopted \gm-NLSy1 and blazar samples in Section \ref{sec:sample}. The multi-wavelength data reduction and analysis procedures are presented in Section \ref{sec:data_red}. In Section \ref{sec:model}, we briefly elaborate the single-zone leptonic emission model adopted to reproduce the broadband emission. The results are presented in Section \ref{sec:results} and they are compared with the Seyfert and blazar populations in Section \ref{sec:dis_seyfert} and \ref{sec:dis_blz}, respectively. We briefly discuss various scenarios proposed to explain the black hole mass discrepancy in NLSy1 galaxies in Section \ref{sec:dis_bh_mass}   and summarize in Section \ref{sec:summary}. Throughout, we use Hubble constant $H_0=67.8$~km~s$^{-1}$~Mpc$^{-1}$, $\Omega_m = 0.308$, and $\Omega_\Lambda = 0.692$ \citep[][]{2016A&A...594A..13P}.

\section{The Sample} \label{sec:sample}
We study all fourteen NLSy1 galaxies detected so far in the \gm-ray band \citep[][]{2018ApJ...853L...2P}. In order to increase the sample size, we also consider two more NLSy1 sources, namely FBQS J1102+2239 and SDSS J124634.65+023809.0 (hereafter SDSS J1246+0238), which are reported as \gm-ray emitters in previous works \citep[][]{2011nlsg.confE..24F}\footnote{A few other compact steep spectrum (CSS) NLSy1 galaxies are also detected in \gm-rays \citep[e.g., 3C 286, SDSS J130522.74+511640.2;][]{2015arXiv151005584L,2017FrASS...4....8B}. However, they most likely belong to the parent population (i.e., with misaligned jets) of flat radio spectrum NLSy1 galaxies \citep[][]{2015A&A...578A..28B}. Since, in this work, we focus on objects with closely aligned jets, \gm-ray detected CSS NLSy1s are not considered. Moreover, the \gm-ray detection of SDSS J164100.10+345452.7, claimed by \citet[][]{2018A&A...614L...1L}, has not been confirmed \citep[see,][]{2018rnls.confE..20C} and hence this object was also not included in our sample.}. The general properties of these sources are given in Table \ref{tab:basic_info}. To compare the physical properties of \gm-NLSy1 galaxies with blazars, we consider the latter population included in the Candidate Gamma-Ray Blazar Survey \citep[][]{2008ApJS..175...97H}, whose broadband properties are studied in \citet[][hereafter P17]{2017ApJ...851...33P}. 

We emphasize that our primary aim is to study the average physical properties of \gm-NLSy1 galaxies rather than any of their specific activity states. Three sources, TXS 0956+326, TXS 1419+391, and TXS 1518+423, did not have any X-ray spectral measurements other than single point {\it ROSAT} All-Sky Survey data \citep[][]{1999A&A...349..389V,2000IAUC.7432....3V}. We acquired target of opportunity (ToO) observations of these objects from the {\it Neil Gehrels Swift} observatory \citep[][]{2004ApJ...611.1005G} to generate a meaningful X-ray spectrum and to complete the X-ray coverage of \gm-NLSy1 galaxies.

There are nine objects in our sample that have good quality X-ray data, from \xmm, \suzaku, or \nustar\ publicly available that we use to perform a detailed X-ray spectral fitting. For the broadband SED modeling, on the other hand, we consider \swift~dataset due to availability of the simultaneous X-ray Telescope (XRT) and Ultraviolet-Optical Telescope (UVOT) observations.

\section{Data Compilation and Reduction}\label{sec:data_red}
\subsection{Gamma-rays}\label{subsec:lat}
We use the \fermi-LAT results published in \citet[][]{2018ApJ...853L...2P} and \citet[][for FBQS J1102+2239 and SDSS J1246+0238]{2011nlsg.confE..24F} and generate bow-tie plots for SED modeling. None of the \gm-NLSy1 sources are detected at very high energies \citep[$>$50 GeV,][]{2016ApJS..222....5A} and since all of them are relatively nearby ($z<1$), the effect of the extragalactic background light on their \gm-ray spectra is negligible.

\subsection{Hard X-rays}\label{subsec:hard-X}
Three sources, 1H 0323+342, PMN J0948+0022\footnote{PMN J0948+0022 is reported as a BAT detected source in the 100-month catalog being developed by the Palermo BAT survey team (http://bat.ifc.inaf.it/).}, and CGRaBS J1222+0413, are detected with the Burst Alert Telescope (BAT) onboard \swift~satellite \citep[][]{2018ApJS..235....4O}. We use publicly available BAT spectra and fit them with a power-law model within XSPEC \citep[][]{Arnaud96} to extract the 14$-$195 keV SEDs. 

For the sources with public \nustar\ data (Table~\ref{table:xray_obs}), we reduce the spectra using the \nustar\ Data Analysis Software (NuSTARDAS) version 6.0. We extract source counts from 30$^{\prime\prime}$ circular regions, and background counts from larger circular regions on the same chip, avoiding contaminating sources. We bin the spectra to a signal to noise ratio of 6, and to oversample the instrumental resolution by a factor of 3. Details of the \nustar\ and other spectra used for X-ray spectral fitting, are given in Table~\ref{table:xray_obs}.

\subsection{Soft X-rays}\label{subsec:soft-X}
We reduce the \xmm\ data using the Science Analysis Software (SAS) version 16.1.0. We use the \emph{epproc} tool to reduce the EPIC-pn data. We use 20--30$^{\prime\prime}$ circular source regions, depending on the brightness of the source, to extract source photons. We extract background photons from larger circular regions on the same chip, avoiding the high-Copper background region of the pn detector. For the sources where we find features in the pn spectra, we also reduce the MOS data using \emph{mosproc} and following the same procedure. Because of the low signal, the MOS spectra are only used to confirm the presence of spectral features, not for detailed fitting. We bin all spectra to a signal-to-noise ratio of 6, and to oversample by a factor of 3.

We reduce the \suzaku~data with the latest HEASOFT software package (v6.18). Filtered event lists for XIS0, XIS1 and XIS3 detectors are created by running the \suzaku~pipeline. The CALDB version we use for XIS detectors is 20160607. We use XSELECT to extract the spectral products. The size of the circular region we use to extract the source regions is 250$''$ in radius and the background regions were selected from the surrounding areas free from the target and the calibration sources. XISRESP is used to generate the corresponding response files with `medium' resolution. The spectra and the response files of the front-illuminated CCDs (XIS0 and XIS3) are combined by using ADDSPEC. Hereafter, the combined spectrum of the front-illuminated CCD XIS0 and XIS3 is called FI spectrum (blue in figures) and the spectrum of the back-illuminated CCD XIS1 is called BI spectrum (green in figures). We analyze the FI and BI data over the 0.7--10~keV and 0.7--9~keV energy range, respectively. The 1.7--2.5~keV band is ignored in both spectra due to calibration issues around the Si K edge\footnote{https://heasarc.gsfc.nasa.gov/docs/suzaku/analysis/sical.html}.

We use Chandra Interactive Analysis of Observations (CIAO v 4.9) and calibration files (v 4.7.8) to analyze the {\it Chandra} observation. The tool {\tt chandra\_repro} is used to reprocess the data. We consider the source region as a circle of 3$^{\prime\prime}$ radius centered at the target \gm-NLSy1 galaxy. On the other hand,  a circle of 10$^{\prime\prime}$ radius is chosen as a background from a nearby source-free region. The spectra are extracted using {\tt specextract}. We bin the data to have at least 1 count per bin and perform the fitting in XSPEC using C-statistic \citep[][]{1979ApJ...228..939C}. The uncertainties are estimated at 90\% confidence level.

\swift~XRT observations are analyzed following the standard methodology. We use the task {\tt xrtpipeline} to clean and calibrate the event files. Light curves are generated using the script {\tt xrtgrblc} which selects the source and background regions based on the count rate of the target. Piled-up observations, if any, point-spread function correction, and vignetting are properly accounted using {\tt xrtlccorr}. To generate the XRT spectrum combining various observations, the filtered event files and exposure maps are summed using {\tt xselect} and {\tt ximage} tools, respectively. We appropriately choose the sizes of the source and background regions depending on the source count rate \citep[][]{2013ApJS..207...28S} to extract the source and background spectra, respectively. The tool {\tt xrtmkarf} is used to generate the ancillary response files. The spectra are binned to have at least 1 or 20 counts per bin depending on the source brightness. For the low count spectrum, we adopt C-statistic and use a simple absorbed power-law model. The X-ray spectra of the bright sources are also fitted with an absorbed broken power-law model for the SED modeling\footnote{We choose the best-fitted model based on the {\tt f-test} and retains a broken power-law model if the null hypothesis probability of the {\tt f-test} is $<$10$^{-4}$.}. The neutral Galactic Hydrogen column density values are taken from \citet[][]{2005A&A...440..775K}. XSPEC is used to perform the fitting and we compute the uncertainties at 90\% confidence level.

\subsection{Optical-UltraViolet}\label{subsec:optical}
\swift-UVOT snapshot observations are first combined with the tool {\tt uvotimsum} to improve the signal-to-noise ratio. To perform the photometry using {\tt uvotsource}, we select the source and background regions as circles of 5$^{\prime\prime}$ (centered at the source) and 30$^{\prime\prime}$ (free from source contamination) radii, respectively. Following \citet[][]{2011ApJ...737..103S}, we de-redden the extracted source magnitude for the Galactic extinction and convert them to flux units using the zero-points of \citet[][]{2011AIPC.1358..373B}.

There are two {\it XMM-Newton} observations of the \gm-NLSy1 galaxy TXS 2116$-$077 \citep[][]{2018MNRAS.477.5127Y}. We use the pipeline {\tt omichain} to analyze the Optical Monitor (OM) data using the default parameter settings. Extracted source magnitudes are then corrected for the Galactic reddening and converted to flux units\footnote{https://www.cosmos.esa.int/web/xmm-newton/sas-watchout-uvflux}.

\subsection{Archival Measurements}\label{subsec:archive}
In addition to analyzing the data, as elaborated above, we also consider archival spectral measurements using ASDC SED Builder\footnote{https://tools.ssdc.asi.it/}. In particular, we are benefited with the {\it Wilkinson Microwave Anisotropy Probe} and {\it Planck}  data covering the high-frequency radio band and the far-IR data enabled by the {\it Wide-field Infrared Survey Explorer}. These data allow us to constrain the synchrotron peak of \gm-NLSy1 galaxies much better than done in the past. Since for many sources there are not good SED coverage available, the archival data points are also useful in getting an estimate about their average activity state which is the focus of our study.

\section{Leptonic Radiative Processes}\label{sec:model}
\subsection{The Model}\label{subsec:lepto}
The broadband SEDs of \gm-NLSy1 galaxies are modeled with a simple one-zone leptonic emission model \citep[e.g.,][]{2009ApJ...692...32D,2009MNRAS.397..985G}, whose main characteristics are briefly described here. We assume the central black hole of mass $M_{\rm BH}$ to be surrounded by a standard accretion disk \citep[][]{1973A&A....24..337S}\footnote{Considering a standard thin-disk model might be an oversimplification of the real accretion disk spectrum. The disk can be slim \citep[instead of thin,][]{1988ApJ...332..646A} which predicts a larger inner disk temperature and hence a bluer spectrum for high-accreting (moderately super-Eddington) systems \citep[][]{1996ApJ...458..474S}. However, none of the \gm-NLSy1 galaxies exhibit super-Eddington accretion. Another simplification is the assumption of the Schwarzschild, i.e., non-rotating, black hole. For a rapidly spinning black hole, the accretion efficiency will be even larger since the innermost stable orbit moves inwards. This again implies the accretion disk spectrum to peak at higher UV frequencies, and therefore it will be fainter at IR wavelengths. In order to match with the IR data, we will need to increase the black hole mass. Thus, it is possible that the mass derived from the disk modeling method could be slightly underestimated, but by an amount which is likely to be smaller than the overall uncertainties associated with the virial estimation.} of luminosity $L_{\rm disk}$ and inner and outer radii of 3 $R_{\rm Sch}$ and 500 $R_{\rm Sch}$, respectively, where $R_{\rm Sch}$ is the Schwarzschild radius. The presence of the X-ray corona, the broad-line region (BLR) and the dusty torus are also considered. The radiative profile of the corona is assumed to have power-law with an exponential cut-off \citep[$L_{\nu}  \propto \nu^{-\alpha_{\rm cor}}  \exp (- h\nu / 150\, {\rm keV)}$,][]{2009MNRAS.397..985G}. On the other hand, we assume both the BLR and torus as spherical blackbodies with spectral profiles peaking at rest-frame Lyman-$\alpha$ frequency and a characteristic torus temperature ($T_{\rm tor}$), respectively. The sizes of the BLR and torus are constrained from the accretion disk luminosity: $R_{\rm BLR}=10^{17}~L_{\rm disk, 45}^{1/2}$ cm and $R_{\rm tor}=2.5\times10^{18}~L_{\rm disk, 45}^{1/2}$ cm, respectively, where $L_{\rm disk,45}$ is the accretion disk luminosity in units of 10$^{45}$ \lum. 

The jet is assumed to have a conical shape with the semi-opening angle of 0.1 radian. The spherical emission region is considered to move with the bulk Lorentz factor $\Gamma_{\rm b}$ and cover the entire jet cross-section. Thus, its size is constrained from its distance from the central engine ($R_{\rm diss}$). The radiating particle population has a broken power-law energy distribution with slopes $p$ and $q$ before and after the break energy $\gamma_{\rm b}$, respectively. In the presence of uniform but tangled magnetic field, electrons radiate via synchrotron and inverse-Compton mechanisms \citep[e.g.,][]{2009herb.book.....D}. For the EC process, we compute the radiative energy densities of various AGN components in the comoving frame following \citet[][]{2009MNRAS.397..985G}. Moreover, various jet powers, i.e., kinetic ($P_{\rm kin}$), radiative ($P_{\rm rad}$), electron ($P_{\rm ele}$), and magnetic ($P_{\rm mag}$), are derived using the prescriptions of \citet[][]{2008MNRAS.385..283C}. A crucial assumption while computing the kinetic jet power, i.e., the power carried by cold protons, is the consideration of the equal number density of radiating electrons and protons. In other words, we assume the jet to be made of pure electron-proton plasma and no pairs.

\subsection{Accretion Disk Luminosity and the Black Hole Mass}
In the SED modeling of the strong-line blazars, two key parameters are $M_{\rm BH}$ and $L_{\rm disk}$. Conventionally, optical spectral line and continuum measurements are used to derive them \citep[see, e.g.,][]{2012ApJ...748...49S}. Furthermore, whenever the big-blue-bump is visible at optical-UV energies, it can be attributed to the accretion disk radiation \citep[e.g.,][]{2009MNRAS.397..985G}. By reproducing this emission with a standard disk model \citep[][]{1973A&A....24..337S,2002apa..book.....F}, both $M_{\rm BH}$ and $L_{\rm disk}$ can be calculated. It is found by studying a large sample of blazars that parameters derived from both methods reasonably agree with each other (e.g., P17).

The $M_{\rm BH}$ derived from the optical spectroscopy is argued to be underestimated due to the radiation pressure effect in highly accreting systems and also possibly due to a flat BLR geometry \citep[][]{2008MNRAS.386L..15D,2008ApJ...678..693M}. The latter effect is proposed to be more pronounced in the sources with a pole-on view, typically blazars and \gm-NLSy1s. The disk modeling approach, on the other hand, is free from these issues and relies mainly on the visibility of the big-blue-bump. In our SED modeling, we always consider the optical-UV bump to originate from the accretion disk. For the sources in which the accretion disk radiation is not visible and the optical-UV spectrum is rather synchrotron dominated, we start with the parameters obtained from the optical spectrum and modify it so as (i) not to overproduce the observations, and (ii) to maintain $0.01L_{\rm Edd}\lesssim L_{\rm disk}<L_{\rm Edd}$. The second constraint is driven by the fact that all \gm-NLSy1 galaxies have rest-frame EWs of emission lines larger than 5\AA~\citep[similar to FSRQs,][]{2017ApJS..229...39R}, thus indicating the underlying accretion process to be radiatively efficient \citep[see also,][]{2011MNRAS.414.2674G}.

\subsection{SED Modeling Guidelines}
The model used in this work does not perform any statistical fitting. Therefore, we cannot claim the uniqueness of the derived SED parameters. However, depending on the quality and availability of the multi-wavelength data set, they are fairly constrained. Below, we briefly explain the strategy adopted to determine the SED parameters from the observations.

We start by fixing $M_{\rm BH}$ and $L_{\rm disk}$ to the values derived from the modeling of the big blue bump or from the optical spectroscopy, keeping $L_{\rm disk}$ in the radiatively efficient regime. These two parameters regulate the behavior of the external photon energy densities and also the size of the BLR and torus. The low frequency ($\lesssim10^{11}$ Hz) radio data cannot be explained by the model due to synchrotron self absorption. Furthermore, it is straightforward to determine the level of the synchrotron emission from the optical-UV spectrum in those objects which do not exhibit the optical-UV bump. On the other hand, if the accretion disk emission is observed, an approximate level of the synchrotron emission can be computed from the high frequency radio data ($\gtrsim$10$^{12}$ Hz) and IR spectral measurements. The peak of the synchrotron spectrum enables us to compute the normalization of the broken power-law electron energy distribution \citep[see, e.g.,][]{2012MNRAS.419.1660S}. To parameterize the remaining variables of the particle population, we use both X-ray and \gm-ray spectral measurements. The shapes of the X-ray and \gm-ray spectra determine the low-and high-energy slopes, respectively, and also control the peak energy ($\gamma_{\rm b}$) of the radiating population. The shape of the optical spectrum, in synchrotron dominated sources, further regulates the high energy slope.

A Compton dominated SED reflects the prevalence of the external radiation energy densities (i.e., the BLR, torus) over magnetic one. Since these energy densities are a function of $R_{\rm diss}$, we are able to constrain the dissipation distance from the observed Compton dominance or CD\footnote{Compton dominance is the ratio of the inverse Compton to synchrotron peak luminosities.}. Moreover, the X-ray emission from \gm-NLSy1 galaxies can be complex due to possible contributions from the corona, SSC, and EC. Though the coronal radiation is soft, spectral behavior of SSC and EC can be contrasting since electrons of different energies are involved along with different seed photons for inverse Compton scattering. The level of SSC radiation dictates the magnetic field and the size of the emission region which are also connected to $R_{\rm diss}$. The observed X-ray to \gm-ray SED also enables us to constrain the bulk Lorentz factor $\Gamma_{\rm b}$ (or equivalently Doppler factor $\delta_{\rm b}$) due to $\Gamma_{\rm b}^2$ dependence of the external photon densities in the emission region frame \citep[][]{1995ApJ...446L..63D}.

\section{Results}\label{sec:results}
We have performed a detailed X-ray spectral fitting of nine \gm-NLSy1 galaxies that have good quality observations from \xmm, \suzaku, and/or \nustar. The primary motivation of this exercise is to throw more light on the {\it Seyfert} nature of these intriguing objects. On the other hand, the physical properties associated with the jet are studied and compared with blazars by reproducing the broadband SEDs with the conventional leptonic radiative model. Below, we present the results derived from these two analyses.

\subsection{X-ray Spectral Modeling}
All X-ray spectra are fitted in \textsc{xspec}, version 12.9.1p. If spectra from different instruments were taken simultaneously we fit them together, otherwise treat them independently.

\begin{figure}[!t]
\hbox{\hspace{2cm}
\includegraphics[width=0.515\linewidth]{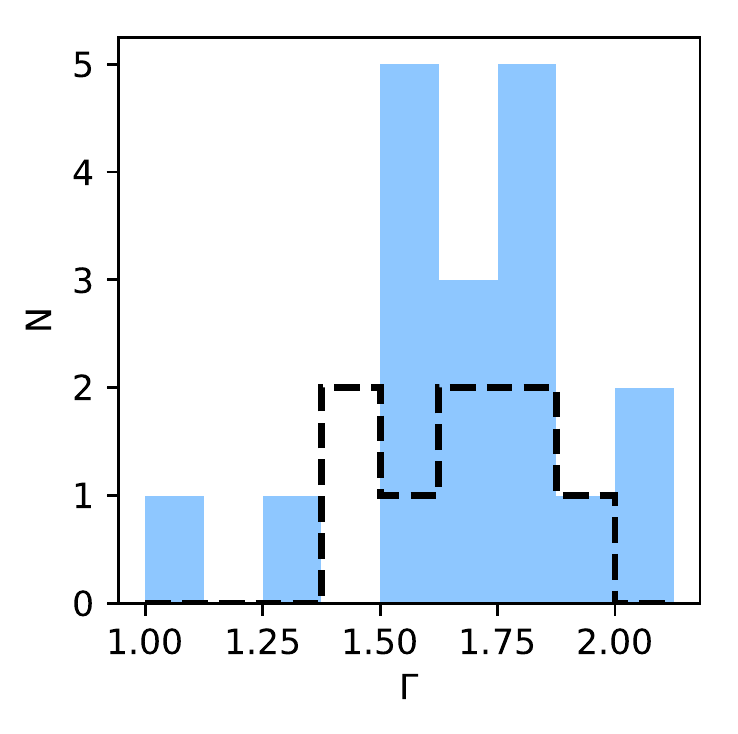}
}
\hbox{
\includegraphics[width=\linewidth]{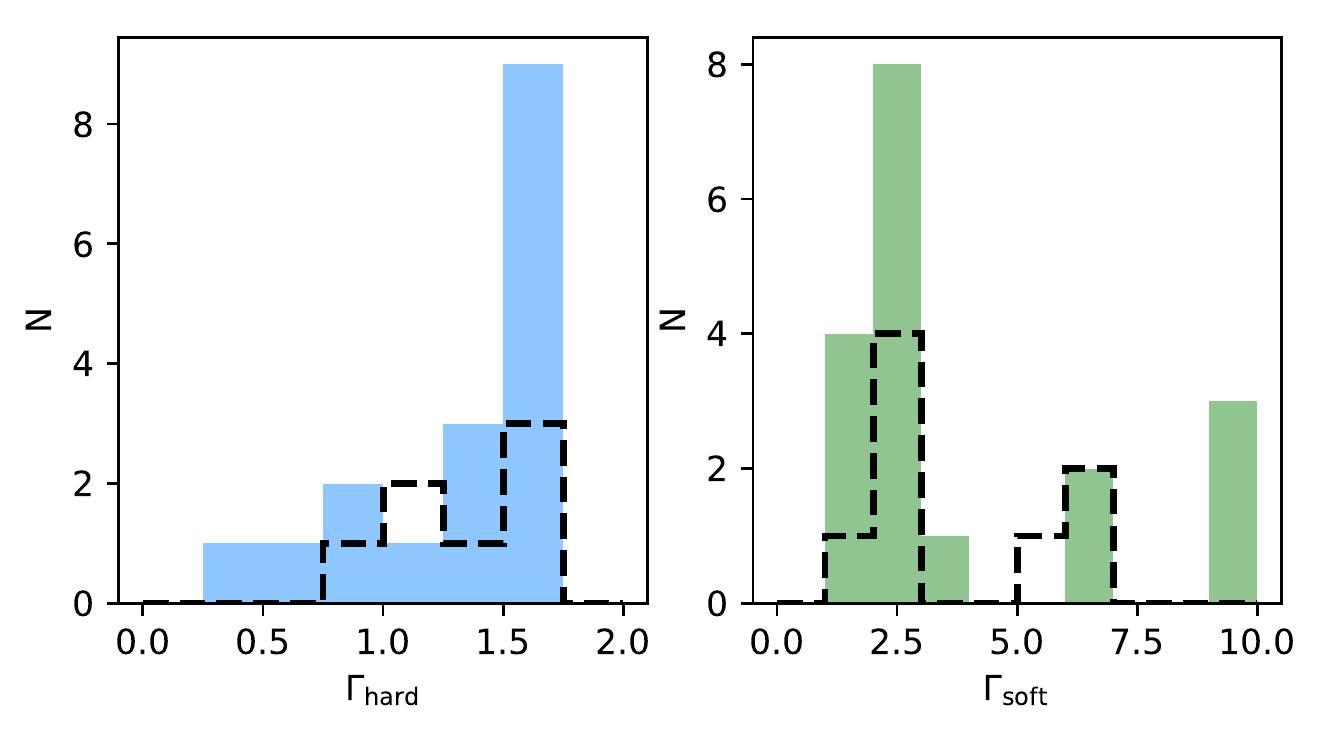}
}
\hbox{
\includegraphics[width=\linewidth]{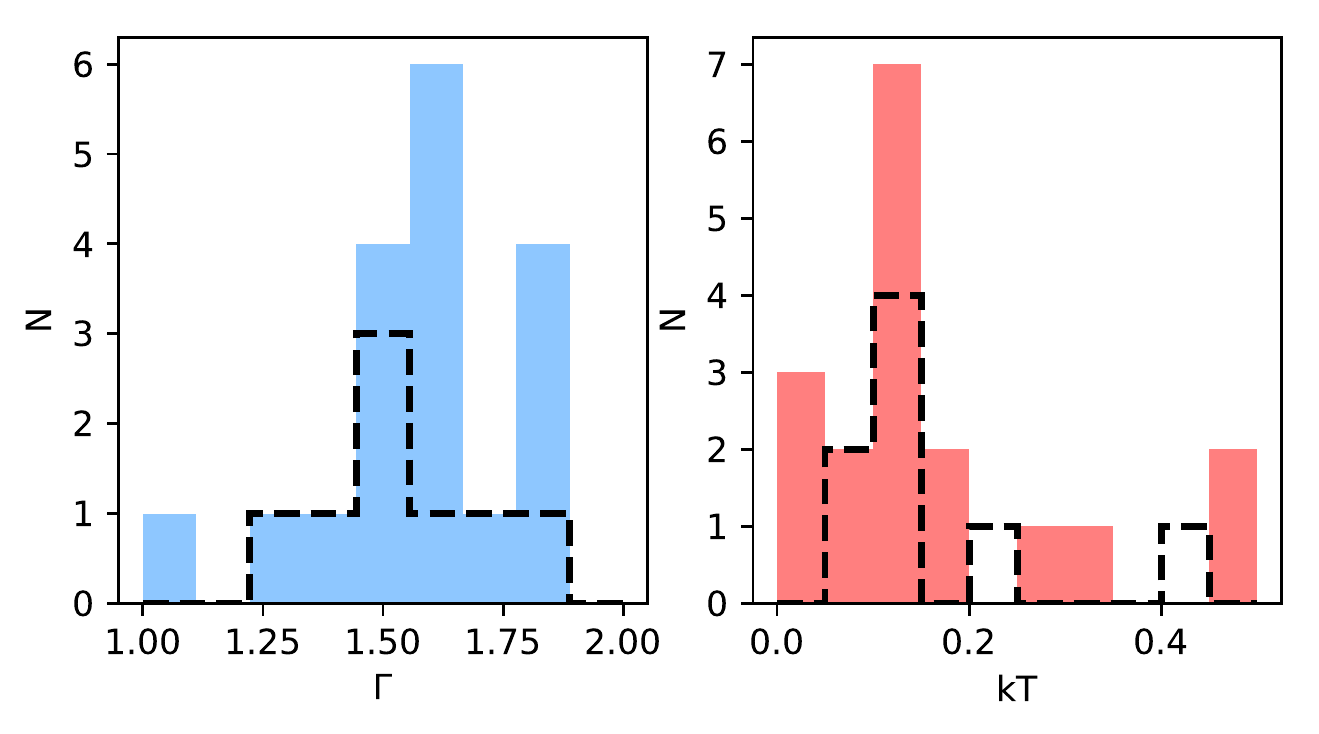}
}
\caption{Parameter distribution for the X-ray sample. The top row shows the photon indices for the single power-law fit, the middle row for the double power-law model, and the bottom row shows the photon index and temperature for the power-law plus black-body model. $\Gamma_{\rm hard}$ and $\Gamma_{\rm soft}$ are the photon indices after and before the break energy in the broken power-law model, respectively. In all cases, the shaded bars show the distribution for all observations, while the dashed line shows the distribution when the average value from all observations of a given source is used.}
\label{fig:xray_distributions}
\end{figure}

For each source, we fit three simple spectral models. We begin with a single power-law, modified by Galactic absorption \citep[modeled with TBnew:][]{Wilms00}. We then explore two additional more complex models: a double power-law, in case the spectrum has a contribution from both the jet and the corona, and a power-law plus black body model in case a strong soft excess is present\footnote{The one exception to this is FBQS J1102+2239, which only has three energy bins and thus cannot constrain models with more than three parameters. In this case, we fit only the single power-law model.}.
If these multi-component models are still insufficient to adequately describe the data, we consider each source individually and explore models based on the data residuals.

The spectra of all sources in the X-ray sample are shown in Figure~\ref{fig_xrayspectra_1} (and A2), and the residuals of the fits with the three simple models are shown in Figure~\ref{fig_xray_residuals} (and A4). We provide the fitted parameters in Table~\ref{table:xrayfitparams}. In most cases, these models do a good job of fitting the spectra: 10/15 have reduced $\chi^2$ values less than 1.2, with no obvious residuals. Of these 10, 4 are best fit with a double power-law, 2 with power-law plus black body, 1 with the single power-law, and 3 have equivalent best fits with the double power-law and power-law/black body models. The distributions of the spectral parameters from these fits are shown in Figure~\ref{fig:xray_distributions}.

For the single power-law fit, the distribution of $\Gamma$ values peaks at $\sim1.6$, with a fairly large scatter. When averaged by source, the distribution is consistent with being the same. When two power-laws are used, the harder $\Gamma$ still peaks at 1.6, with a tail down to very low values. These extremely low values are from the observations where a second power-law is not strongly required, and are very weakly constrained. The softer power-law distribution peaks at $\sim2.5$, with a few unconstrained values above 4. Finally, in the case of the power-law/black-body fits the powerlaw index again peaks at 1.6, and there is a broad distribution of temperatures, with the largest peak just above 0.1~keV.

\subsubsection{Individual sources}
The remaining five spectra are the 2009 and 2013 \suzaku\ spectra and the 2015 \xmm\ spectrum of 1H~0323+342, the 2012 \xmm\ spectrum of PKS~1502+036, and the 2012B \xmm\ spectrum of PKS~2004$-$477. In all three cases, the reason for the poor fit is an absorption-like feature around 0.5~keV. We note that a similar feature is visible in the 2016 \xmm\ spectrum of PKS~2004$-$477 as well, although all three models give a formally acceptable fit to this spectrum. We examine the spectra of these sources in more detail, using the double power-law model as a baseline.

\noindent\emph{1H~0323+342:}\\
\citet{2018MNRAS.475..404K} discuss the feature in 1H~0323+342, and show that the fit can be greatly improved by allowing a higher Galactic absorption column. We therefore re-fit the data, allowing for a higher column. This improves the fit to $\chi^2$/dof of 205/165, but leaves a narrow residual at 0.66~keV. Fitting this with a narrow Gaussian absorption line improves the fit by $\Delta\chi^2$ of 18, for 2 additional degrees of freedom. As the EPIC-pn count rate for this observation ($\sim4$~s$^{-1}$) is above the threshold for pileup in large window mode, we extract a spectrum from an annular source region, excluding the central 9 pixels. When we fit this annular spectrum with the same model, the feature remains, so we conclude that it is not due to pileup (and we note that \citeauthor{2018MNRAS.475..404K} found no evidence for strong pileup effects in the pn data). We also reduce the MOS spectra for this source for comparison. The two MOS spectra show no such feature, and while similar low energy residuals are also found in the Suzaku spectra,they are not at the same energy as each other or the pn so we conclude that this line is most likely an instrumental effect in the pn spectrum. 

\begin{figure*}
\centering
\includegraphics[width=0.4\linewidth]{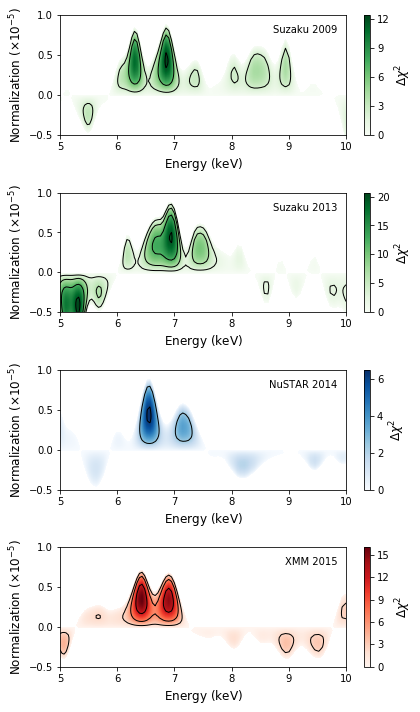}
\includegraphics[width=0.4\linewidth]{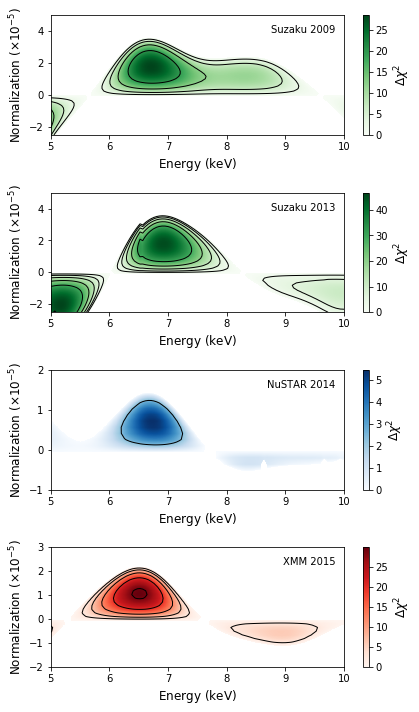}
\caption{Results of stepping a narrow Gaussian across the 5--10~keV band for the four X-ray observations of 1H~0323+342. The black lines correspond to 1--5$\sigma$. The left plots are for a scan with a Gaussian of line width $\sigma=0.01$~keV and the right plots are for a Gaussian of line width $\sigma=0.5$~keV.}
\label{fig:linescan}
\end{figure*}

In addition to the low energy residuals, 1H~0323+342 shows weak Fe~K emission, and is the only source in our sample which does so. An emission feature was found in the 2009 \suzaku\ spectrum \citep{Walton13}, time-averaged \swift\ XRT spectrum \citep{2014ApJ...789..143P}, and in both \suzaku\ spectra and the \nustar\ spectrum \citep{2018MNRAS.479.2464G} and interpreted as a relativistic line from the accretion disk. \citet{2018MNRAS.475..404K} found residuals in the 2015~\xmm\ spectrum consistent with two narrow emission lines at 6.4 and 6.95~keV (the neutral K$\alpha$ line and the Fe\textsc{xxvi} K$\alpha$ line).
 
We investigate the residuals in this band with a blind line scan \citep{Miniutti06}. We step a narrow Gaussian across the 5--10~keV band in 100 steps and across a range of normalizations and record the $\Delta\chi^2$ at each point. We then plot contours corresponding to the 1,2 and 3$\sigma$ confidence levels for two free parameters (Figure~\ref{fig:linescan}). All four observations show emission lines between 6--7~keV, although the line significance for \nustar\ is only $\sim2\sigma$. Interestingly, the line shapes differ between the observations: the 2009 \suzaku\ and 2015 \xmm\ spectra show two peaks, while the 2013 \suzaku\ spectrum shows a single broader feature peaking at 7~keV and the \nustar\ spectrum shows a single feature at $\sim6.6$~keV. While this is suggestive of complex line structure and variability, this should be treated with caution. As an additional test, we repeat the line scan with a much broader $\sigma=0.5$~keV line, and in this case an equivalent or better best-fit is returned in all cases and the line energies are consistent between the different instruments.
 
While it is still possible that there are multiple lines and significant variability, the simplest explanation is a single fairly broad line, and the appearance and disappearance of double line is due to noise. A recent 300~ks \xmm/\nustar\ campaign shows a single broad line (Mundo et al., in prep), and is by far the highest signal X-ray dataset for this source. 

Based on this result, we jointly fit all four spectra with a relativistic line profile \textsc{relline} \citep{Dauser10}. We fit all the spectra simultaneously with the double power-law model, with a free Galactic column, and \textsc{relline}. We allow the power-law parameters and the line normalization to vary between spectra, but tie the relativistic blurring parameters to get the best constraints out. We fix the emissivity to the classical value of 3, and require that the inclination be $<20$\textdegree. We fix the spin to the maximal value and fit for the inner radius of the disk, $R_\mathrm{in}$, to allow for the possibility that the disk is truncated. The power-law parameters for each spectrum are consistent with the values given in Table~\ref{table:xrayfitparams}. The parameters of \textsc{relline} are given in Table~\ref{tab:relline}. The inclination is low, consistent with the AGN being viewed down the jet axis, and the inner radius suggests that the disk is moderately truncated. The remaining residuals are largely due to weak features in the \suzaku\ spectra, which are not present in the other instruments and are likely due to calibration features.

\noindent\emph{PKS~1502+036:}\\
In this case, there is no improvement to the fit by freeing the column density of the Galactic absorption, and the MOS spectra are consistent with the pn. It is likely, therefore, that there is a genuine absorption feature in the spectrum of PKS~1502+036. Fitting with a narrow Gaussian line improves the fit by $\Delta\chi^2$ of 17, for 2 additional degrees of freedom, and gives an energy of $0.57\pm0.01$~keV ($\sim22$~\AA) in the observed frame. This is the rest frame energy of the 1s--2p line of O\textsc{vii}, so the feature is most likely produced by additional Galactic O\textsc{vii} absorption not accounted for in the model.

\noindent\emph{PKS~2004$-$477:}\\
This object shows a similar low energy feature as the other two sources, and the pn spectrum shows some additional possible features around 2--3~keV. These higher energy residuals are not present in the MOS data however, so we focus on the low energy feature.
As with PKS~1502+036, freeing the column density does not improve the overall fit. Including a narrow line with the energy fixed at 0.57~keV improves the fit by $\Delta\chi^2$ of 6, for 1 degree of freedom, but does not remove the residuals and the fit is still not acceptable ($\chi^2_\nu=1.38$), so the residual cannot be due simply to Galactic O\textsc{vii}. If we free the energy and width of the line, the fit improves to $\chi^2$/dof $=66/62$, and no strong residuals remain, but the energy of the line is constrained only to $E<0.55$~keV and the line width to $\sigma=0.45_{-0.25}^{+0.05}$~keV. These parameters are not consistent with a single absorption line.

Freeing both the column density and oxygen abundance of the Galactic absorption gives an acceptable fit ($\chi^2/$dof of 73/63), but the required column density is $(9\pm2)\times10^{20}$~cm$^{-2}$ (three times the literature value) and the oxygen abundance is $4\pm1$ times solar, so this is unlikely to be the correct interpretation.

The other possibility is that the absorption is intrinsic to the source. Because of the source redshift ($z=0.24$) the feature is too high energy to be due to neutral absorption, but absorption from ionized gas could produce such a feature. Fitting for warm absorption in the rest frame of the source with a general \textsc{xstar} grid from \citet{Walton13}, we find that the fit is greatly improved ($\chi^2/$dof of 72/63) relative to just Galactic absorption. 

\subsection{Broadband SED and Modeling}
\begin{figure*}[!t]
\includegraphics[scale=0.47]{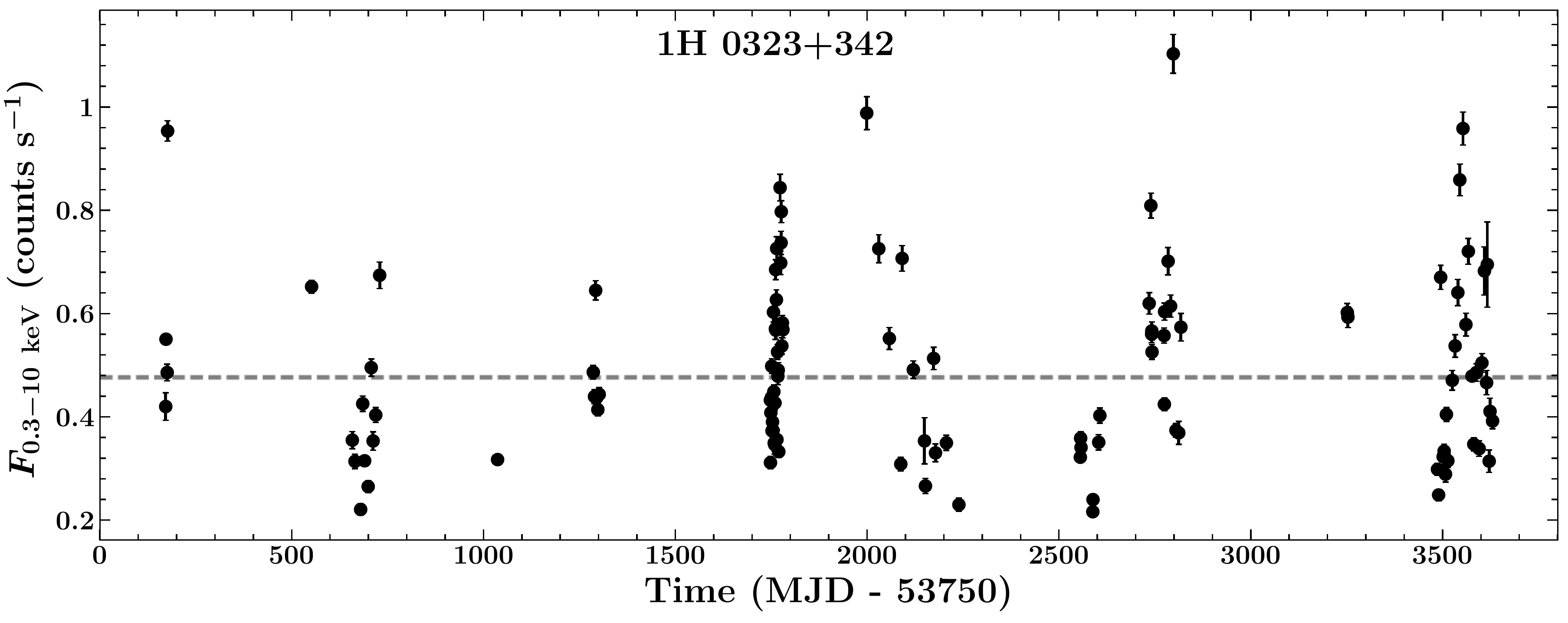}
\caption{\swift-XRT (0.3$-$10 keV) light curve of the \gm-NLSy1 galaxy 1H 0323+342. Horizontal dashed line represents the mean flux level of the source. Only those observations are combined for SED modeling in which the source remain below the mean flux value.\label{fig:xrt_lc}}
\end{figure*}

The \gm-ray spectral parameters of \gm-NLSy1 galaxies provided in \citet[][]{2018ApJ...853L...2P} are derived from the time-averaged analysis of the LAT data covering first $\sim$8.5 years of the \fermi~mission. Though a few \gm-NLSy1 sources have shown GeV flares during this period \citep[e.g.,][for 1H 0323+342]{2014ApJ...789..143P}, they remain in an elevated flux state for a very short period of time ($<$a week). Therefore, results obtained from the overall \fermi-LAT data analysis can be considered as representatives of the average flux state of the sources. On the other hand, compared to all-sky survey mode operation of the \fermi-LAT, low frequency X-ray observations are taken in the pointed mode from \swift-XRT. Therefore, one should also consider the possible impact of variability when combining many XRT observations. Among sixteen \gm-NLSy1 galaxies, only four (1H 0323+342, PMN J0948+0022, PKS 1502+036, and PKS 2004$-$447) have more than 15 \swift~observations so far. For these objects, we generate 0.3$-$10 keV light curves and select only those data sets in which the source fluxes remain below the mean flux level derived from the light curve analyses. The selected observations are then combined to generate the X-ray SED representing an average flux state of the \gm-NLSy1 galaxy. We demonstrate this approach for 1H 0323+342 in Figure \ref{fig:xrt_lc}. Remaining objects have a few and sporadic X-ray observations with no reported flares. Therefore, we combine all of their XRT pointings to generate a meaningful spectrum. TXS 2116$-$077 was observed twice with {\it XMM-Newton} and have no existing XRT measurements. We use the second {\it XMM-Newton} data (obs id: 0784090301) as it represents a relatively low activity state of the source \citep[][]{2018MNRAS.477.5127Y}.

\begin{figure}[t!]
\includegraphics[width=\columnwidth]{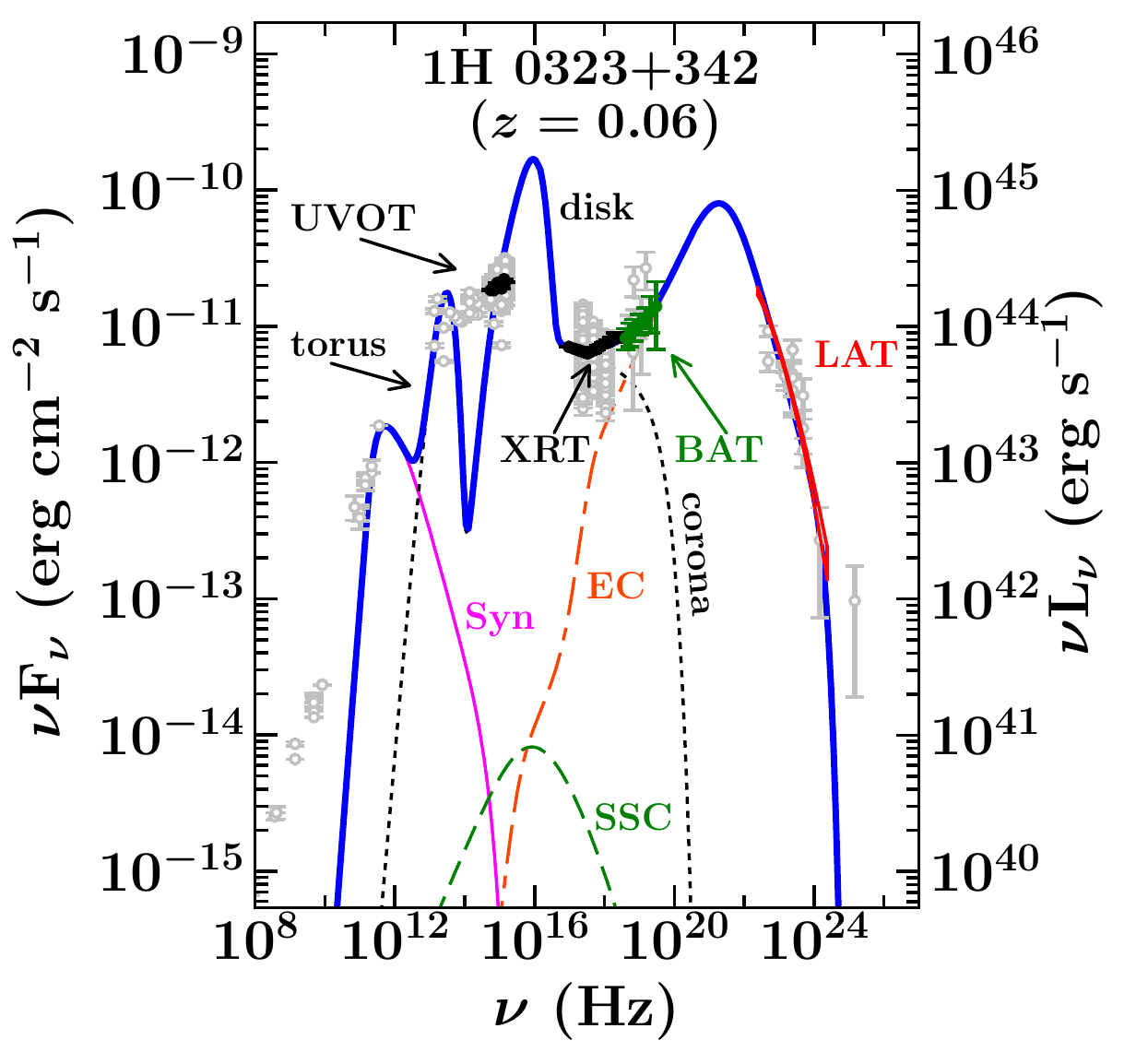}
\caption{The modeled broadband SED of 1H 0323+342. The grey filled circles represent the archival observations. Black and green data points correspond to the observations analyzed by us. The instruments used are also labeled. At the \fermi-LAT energy range, the red bow-tie plot is generated using the spectral parameters reported in \citet[][]{2018ApJ...853L...2P}. The combined thermal radiations from the torus, the accretion disk, and the X-ray corona are displayed with the black dotted line. The synchrotron, SSC, and EC emissions are denoted with pink thin solid, green long-dashed, and orange dash-dash-dot lines, respectively. The summed radiation is shown with the blue thick solid line. The modeled SEDs of other \gm-NLSy1 galaxies are provided in the appendix Figure \ref{fig:app_SED1} and \ref{fig:app_SED2}.\label{fig:SED}}
\end{figure}

For objects with both the \swift-BAT and \nustar~data, we prefer the former as it provides a hard X-ray spectrum averaged over a long-period of time. We use the \nustar~data for the SED modeling of those sources which remain below the \swift-BAT detection threshold. The \gm-NLSy1 galaxy PKS 2004$-$447 was observed twice with \nustar. We use the second observation (obs id: 60201045002) for the SED modeling as it has a longer exposure and a relatively lower 3$-$79 keV flux with respect to the first pointing.

The X-ray spectral parameters associated with the SED modeling are provided in Table \ref{tab:x-ray_sed} and \ref{tab:hard-x-ray_sed}. CGRaBS J0932+5306 and FBQS J1102+2239 do not have good quality XRT data. Therefore, we use their {\it Chandra} observations. 

The modeled broadband SED of one of the \gm-NLSy1 source, 1H 0323+342, is shown in Figure \ref{fig:SED}. Remaining SEDs are presented in Figure \ref{fig:app_SED1} and \ref{fig:app_SED2} in appendix. We provide the SED parameters and jet powers computed from the modeling in Table \ref{tab:sed_param} and \ref{tab:jet_param}, respectively.

The \gm-ray spectra of all sources are steep, implying their inverse-Compton peak to be located below 100 MeV. Availability of the hard X-ray measurements for a few \gm-NLSy1 galaxies, either from the \nustar~or BAT (Table \ref{tab:hard-x-ray_sed}), along with the LAT spectrum ensures a rather precise estimation of the high-energy peak. The hard X-ray spectra are found to be dominated by the jet emission, whereas, the soft X-ray band often shows an additional contribution from the X-ray corona emission, particularly below 2 keV. The optical-UV spectrum of a majority of the sources (12 out of 16) exhibits a bump which we interpret due to accretion disk radiation. For the remaining sources, synchrotron emission dominates.

\section{A Comparison with Seyferts}\label{sec:dis_seyfert}
\begin{figure}
\includegraphics[width=\linewidth]{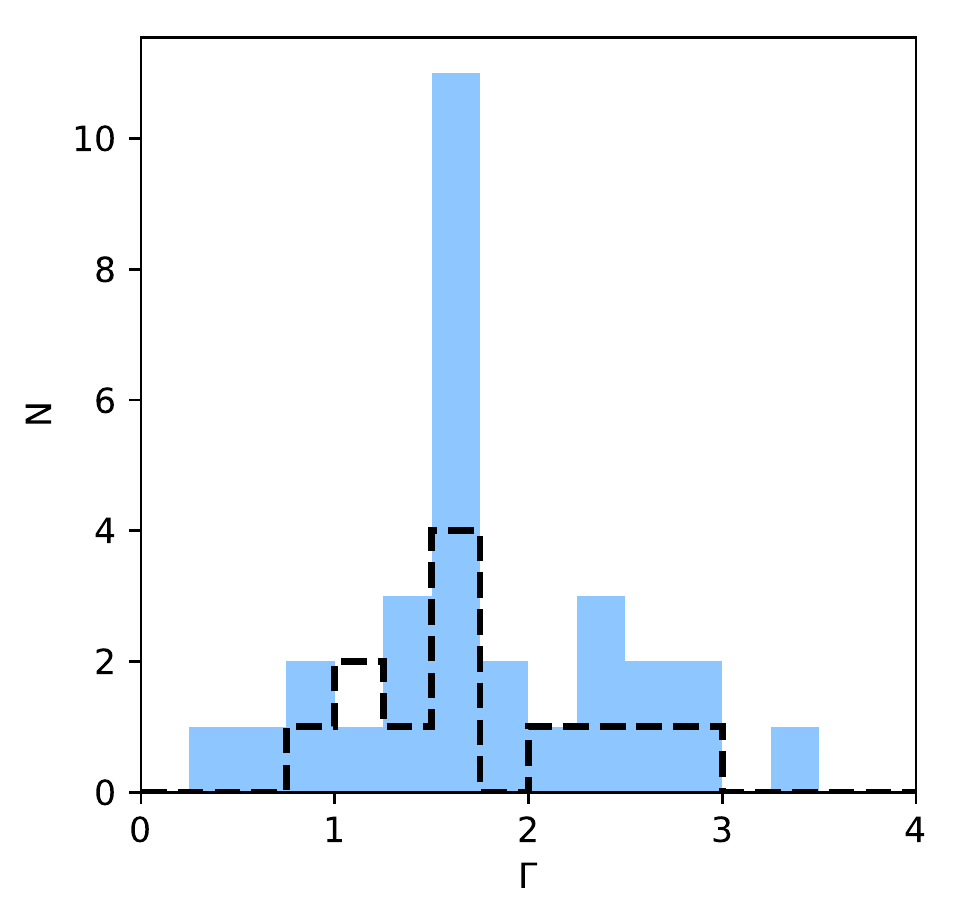}
\caption{Distribution of photon indices for the double power-law fit, as in the middle row of Figure~\ref{fig:xray_distributions}, but showing all values on a single plot and excluding the unconstrained and unphysical values of $\Gamma>4$. This clearly illustrates two peaks, one at $\sim1.5$, corresponding to the jet emission, and one at $\sim2.5$, most likely corresponding to classical NLSy1 X-ray emission.}
\label{fig:2pl_zoom}
\end{figure}

The largest difference between the X-ray properties of the \gm-NLSy1s and the regular NLSy1s is the presence of jet emission in the X-ray spectrum of the former. This results in extremely hard, power-law dominated spectra. 
Fitting with a single power-law gives a distribution of indices peaking at $\Gamma\sim1.6$ (Figure~\ref{fig:xray_distributions}). This is much harder than the typical value for NLSy1s of $\sim2.5$, significantly softer than the general AGN population \citep[e.g.,][]{Brandt97, Grupe10}. 
When we fit with two power-laws, we find a bimodal distribution, with one peak at $\sim1.6$ and a second weaker peak at $\sim2.5$ (Figure~\ref{fig:2pl_zoom}). 

This is most likely due to the simultaneous presence of emission from the corona and jet, and the second peak is weaker because the coronal emission is not detected in all sources and observations. The second peak is consistent with the unusually soft spectra of normal NLSy1s, which suggests that the underlying spectrum is consistent between the \gm-ray loud and quiet sources. In almost every case where the second power-law improves the fit and is well constrained, the index lies in the range 2.3--2.9, with the sole exception of the 2013 \suzaku\ observation of 1H~0323+342, where it is softer ($\Gamma=3.3\pm0.1$). 

The distribution of black body temperatures is fairly narrow. We can discount the obvious outliers with $kT>0.2$~keV, as in all cases these are unconstrained and not required by the data. The remaining values are typically in the range 0.5--0.15~keV, which is somewhat softer than the typical value of 0.1--0.2 found in other AGN, independent of mass or accretion rate \citep[][]{2004MNRAS.349L...7G}. In reality, the underlying AGN continuum should be a combination of a soft excess and a power-law, but given the limited data quality and jet dominated spectra it is not possible to disentangle these two components. In most cases, the fits are equivalent between the double power-law and power-law black body models, but we note that the 1H~0323+342 2009 and 2013 \suzaku\ spectra are significantly better fit with a black body than a power-law. This may explain the unusually soft photon index recovered for the double power-law model in the 2013 spectrum mentioned above, as this is likely the result of fitting a power-law to a spectrum with a strong soft excess. In both of these cases, and in the 2015 \xmm\ spectrum, the temperatures we find are in the range 1.3-1.5, consistent with the values from other AGN. Conversely, the lower than normal temperatures we find for the other sources may be due to fitting a power-law continuum with a black body. We do not have sufficient evidence in general to argue for a difference in the soft excess temperatures between our sample and AGN/NLSy1s in general.

\subsection{Excess absorption}
In all the cases where we find a poor fit ($\chi^2_\nu>1.2$) this is attributable to extra absorption, not included in the model. This is likely more obvious than in standard NLSy1s, which have more complex spectra and are therefore less sensitive to the Galactic column due to degeneracies with other parameters. For 1H~0323+342 and PKS~1502+036, the reason for the low energy residuals in the spectrum is likely attributable to the Galactic column being different from the assumed literature value. 

For the third source, PKS~2004$-$477, the absorption appears to be intrinsic, and due to a small amount of warm absorption. This obviously requires that the line of sight to the source intersect with the absorbing gas, and also that there is enough of a contribution from the `standard' Seyfert X-ray spectrum to show the absorption (assuming the absorption affects the softer emission from the accretion disk/corona and not the emission from the jet). With further data, particularly with future missions such as \emph{Athena} and \emph{XRISM}, it should be possible to fully explore the nature of this absorption and probe the connection between it and the jet.

\subsection{Fe emission}
Only one source in our X-ray sample shows evidence for Fe~emission, 1H~0323+342. The origin of this emission is not clear. Previous authors have interpreted this weak emission as a relativistic disk line \citep{Walton13, 2014ApJ...789..143P, 2018MNRAS.479.2464G} and as multiple narrow lines \citep{2018MNRAS.475..404K}. We used blind line scans to examine all the publicly available data, and find that the data can be equivalently well fit with either two narrow lines that are transient and move in energy, or a single broad line. We favor the latter interpretation, because in this case the line is consistent between the different observations, but cannot rule out a complex combination of variable narrow lines.
 
Assuming that this is correct, we fit a relativistic disk line model to the data, with a fixed high spin, and find a low inclination ($<7$\textdegree, although note that we restrict the parameter range to $<20$\textdegree) and evidence for moderate truncation of the disk. This differs from the results of \citet{2014ApJ...789..143P} and \citet{2018MNRAS.479.2464G}, largely because in this case the soft excess is fit with a second power-law, whereas in previous reflection models it has generally been assumed that the soft excess is due to reflection. It is hard to say which of these is the correct approach, but this will likely be resolved with further data, such as the 300~ks \xmm\ campaign in 2018 (PI Kara; Mundo et al., in prep).

\subsection{Variability}
Although it is difficult to be sure given the limited X-ray data available, our sample appears to be less variable than the other NLSy1s, which are famous for their extreme X-ray variability. Because of their low masses and high accretion rates, NLSy1s have been observed to change X-ray flux by over an order of magnitude in a few hours \citep[e.g.,][]{Boller97,Komossa17}

Of the sources presented here, the largest change in flux is a factor of 3 between the 2004 and 2012A spectra of PKS~2004-477, and there is no significant change in spectral shape associated with this, compared to the drastic changes observed in other NLSy1s. The other sources in our sample show at most a factor of 2 in spectral variability, with no large changes in spectral shape.
There are two possibilities for this behavior: either the classical Seyfert part of the X-ray spectrum is as variable as in other NLSy1s but is not observed because of the relatively constant jet emission, or there is a qualitative difference between these and other Sy1s, which may be related to the jet launching process.

In the former case, we would therefore expect to primarily see variability in the soft band, where there is a stronger contribution from the corona. This is generally not the case: the sources with multiple observations show uniform flux changes through the bandpass, although PMN~J0948+0022 is slightly harder in the lower flux observation. However, if the spectrum is totally jet dominated then there should be no such energy dependence.

If there is a qualitative difference in the variability between these and classic NLSy1s then it would indicate that the \gm-ray loudness is not simply an inclination effect (i.e., not all NLS1s have jets). One explanation for the reduced variability would be truncation of the accretion disk, associated with the jet launching \citep[e.g.,][]{Lohfink13}. This is consistent with the large inner radius we measure from the iron line, but both that result and the limited variability are very tentative and further study is needed.

\begin{figure*}[ht!]
\hbox{
\includegraphics[scale=0.4]{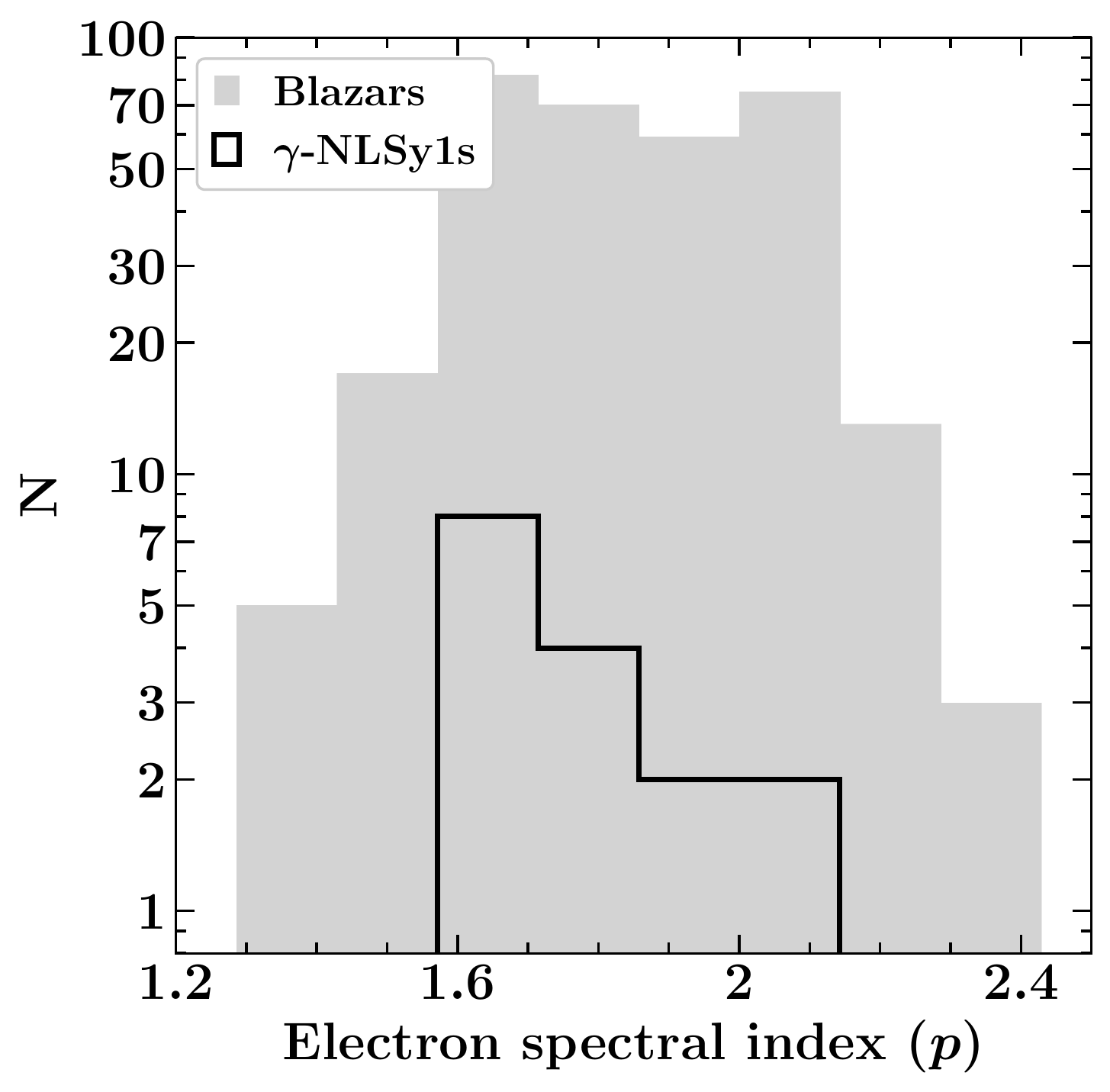}
\includegraphics[scale=0.4]{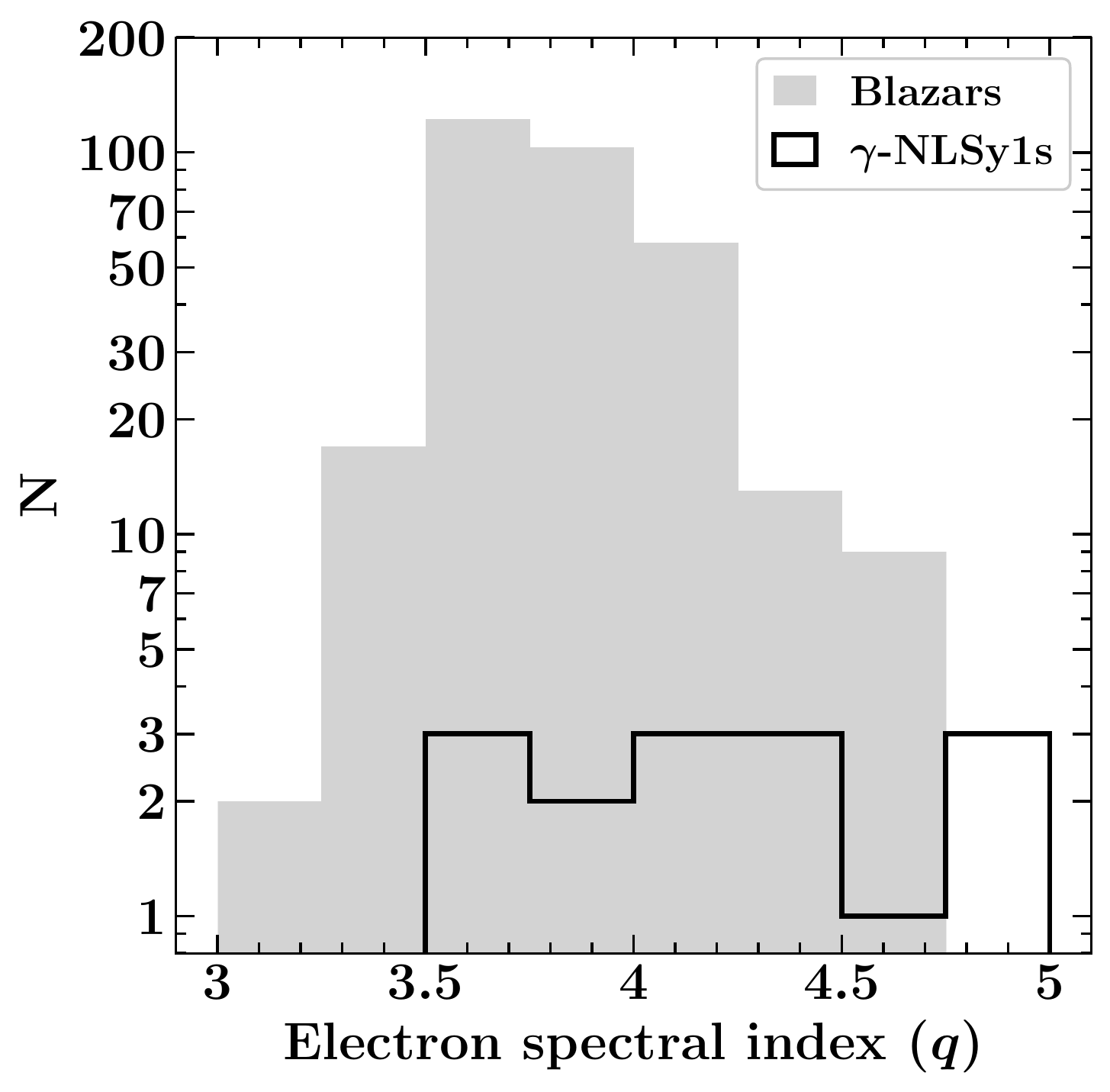}
\includegraphics[scale=0.4]{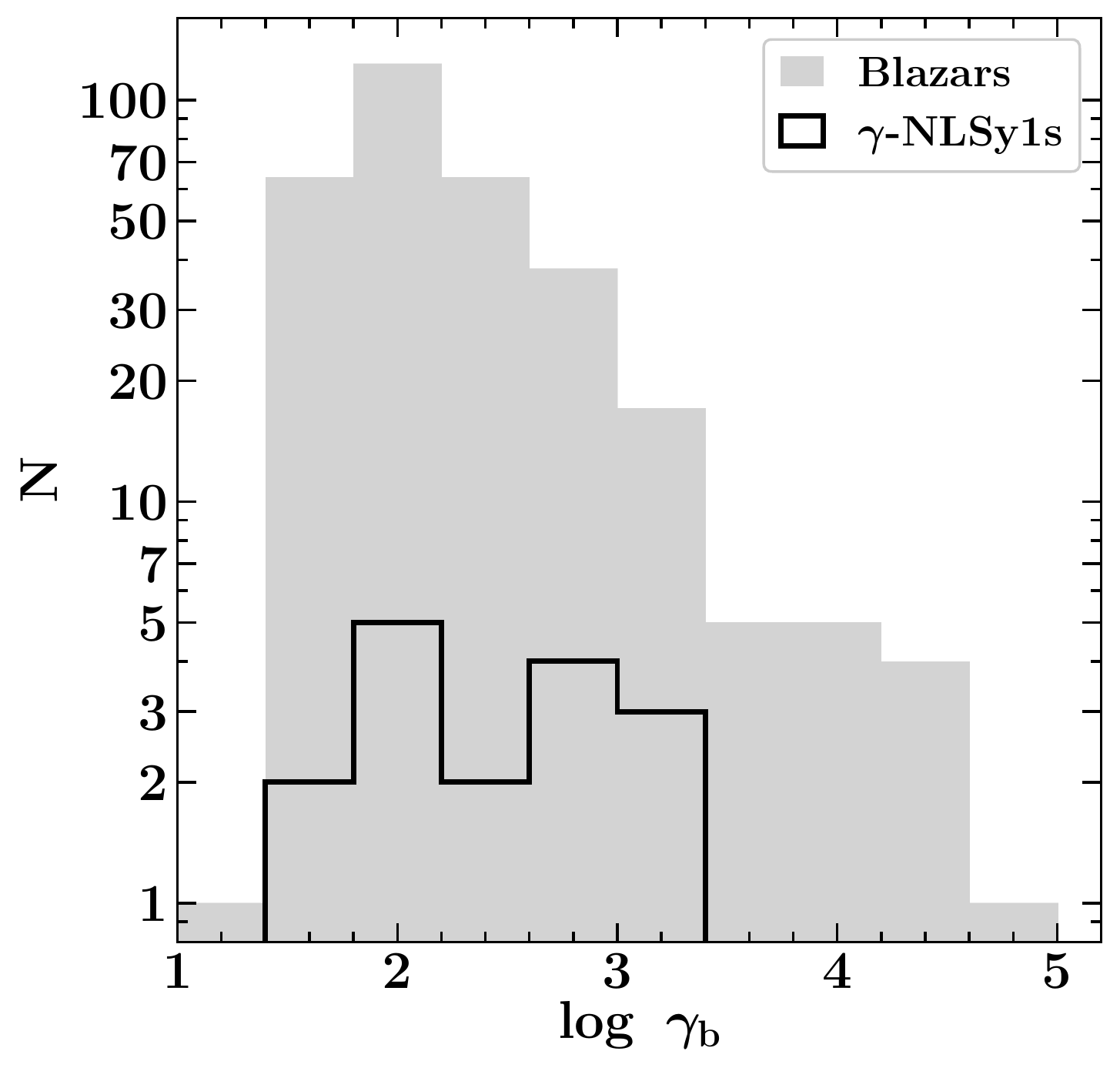}
}
\hbox{
\includegraphics[scale=0.4]{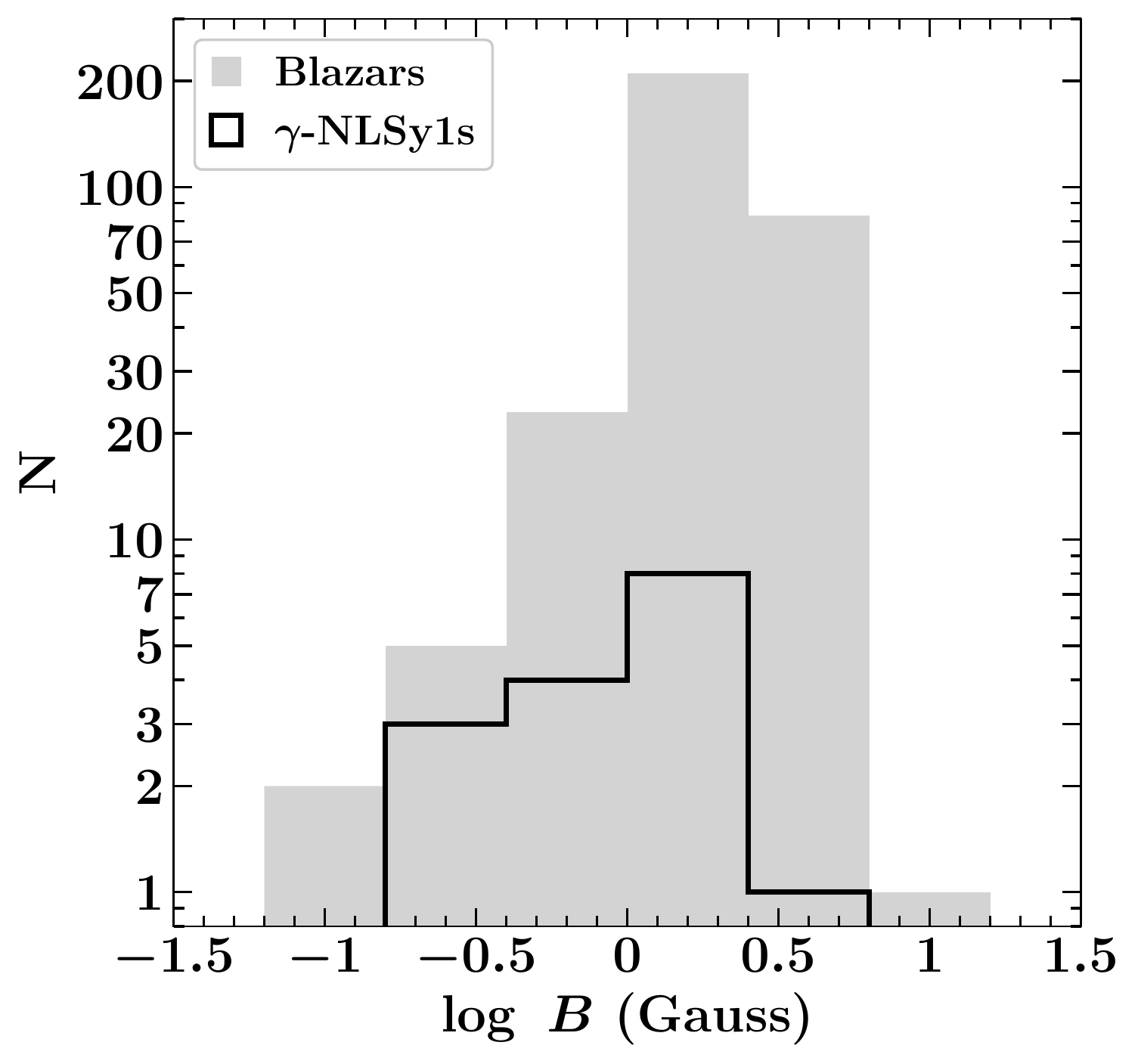}
\includegraphics[scale=0.4]{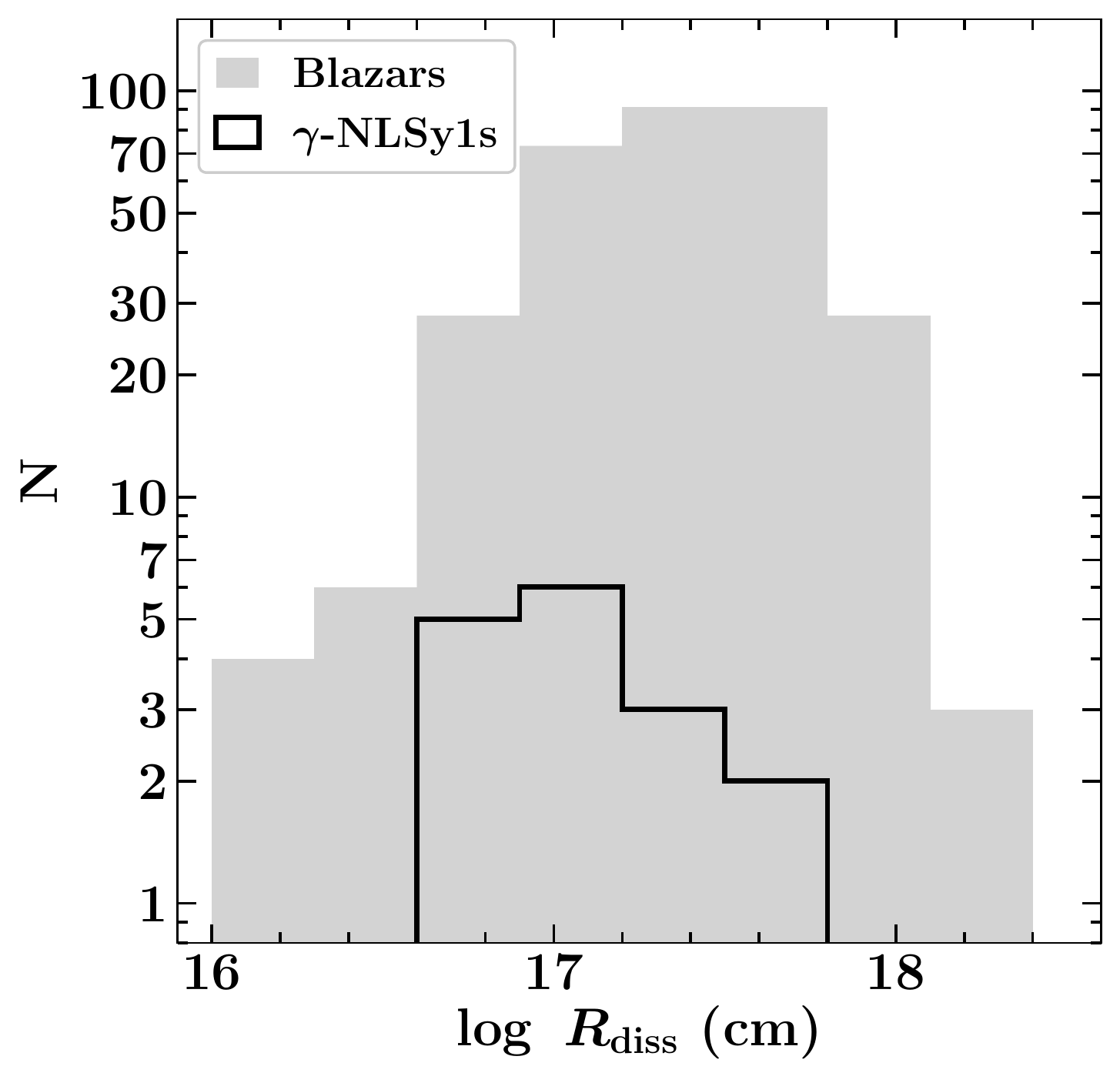}
\includegraphics[scale=0.4]{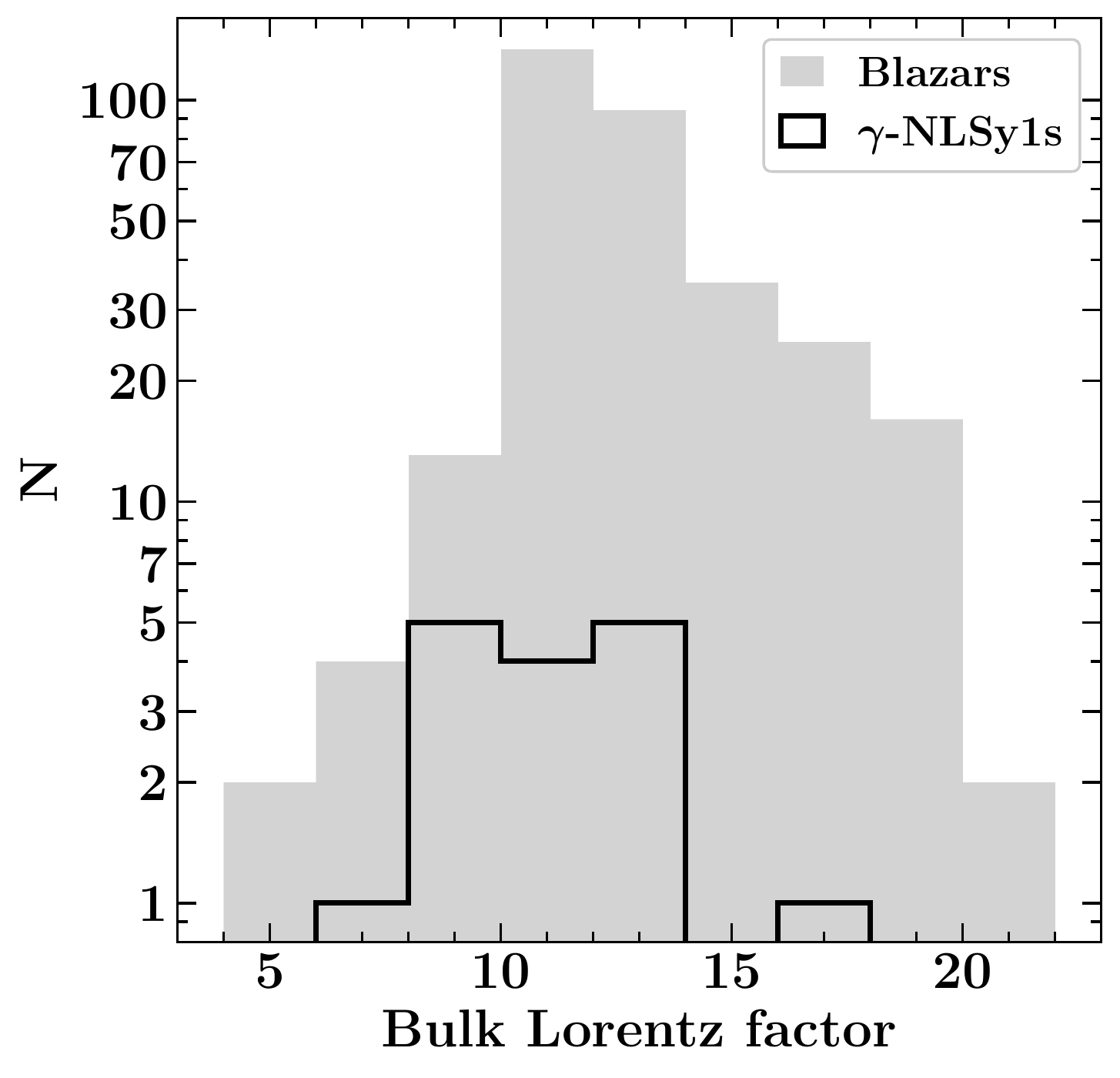}
}
\hbox{\hspace{3.0cm}
\includegraphics[scale=0.4]{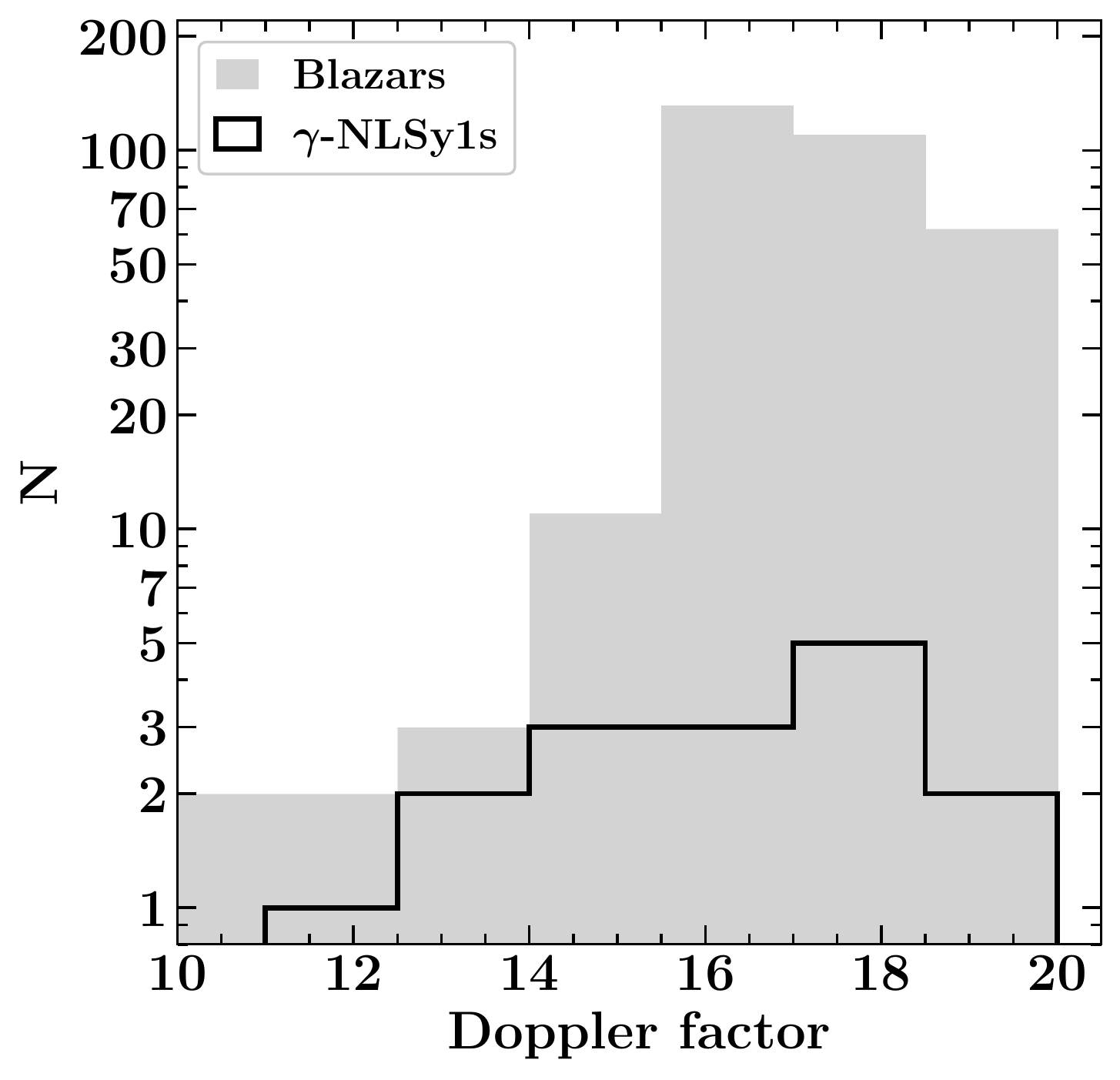}
\includegraphics[scale=0.4]{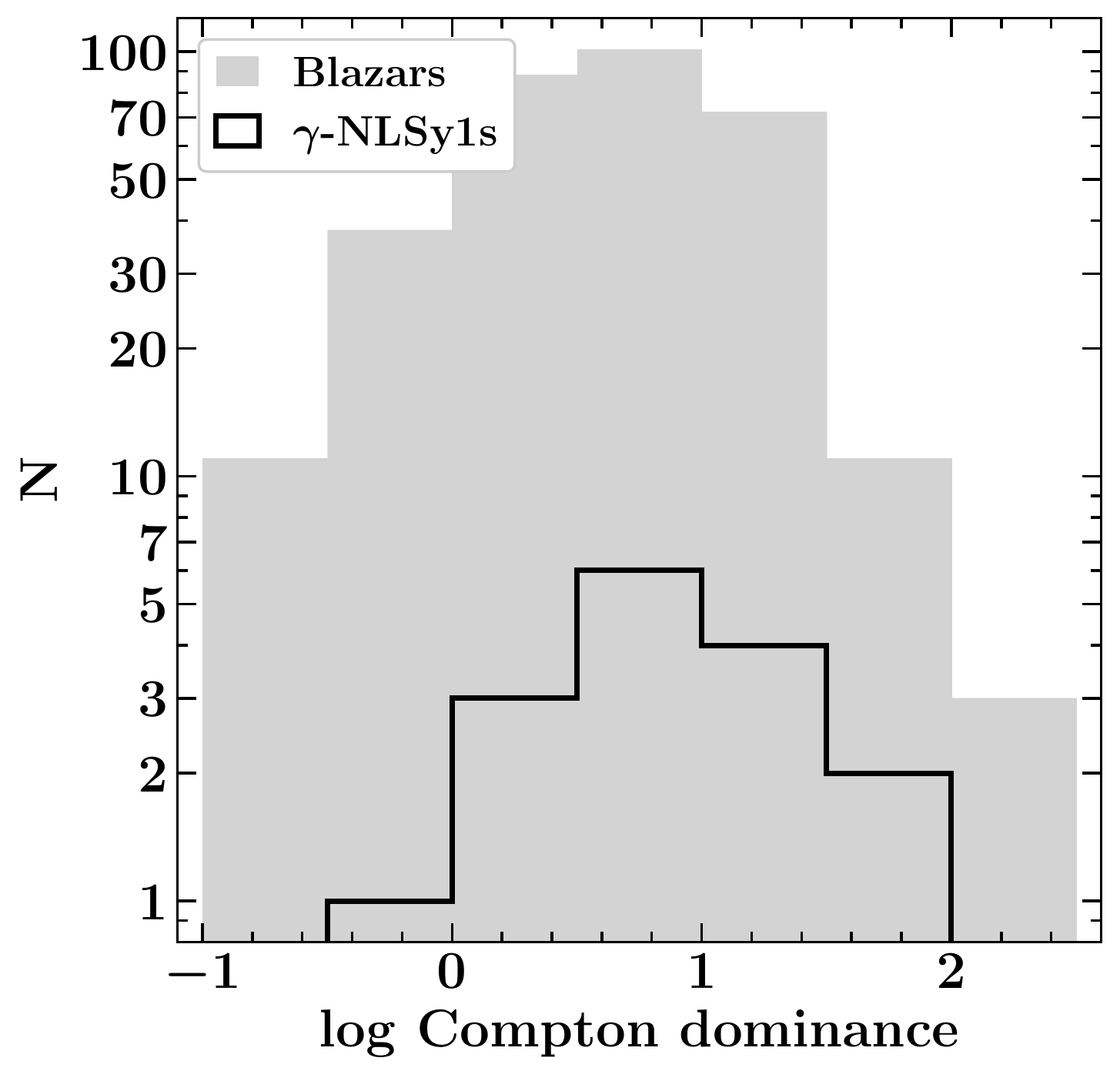}
}
\caption{Histograms of the parameters derived from the SED modeling of \gm-NLSy1 galaxies. Distributions of the electron spectral indices before ($p$, top left) and after ($q$, top middle) the break energy ($\gamma_{\rm b}$, top right), magnetic field (middle left), distance of the emission region in cm (middle center) and the bulk Lorentz factor (middle right). Bottom panel refers to the Doppler factor (left) and the Compton dominance (right). Grey shadow histograms correspond to \gm-ray emitting blazars studied in P17.\label{fig:sed_param}}
\end{figure*}

\section{A Comparison with Blazars}\label{sec:dis_blz}
\subsection{Broadband Spectral Parameters}\label{subsec:dis_sed}
The histograms of the slopes of the broken power-law indices, $p$ and $q$, for \gm-NLSy1 galaxies are shown in the top left and middle panels of Figure \ref{fig:sed_param} with black solid lines. For a comparison, we also show the distributions for \gm-ray emitting blazars studied in P17 with gray-filled histograms. On average, $p$ has a similar value for both \gm-NLSy1 and blazar populations: $\langle p \rangle=1.77\pm0.16,~1.83\pm0.19$, respectively. On the other hand, former exhibits a relatively steep high energy slope ($\langle q \rangle=4.27\pm0.49)$ compared to $3.80~\pm~0.27$ obtained for blazars. Since $q$ is mainly constrained from the \gm-ray spectrum, it suggests a steeper spectrum in the \gm-ray band for \gm-NLSy1 objects. This could also be one of the reasons for the \fermi-LAT detection of a very few radio-loud NLSy1 galaxies as the hard \gm-ray spectrum objects are easier to detect up to very high energies \citep[][]{2016ApJS..222....5A}. Furthermore, the peak of the particle energy distribution or $\gamma_{\rm b}$ (Figure \ref{fig:sed_param}, top right panel) reflects a similar averaged value of $\langle \log\gamma_{\rm b} \rangle=2.42\pm0.54$ and $2.28\pm0.64$ for \gm-NLSy1 and blazar populations, respectively. Interestingly, the largest \gm$_{\rm b}$ for \gm-NLSy1 objects is $<$2000, whereas, the blazar distribution extends beyond 10000. This is due to the inclusion of a few BL Lac objects in the blazar sample that have synchrotron peaks located at very high frequencies ($>$10$^{15}$ Hz, P17) and thus a large \gm$_{\rm b}$. It is important to emphasize that a steep $q$ and a low value of \gm$_{\rm b}$ obtained for \gm-NLSy1 galaxies indicates them to be low-synchrotron peaked beamed AGNs similar to FSRQ class of blazars.

\begin{figure*}[t]
\hbox{\hspace{3.0cm}
\includegraphics[scale=0.4]{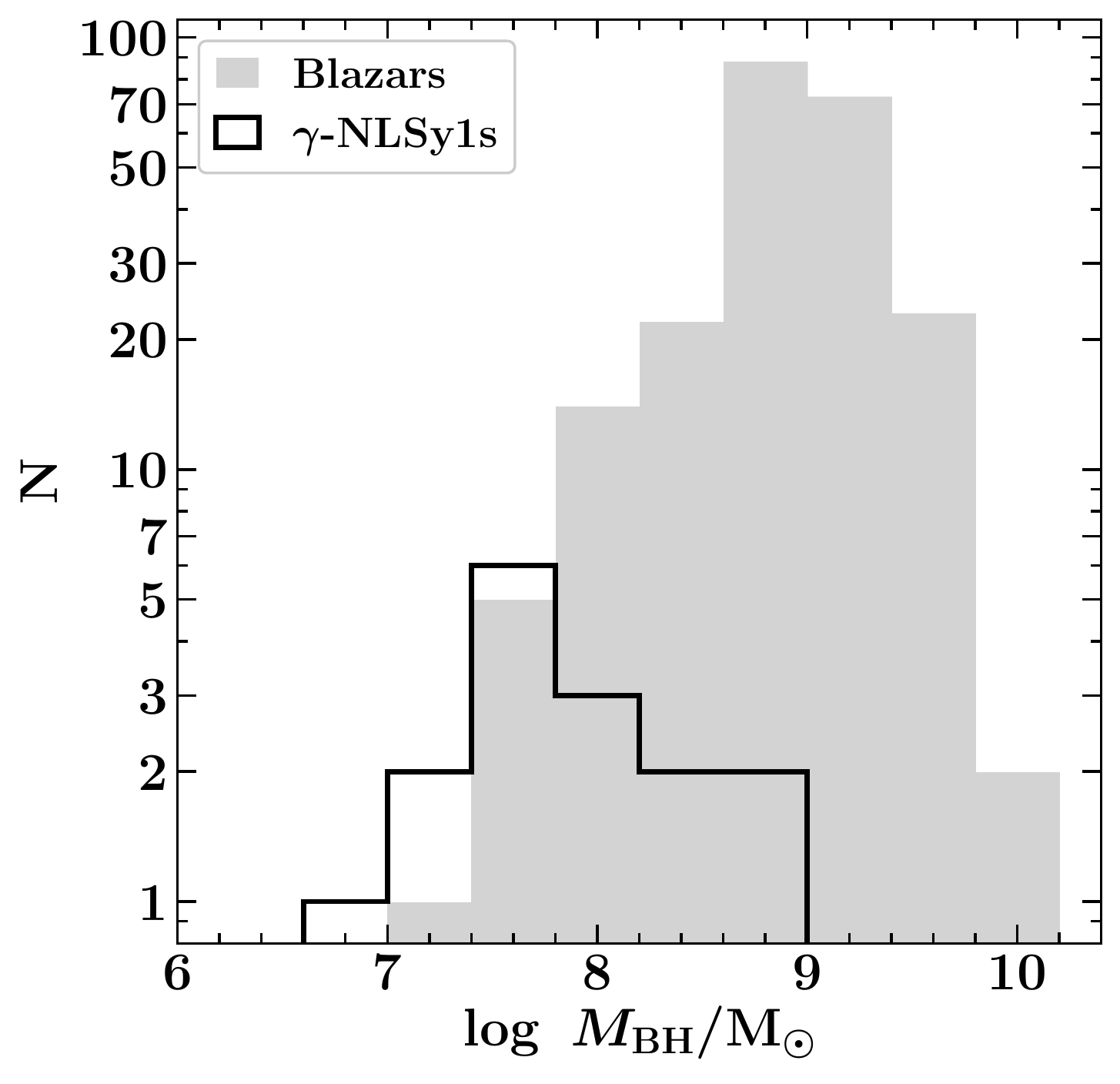}
\includegraphics[scale=0.4]{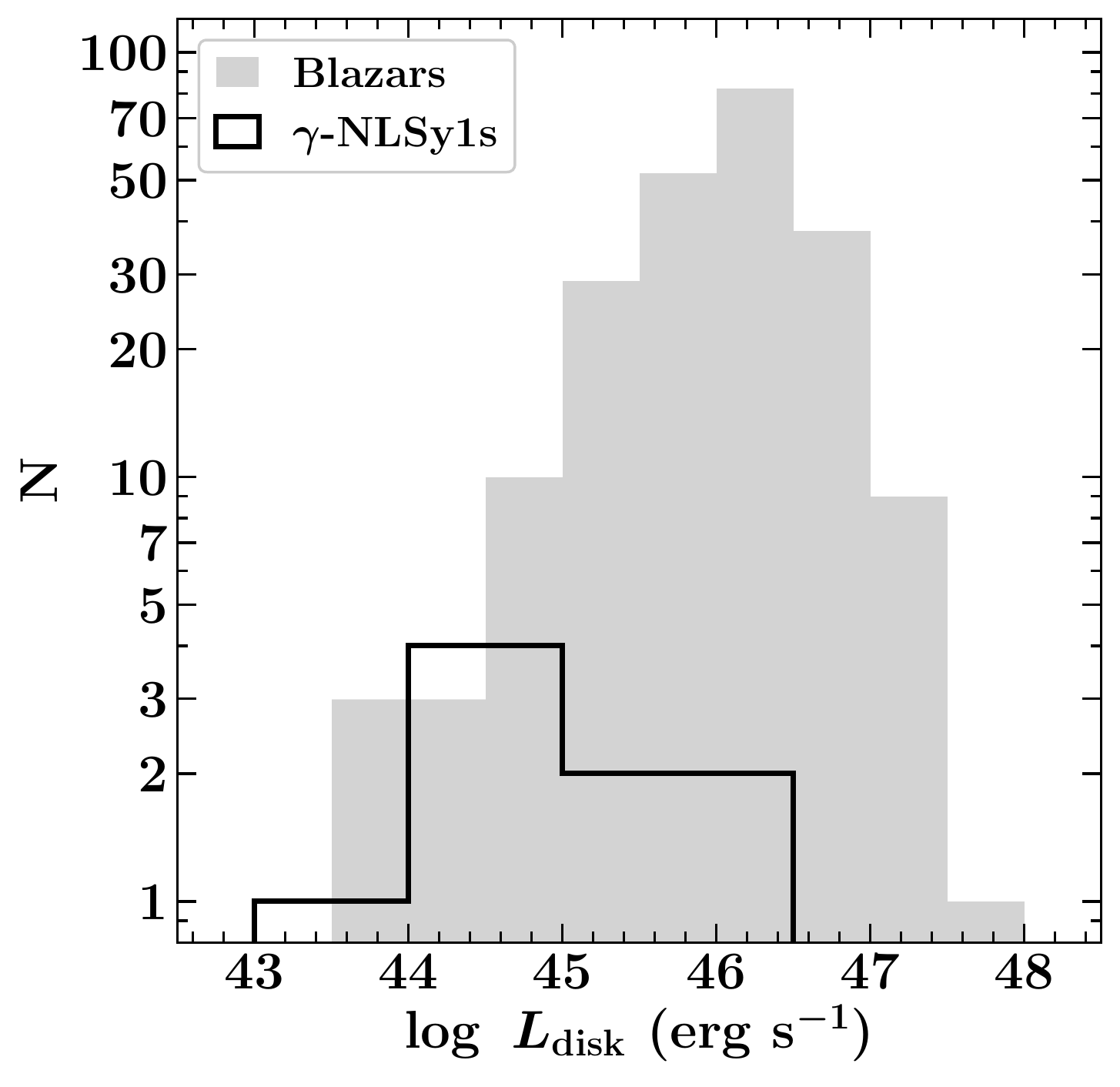}
}
\caption{Histograms of $M_{\rm BH}$ (left) and $L_{\rm disk}$ (right). Only those blazars are considered from P17 that have measurements either from the disk modeling or optical spectroscopic approaches. Other information are same as in Figure \ref{fig:sed_param}. See the text for details.\label{fig:mbh_ld}}
\end{figure*}

The average magnetic field derived for \gm-NLSy1 galaxies is relatively lower ($\langle B \rangle=0.91\pm0.33$ Gauss) compared to blazars ($1.83\pm0.25$ Gauss), though the overall distribution of the former lies within the range of the latter (Figure \ref{fig:sed_param}, middle left panel). The distance of the emission region or $R_{\rm diss}$ (cm), in logarithmic scale, peaks at  $\langle R_{\rm diss} \rangle=17.11\pm0.30$ and $17.35\pm0.38$ for \gm-NLSy1s and blazars, respectively. The emission region is found to lie within the BLR for 7 objects and for remaining it lies outside the inner boundary of the BLR and inside the torus. The corresponding distributions are shown in the middle center panel of Figure \ref{fig:sed_param}.

The distributions of the bulk Lorentz factor $\Gamma_{\rm b}$ and the Doppler factor $\delta_{\rm b}$ are displayed in the middle right and bottom left panels, respectively, of Figure \ref{fig:sed_param}. Both $\Gamma_{\rm b}$ and $\delta_{\rm b}$ for \gm-NLSy1 sources ($\langle \Gamma_{\rm b} \rangle=10.8\pm2.4$ and $\langle \delta_{\rm b} \rangle=16.1\pm1.9$) are found to smaller than that derived from blazars ($\langle \Gamma_{\rm b} \rangle=12.3\pm2.5$ and $\langle \delta_{\rm b} \rangle=17.1\pm1.5$), though with a large dispersion. The smaller, yet relativistic, $\delta_{\rm b}$ in \gm-NLSy1 galaxies is aligned with the low jet bulk velocity found from radio studies \citep[][]{2015ApJS..221....3G,2018MNRAS.tmp.1742S}. Note that one can argue that a low $\delta_{\rm b}$ can also be due to a larger viewing angle in \gm-NLSy1s, however, it is unlikely. This is because the external Compton scattering mechanism, which is responsible for the \gm-ray emission, is very sensitive to the angle of the line-of-sight with the jet axis due to anisotropy of the external photon field even in the jet-frame \citep[cf.][]{2001ApJ...561..111G}. For a large viewing angle (e.g., $>$10$^{\circ}$), the comoving-frame radiative energy densities of the external AGN components will not be sufficient to explain the observed \gm-ray emission. Hence, it appears that the low Doppler factor is possibly intrinsic to the low-power jets in \gm-NLSy1 sources. Furthermore, the low-level of the Doppler boosting also explains why there are only a few NLSy1s are found in \gm-rays and even fewer their parent population members \citep[due to 2$\Gamma_{\rm b}^2$ dependence;][]{2015A&A...578A..28B}.

\begin{figure}
\includegraphics[scale=0.6]{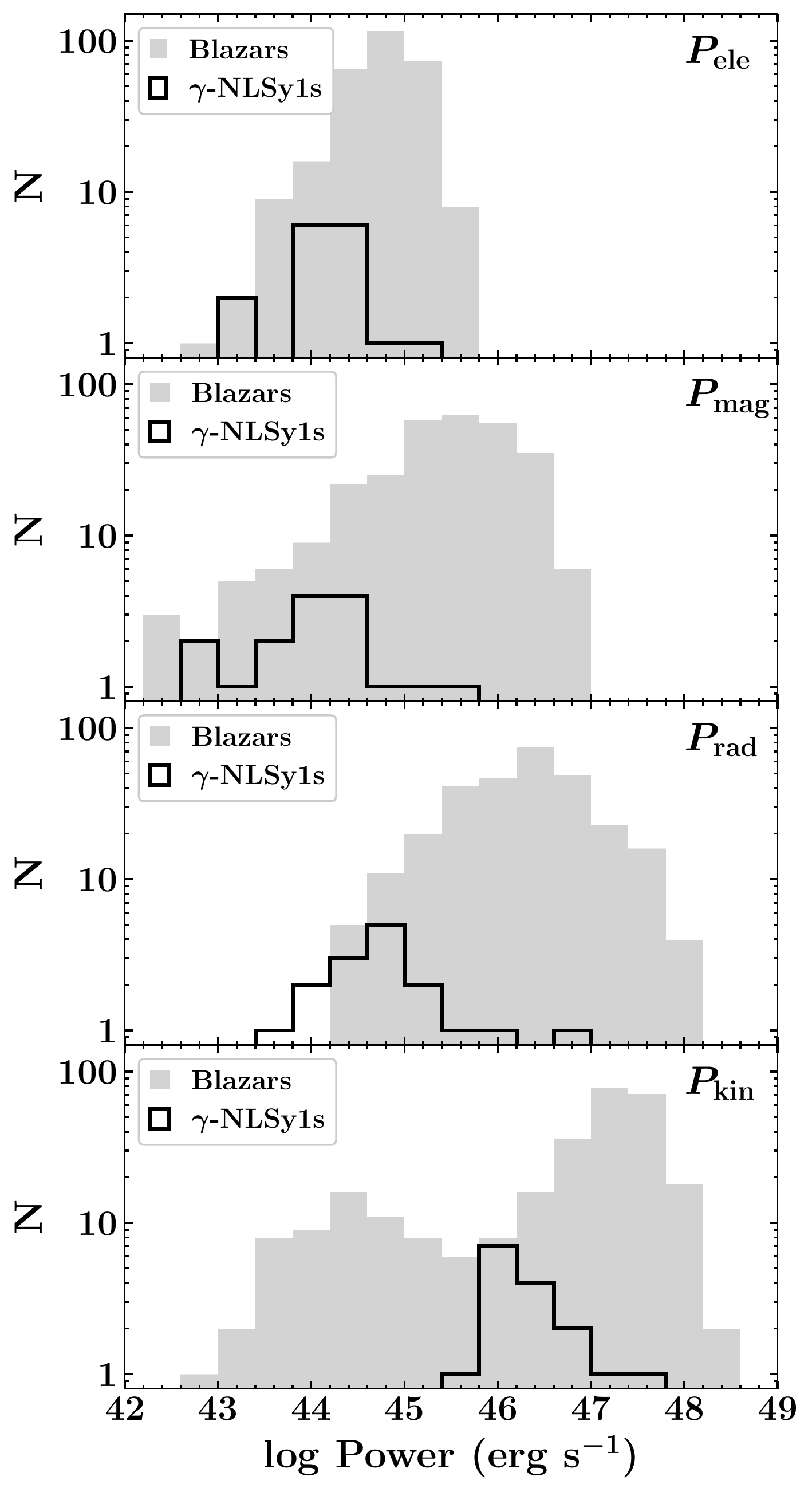}
\caption{This plot shows the distributions of various jet powers, as labeled. For \gm-NLSy1s and blazars, respectively, the average values, in \lum~and logarithmic scale, are as follows: $\langle \log~P_{\rm ele}\rangle=44.2\pm0.5, 44.7\pm0.4$, $\langle \log~P_{\rm mag}\rangle=44.1\pm0.8, 45.4\pm0.8$, $\langle \log~P_{\rm rad}\rangle=44.9\pm0.7, 46.3\pm0.7$,  and $\langle \log~P_{\rm kin}\rangle=46.3\pm0.5, 46.7\pm1.2$. Other information are same as in Figure \ref{fig:sed_param}.\label{fig:jet}}
\end{figure}

A large Compton dominance or CD reflects the prevalence of the high-energy emission over synchrotron radiation. In other words, a major fraction of the bolometric output of a Compton dominated object (e.g., FSRQs) radiates in the X-ray to \gm-ray bands. We show the histogram of CD in the bottom right panel of Figure \ref{fig:sed_param}. As can be seen, all but one \gm-NLSy1 galaxies have CD larger than unity which again indicates their similarity with FSRQs. The Compton dominated SED of \gm-NLSy1 galaxies also implies the dominance of the external photon energy densities over the magnetic one, which, in turn, suggests an efficient accretion process illuminating the surrounding environment of the relativistic jet.

The distributions of the mass of the central black hole and accretion luminosity are shown in Figure \ref{fig:mbh_ld}. On a comparison with blazars\footnote{We consider only those blazars whose $M_{\rm BH}$ and $L_{\rm disk}$ were computed either from the optical spectroscopy or from the disk modeling (P17).} ($\langle \log M_{\rm BH}, \msun \rangle=8.9\pm0.5$), we find that \gm-NLSy1s are powered by a relatively low mass black holes ($\langle \log M_{\rm BH}, \msun \rangle=7.8\pm0.6$), similar to that reported in earlier studies \citep[][]{F15}. However, a few \gm-NLSy1 objects do host massive black holes with $M_{\rm BH}>10^8~\msun$. Similarly, they are found to lie at the low-end of the $L_{\rm disk}$ distribution with $\langle \log L_{\rm disk}, {\rm erg~s^{-1}} \rangle=44.8\pm0.9$ and $46.0\pm0.7$ for \gm-NLSy1s and blazars, respectively. These findings strengthen the idea about \gm-NLSy1 being the low-$M_{\rm BH}$, low-power counterpart of FSRQs.

\subsection{Jet Powers}\label{subsec:dis_jet}
The distributions of various jet powers computed from the SED modeling are shown in Figure \ref{fig:jet}. On average, \gm-NLSy1 galaxies host low power jets, however, their kinetic jet power, $P_{\rm kin}$, is comparable to blazars. Note that the assumption of one proton per electron is crucial in the calculation of $P_{\rm kin}$ and often leads it to exceed $L_{\rm disk}$ and $L_{\rm Edd}$. There could be a few ($\lesssim$10) electron-positron pairs present in the jet which would reduce the budget of the jet power \citep[cf.][]{2016ApJ...831..142M,2017MNRAS.465.3506P}.

The most robust estimate of the jet power is the luminosity that jet produces in the form of radiation ($P_{\rm rad}$), since it is directly proportional to the observed bolometric luminosity $L_{\rm bol}$. In particular:

\begin{equation}
P_{\rm rad} \propto \left\{ \begin{array}{ll}
	\,\frac{\Gamma_{\rm b}^2}{\delta^4}~L_{\rm bol} & {\rm for~syn/SSC} \\
		\,{\Gamma_{\rm b}^4\over \delta^6}~L_{\rm bol} & {\rm for~EC} \\
	\end{array}
	\right. \ ,
\end{equation}
The $P_{\rm rad}$ derived for \gm-NLSy1s is found to be more than an order of magnitude lower than that computed for blazars (Figure \ref{fig:jet}). The low jet power of \gm-NLSy1 galaxies explains the reasons of the detection of only a small number of radio-loud NLSy1 sources in \gm-rays and also their various characteristics observed in the radio band \citep[][]{2015ApJS..221....3G}. This could also account for the lack of diffuse emission on kpc-scale found with Karl J. Jansky Very Large Array \citep[JVLA,][]{2018A&A...614A..87B}.

\subsection{Accretion-Jet Connection}\label{subsec:dis_acc_jet}
\begin{figure*}[t]
\hbox{
\includegraphics[scale=0.57]{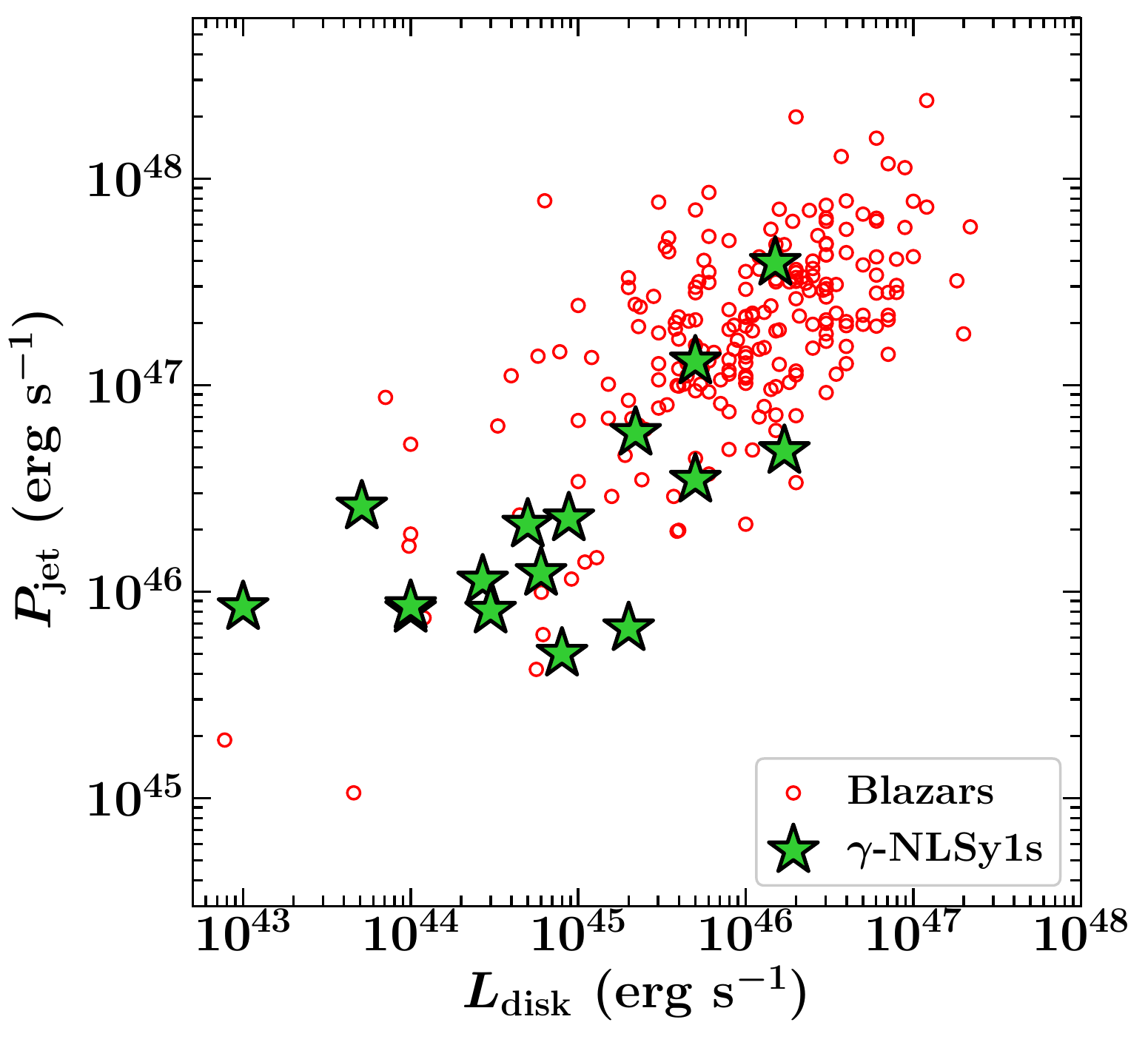}
\includegraphics[scale=0.57]{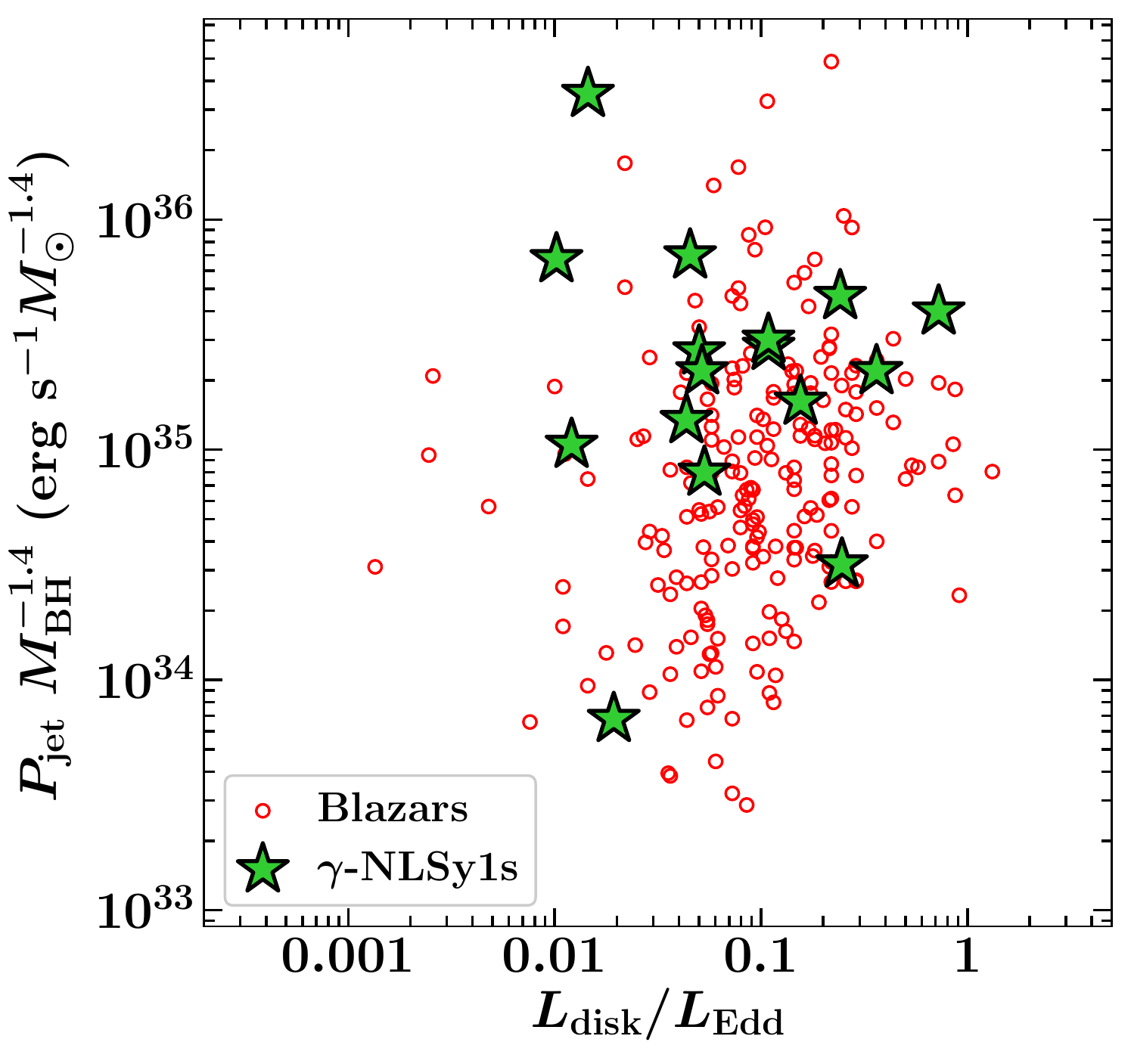}
}
\hbox{\hspace{4.5cm}
\includegraphics[scale=0.57]{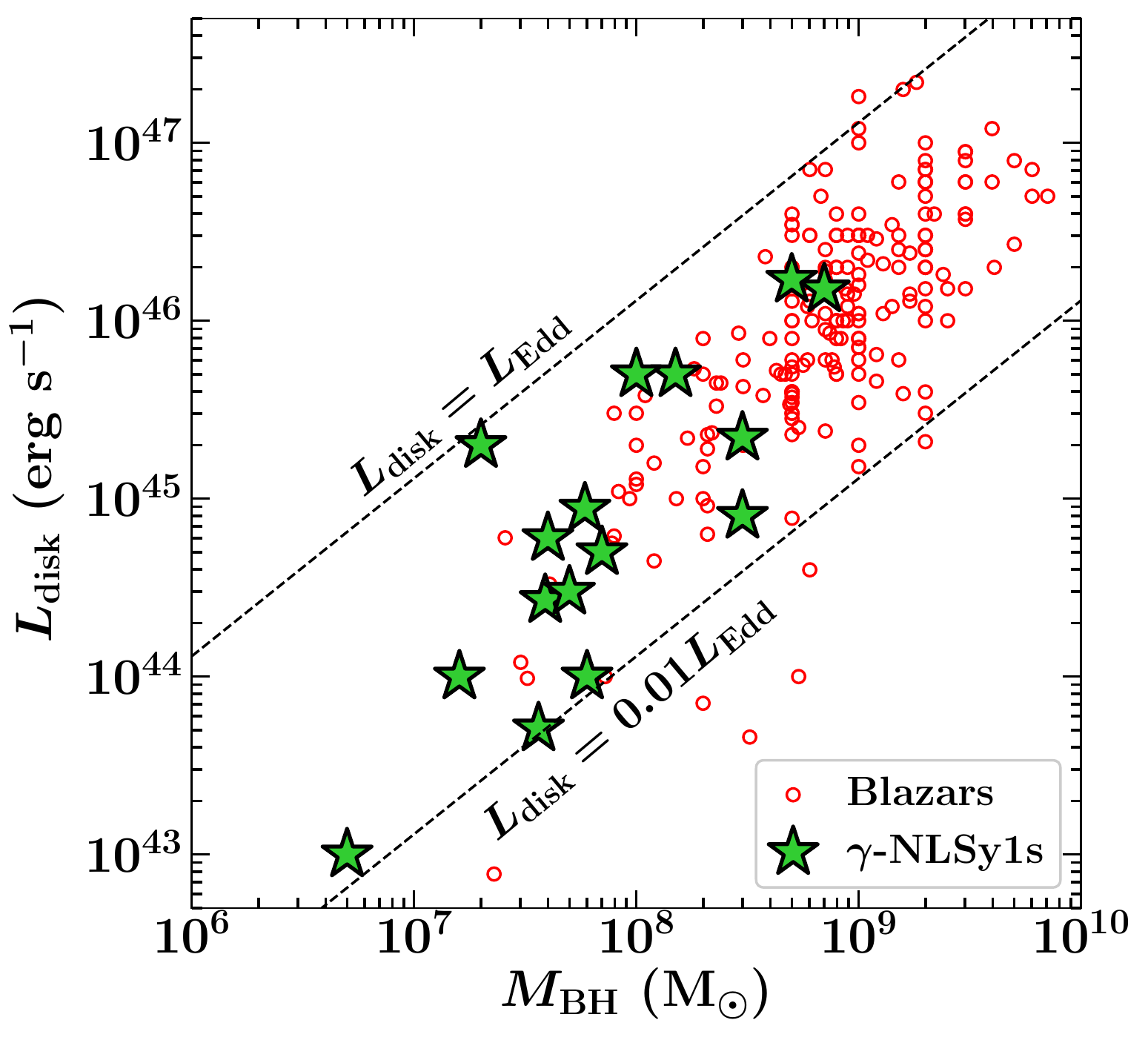}
}
\caption{Top left: A plot of the total jet power ($P_{\rm jet}$ = $P_{\rm ele}$ + $P_{\rm mag}$ + $P_{\rm kin}$) versus luminosity of the accretion disk. Top right: Same as left, but after normalizing the plotted quantities for the mass of the central black hole. Bottom: Variation of $L_{\rm disk}$ as a function of $M_{\rm BH}$. The dashed lines represent $L_{\rm disk}$ in Eddington units, as labeled. Throughout, blazars are shown with red empty circles and green-filled stars correspond to \gm-NLSy1 galaxies.\label{fig:disk_jet}}
\end{figure*}

It is well-known that the accretion luminosity and the jet power in jetted AGNs correlate \citep[e.g.,][]{1991Natur.349..138R}. It is interesting to verify whether \gm-NLSy1 galaxies also follow the same pattern. Another reason to study the accretion disk-jet connection in \gm-NLSy1 sources is the fact that they host relatively low-mass black holes. A comparison of the jet-disk properties of the low-mass systems with high-mass ones (i.e., blazars) will enable us to understand the similarities/differences of the central engine powering these two populations.

In the top left panel of Figure \ref{fig:disk_jet}, we show the variation of the total jet power ($P_{\rm jet}$ = $P_{\rm ele}$ + $P_{\rm mag}$ + $P_{\rm kin}$) as a function of $L_{\rm disk}$.   
As can be seen, a majority of \gm-NLSy1 galaxies (green-filled stars) occupy the low $P_{\rm jet}$ and low $L_{\rm disk}$ regime similar to low-luminosity blazars. In order to further investigate the cause of the low jet power in \gm-NLSy1 objects, i.e., whether it is due to low $M_{\rm BH}$ or the central engine being inefficient in powering the jet, we plot both $L_{\rm disk}$ and $P_{\rm jet}$ independent of $M_{\rm BH}$ in the top right panel of Figure \ref{fig:disk_jet}. This is done by deriving $L_{\rm disk}$ in units of $L_{\rm Edd}$ and scaling $P_{\rm jet}$ by a factor $M^{-1.4}_{\rm BH}$ \citep[][]{2003MNRAS.343L..59H,2012AIPC.1505..574F}. It can be immediately noticed that \gm-NLSy1 galaxies host some of the most powerful jets, after correcting for the mass of the central black hole. This suggests that the observed low jet power of \gm-NLSy1 sources is likely due to their low $M_{\rm BH}$. 

The bottom panel of Figure \ref{fig:disk_jet} shows the variation of $L_{\rm disk}$ as a function of $M_{\rm BH}$. In this diagram, a majority of the \gm-NLSy1 galaxies are found to host relatively low-mass black holes and less-luminous disks. In other words, the central engine of \gm-NLSy1 objects is less powerful compared to blazars. Interestingly, it was postulated within the first year of the \fermi-LAT operation that as the \gm-ray flux threshold of the LAT will decrease,  due to deeper exposure as the time progresses, more `red' FSRQs with low power jets and low $M_{\rm BH}$ will be detected with the \fermi-LAT \citep[][]{2009MNRAS.396L.105G}. The broadband properties of all of the \gm-NLSy1 sources resemble powerful FSRQs and thus these objects well-fit in the proposed scheme. These findings consolidate our current understanding about \gm-NLSy1 systems which host rapidly accreting low-mass black holes at their centers and are still in the early stage of evolution and launching powerful relativistic jets \citep[][]{2000MNRAS.314L..17M,2004ApJ...606L..41G}.

\subsection{Optical Spectroscopic Properties}\label{subsec:opt_spec}
\begin{figure*}[t]
\hbox{
\includegraphics[scale=0.57]{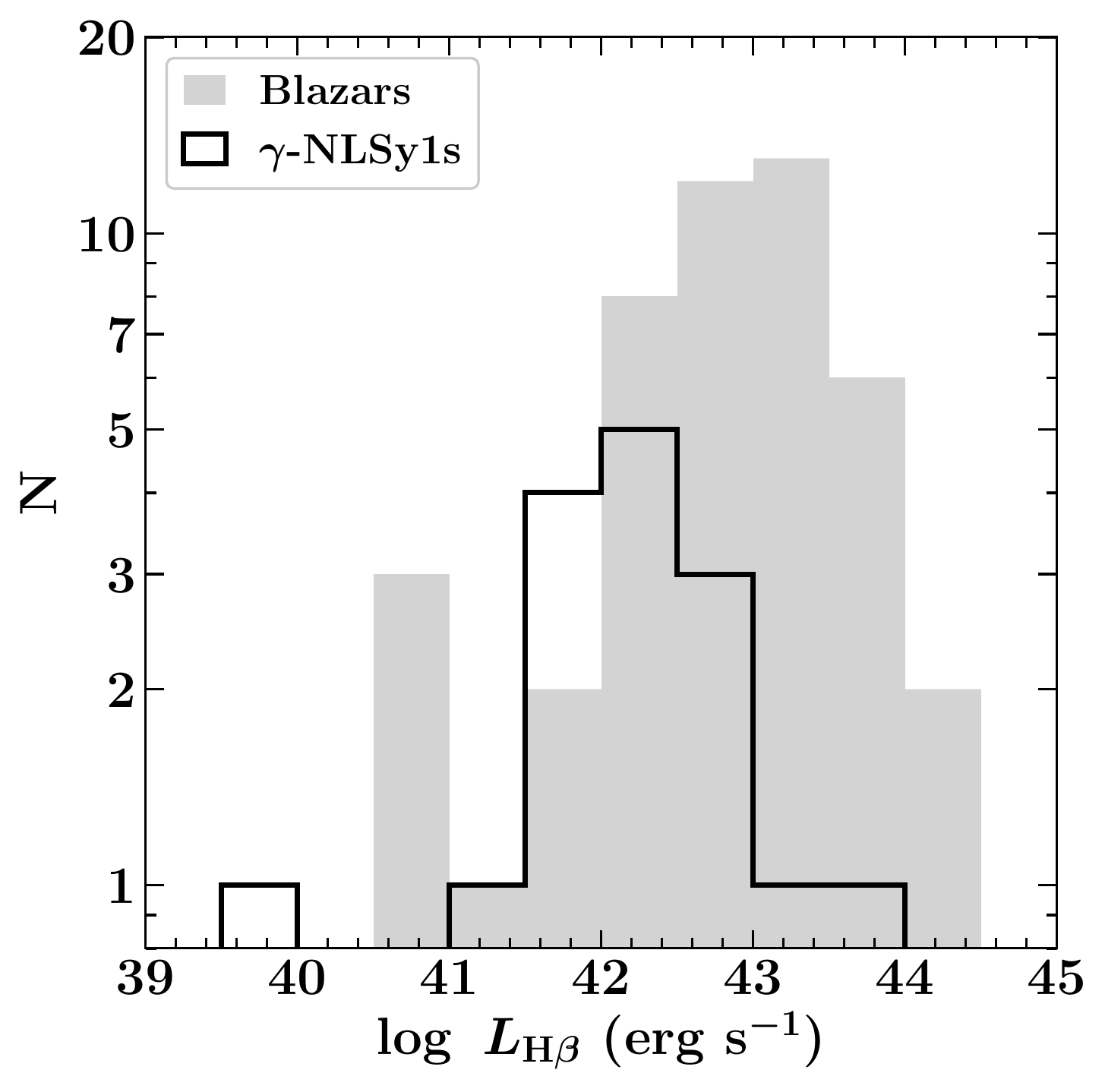}
\includegraphics[scale=0.57]{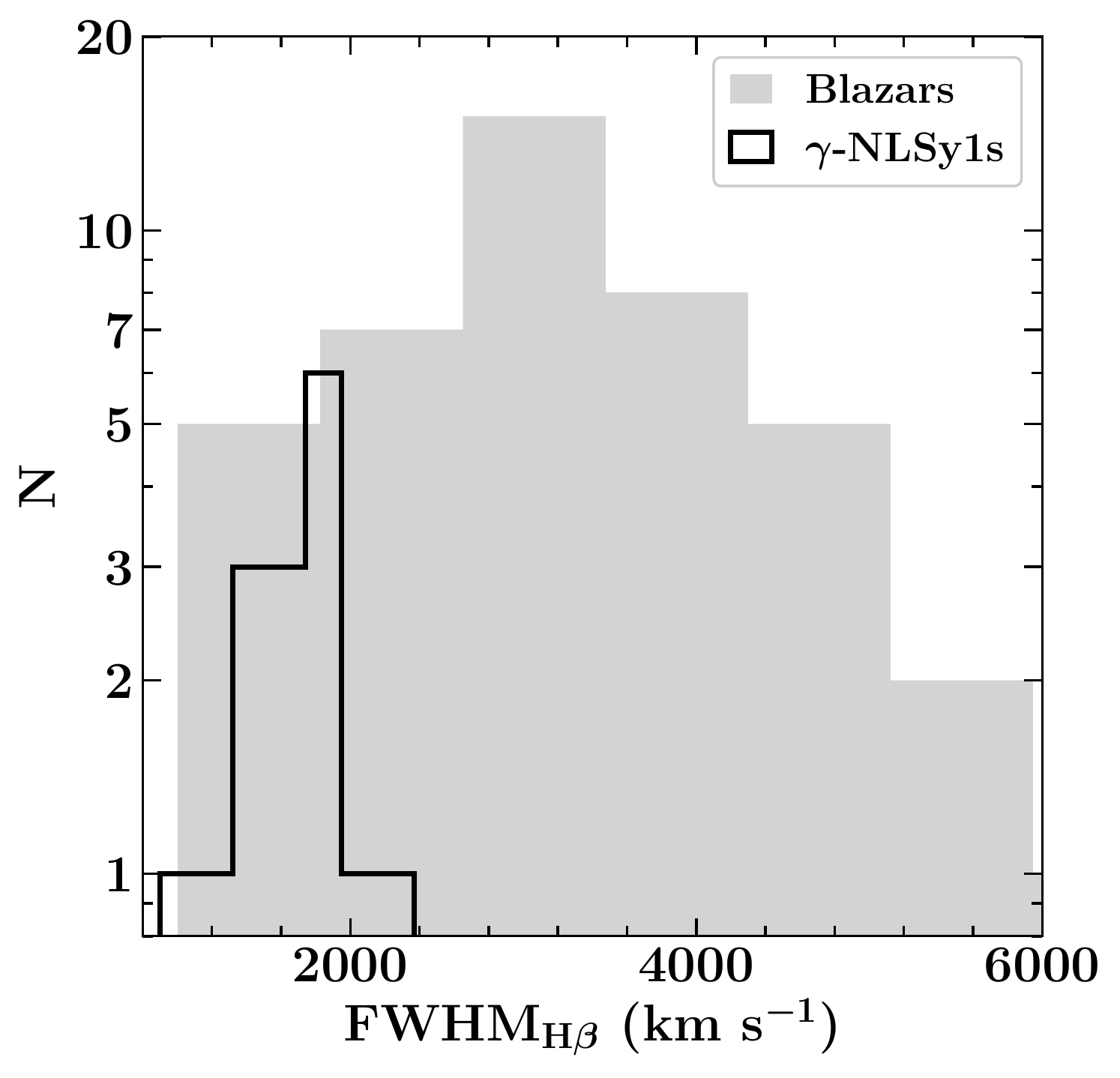}
}
\caption{The distributions of the H$_{\beta}$ emission line luminosity (left), in logarithmic scale, and its FWHM (right). The emission line parameters for blazars are taken from \citet[][]{2012ApJ...748...49S}. We adopt archival values as reported in the literature \citep[e.g.,][]{2001ApJ...558..578O,2007ApJ...658L..13Z,2015MNRAS.454L..16Y,2017ApJS..229...39R} for \gm-NLSy1 galaxies.\label{fig:opt_spec}}
\end{figure*}

In Figure \ref{fig:opt_spec}, we show the distributions of the H$_{\beta}$ line luminosity and full-width-at-half-maximum (FWHM), for both populations. Since H$_{\beta}$ emission line remains visible in the optical spectrum for $z\lesssim$1, this ensures that we compare two populations located at similar distances and hence avoided a possible bias due to the redshift dependent evolution. The emission line parameters for blazars are collected from \citet[][]{2012ApJ...748...49S} and for the most of the \gm-NLSy1s we use the data supplied by \citet[][]{2017ApJS..229...39R}. The H$_{\beta}$ line information for remaining \gm-NLSy1 galaxies are collected from the literature \citep[][]{2001ApJ...558..578O,2007ApJ...658L..13Z,2015MNRAS.454L..16Y}. Comparing both distributions, one can notice that \gm-NLSy1 galaxies exhibit relatively less luminous H$_{\beta}$ lines ($\langle \log L_{\rm H\beta}, {\rm erg~s^{-1}} \rangle=42.1\pm0.8$) with narrow-width ($\langle {\rm FWHM} \rangle= 1695.8~\pm~293.1$ km~s$^{-1}$) with respect to blazars ($\langle \log L_{\rm H\beta}, {\rm erg~s^{-1}} \rangle=42.8\pm0.8$ and FWHM $= 3791.5~\pm~2171.5$ km~s$^{-1}$), though with a large dispersion. What is more peculiar is the fact that many blazars lie in the range covered by \gm-NLSy1 galaxies. This observation suggests that there could be many more \gm-NLSy1 galaxies which are presently classified as blazars \citep[see,][for a similar finding]{2015MNRAS.454L..16Y}. The noteworthy fact is that the \gm-NLSy1s are classified solely based on their optical spectral characteristics, whereas, a quasar is identified as a blazar from its broadband behavior, e.g., a flat radio spectrum, large-amplitude flux and spectral variability, and double-hump SED. The \gm-NLSy1 sources exhibit almost all of the multi-frequency properties observed from blazars, however, whether vice versa is true, is unclear. This is because a careful analysis of the optical spectral properties of blazars has not been done in detail and it is likely that we may find many more blazars fulfilling the criteria of being a \gm-NLSy1. Particular attention can be given to blazars with low-mass black holes and $L_{\rm disk}\gtrsim 0.01L_{\rm Edd}$.

\section{Black hole mass in \gm-NLS\MakeLowercase{y}1 Galaxies}\label{sec:dis_bh_mass}
Historically, the multi-frequency properties of NLSy1 sources are explained by highly accreting black holes whose masses are primarily measured from the optical spectroscopy and found to be comparatively lower with respect to their broad line counterparts \citep[e.g.,][]{2004ApJ...606L..41G,2006AJ....132..321D,2011nlsg.confE..32P}. However, it has been proposed that $M_{\rm BH}$ derived from the optical spectroscopy or reverberation mapping is severely underestimated due to projection effects \citep[][]{2008MNRAS.386L..15D} and radiation pressure effect \citep[][]{2008ApJ...678..693M}. The disk modeling approach adopted in \citet[][]{2013MNRAS.431..210C} also predicts consistently massive black holes in \gm-NLSy1 sources. Below, we discuss predictions of these hypotheses and compare them with the observed properties of \gm-NLSy1s, and also with our findings.

\citet[][]{2008MNRAS.386L..15D} proposed that NLSy1 objects are mostly observed more pole-on compared to broad-line Seyfert 1 (BLSy1) galaxies. Considering a disk-shaped geometry of the BLR, they were able to explain the observed narrowness of the Balmer emission lines and once corrected for the BLR geometrical factor, both NLSy1 and BLSy1 populations exhibit similar $M_{\rm BH}$ and $L_{\rm disk}/L_{\rm Edd}$ values. \citet[][]{2011MNRAS.413...39D} extended it further to the most common member of the jetted AGN family viewed pole-on, i.e., blazars. Interestingly, they explained the observed narrow emission lines in BL Lac objects mainly due to small viewing angle with respect to the jet axis and thus exhibiting the largest BLR geometrical correction. However, this model is unable to explain a number of observations. First, if all NLSy1 galaxies are viewed pole-on, the fraction of the radio-loud NLSy1 should be larger, or at least comparable, to that known for the quasar population, due to beaming effects. Instead, $<$4\% of NLSy1 galaxies are found to be radio-loud \citep[][]{2017ApJS..229...39R}. Second, radio-loud NLSy1 sources should also exhibit larger variability with respect to the BLSy1 population which is not supported by observations \citep[][]{2017ApJ...842...96R}. Moreover, in a more physically motivated scenario, weak and narrow emission lines observed from BL Lac objects are likely due to underlying radiatively inefficient accretion \citep[$L_{\rm disk}/L_{\rm Edd}<0.01$,][]{2017MNRAS.469..255G} and may not be due to a flat BLR geometry. Even in a disk-shaped BLR causing the narrowness of the emission lines, a luminous accretion disk implies an intense BLR and torus photon fields which lead to the observation of a Compton dominated SED \citep[see, e.g.,][]{2014PASJ...66...92L}. Instead, the SEDs of BL Lac sources are generally synchrotron dominated \citep[][]{2010MNRAS.401.1570T}. Finally, there are many BL Lacs that have exhibited broad emission lines in their low activity states \citep[e.g., PKS 0426$-$380;][]{2005AJ....129..559S}. The variation in the emission line widths demands a significant change in the BLR geometry and/or the viewing angle, which are unlikely, if not impossible. Therefore, it appears that orientation effects may not play a major role in determining $M_{\rm BH}$ for all beamed AGNs.

\citet[][]{2008ApJ...678..693M} studied the effect of radiation pressure on $M_{\rm BH}$ computation in highly accreting systems such NLSy1 galaxies. It is argued that a larger gravitational force, and hence more massive black hole, is needed to account for the radiation pressure exerted by the ionizing photons and keep the system gravitationally bound. According to our calculation, all of the \gm-NLSy1 galaxies, except 1H 0323+342, have $L_{\rm disk}$ $<40\%$ of $L_{\rm Edd}$ and thus the radiation pressure may not be too strong in them. Moreover, the mass of the central black hole in 1H~0323+342 has been measured using various techniques, e.g., reverberation mapping \citep[][]{2016ApJ...824..149W}, X-ray variability \citep[][]{2018ApJ...866...69P}, SED modeling \citep[][]{2009ApJ...707L.142A}, and single epoch optical spectroscopy \citep[cf.][]{2007ApJ...658L..13Z}. They all agree with a $M_{\rm BH} \sim10^7$ \msun. Therefore, it is likely that some of the optical spectral parameters \citep[e.g., continuum luminosity contaminated by the jet,][]{2004A&A...424..793W} can compensate for the underestimation of $M_{\rm BH}$ caused by neglecting the radiation pressure effect and hence the overall results from the optical spectroscopy remain valid. Another supporting argument in this regard is the discovery of quasars within the first Gyr of the Universe ($z>6$) hosting more than a billion-solar-mass black holes at their centers \citep[see, e.g.,][]{2015Natur.518..512W}. Many of them are found to be accreting close to or even larger than the Eddington limit. In fact, one of the hypotheses proposed to explain the existence of massive black holes at high-redshifts is the hyper-Eddington accretion that can help the black hole to grow rapidly within a short period of time \citep[cf.][]{2016MNRAS.459.3738I}. If $M_{\rm BH}$ derived from the optical spectroscopy is severely underestimated, it suggests even larger black holes in high-redshift quasars since they have a very high accretion rate. Keeping in mind the maximum possible $M_{\rm BH}$ that a black hole can achieve via accretion \citep[a few $\times10^{10}$ \msun;][]{2016MNRAS.456L.109K,2016ApJ...828..110I}, it can be concluded that the radiation pressure may have only a minor effect, well within the associated uncertainties, on the virial $M_{\rm BH}$ estimation.

\citet[][hereafter C13]{2013MNRAS.431..210C} modeled the NIR-UV spectrum of a sample of radio-loud NLSy1 galaxies with a standard \citet[][]{1973A&A....24..337S} accretion disk spectrum and estimated a consistently higher $M_{\rm BH}$ than that derived from the optical spectroscopy. Among 23 sources studied in C13, three of them, PMN J0948+0022, PKS 1502+036, and RGB J1644+263, are included in our sample. Their Eddington ratios, according to  C13, are 0.006, 0.009, and 0.007, respectively. Such a low Eddington ratios are problematic to explain a number of observations. For example, below $\sim$1\% of $L_{\rm Edd}$, the accretion disk is expected to become radiatively inefficient \citep[][]{2017MNRAS.469..255G} and hence it may not be possible to explain the strong emission lines observed from \gm-NLSy1 galaxies\footnote{We stress that all \gm-NLSy1 galaxies have rest-frame EW of broad emission lines $>$5\AA~\citep[][]{2017ApJS..229...39R}, thus demanding an efficient accretion process.}. Furthermore, a radiatively inefficient accretion disk may not photo-ionize the BLR and consequently, due to lack of the photon fields needed for external Compton process, the broadband SED would be synchrotron dominated similar to that typically observed from BL Lac objects. Contrarily, \gm-NLSy1 sources exhibit a Compton dominated SED (Figure \ref{fig:SED}), thus implying a severe cooling of relativistic electrons by the external radiative environment which ultimately suggests an efficient accretion with a high Eddington ratio. 

\begin{figure*}[t]
\hbox{
\includegraphics[scale=0.4]{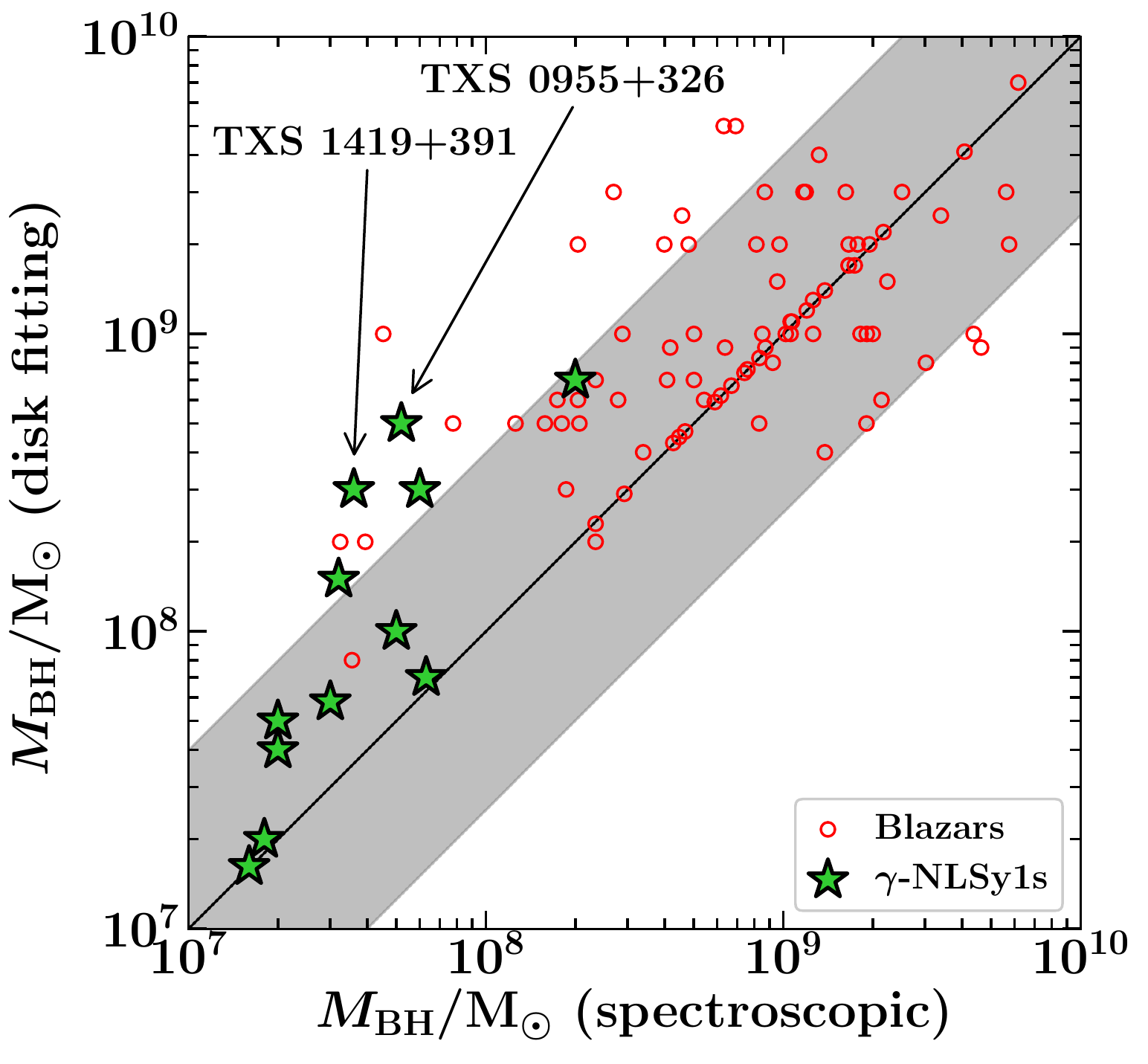}
\includegraphics[scale=0.4]{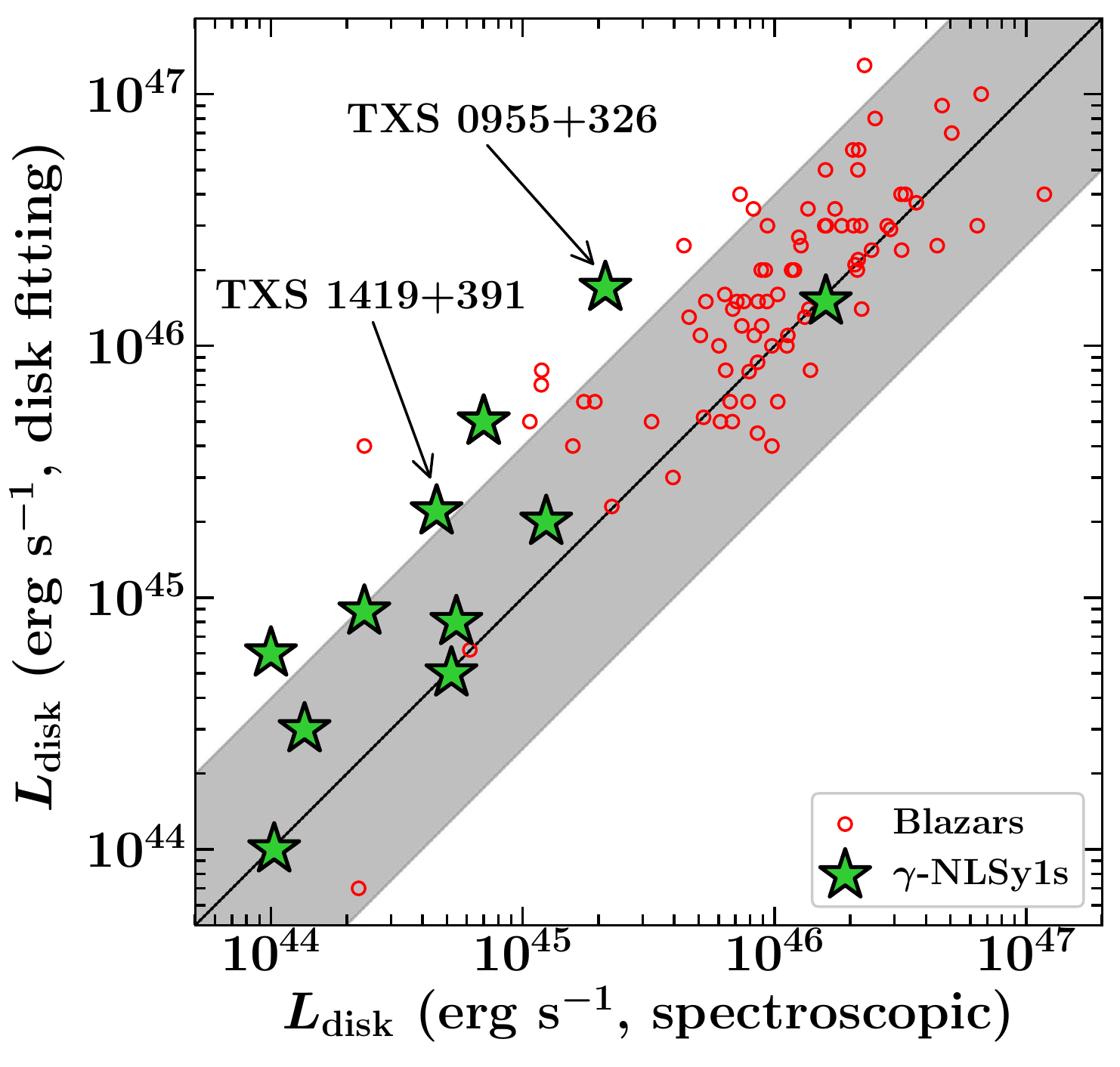}
\includegraphics[scale=0.4]{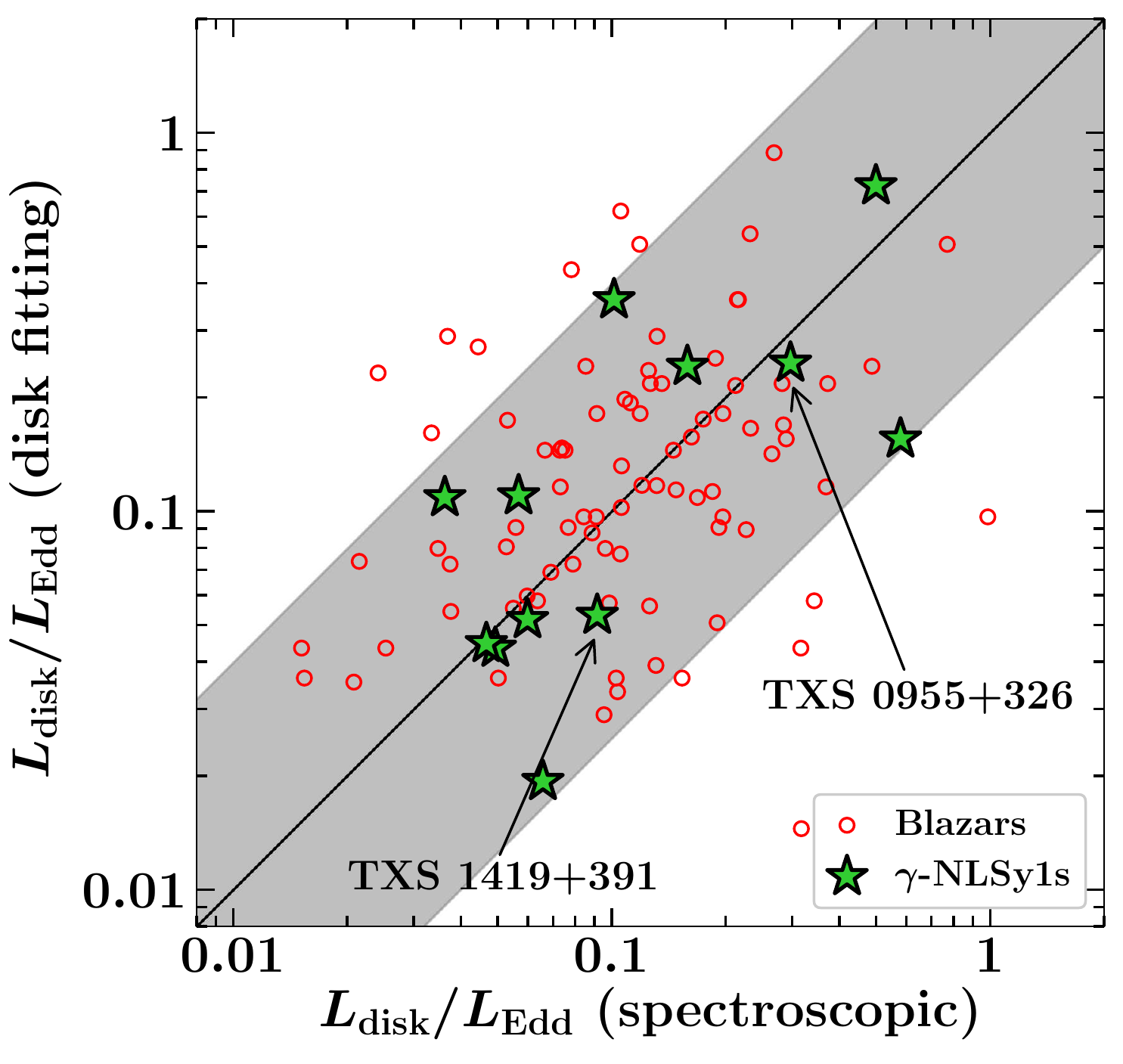}
}
\caption{ A comparison of $M_{\rm BH}$ (left) and $L_{\rm disk}$ (middle) derived from the optical spectral line fitting and disk modeling techniques. Right panel refers to $L_{\rm disk}$ in Eddington units. In all the plots, the grey solid line is the equality line and the shaded area denotes an uncertainty of a factor of four in the optical spectroscopic measurements \citep[e.g.,][]{2011ApJS..194...45S}. See the text for details.\label{fig:disk_spec}}
\end{figure*}

P17 compared $M_{\rm BH}$ and $L_{\rm disk}$ derived from the disk modeling and the optical spectroscopy for a large sample of blazars and found a rather good agreement. There are 12 \gm-NLSy1s whose NIR-UV spectra exhibit a bump and can be explained by a standard accretion disk model (see Figure \ref{fig:SED}, \ref{fig:app_SED1}, and \ref{fig:app_SED2}). Since all of them also have $M_{\rm BH}$ and $L_{\rm disk}$ information estimated from the virial approach, we compare these two methods in Figure \ref{fig:disk_spec}. As can be seen in the left and middle panels, $M_{\rm BH}$ and $L_{\rm disk}$ derived from the two methods reasonably agree for most of the \gm-NLSy1 galaxies within the uncertainties associated with the virial method. This observation suggests that $M_{\rm BH}$ computed from the optical spectrum is reliable. Notably, two main outliers in the left panel of Figure \ref{fig:disk_spec}, TXS 0955+326 and TXS 1419+391, have a larger $M_{\rm BH}$ calculated by disk modeling approach. However, the same technique also predicts a more luminous accretion disk in them (Figure \ref{fig:disk_spec}, middle panel), thus ensuring that the accretion remain radiatively efficient. This is demonstrated in the right panel of Figure \ref{fig:disk_spec} where we plot $L_{\rm disk}$ (or equivalently accretion rate) in Eddington units. In this diagram, both sources lie well within the measurement uncertainties. This brings us to the conclusion that the driving factor to define the jet properties of \gm-NLSy1 galaxies is the accretion rate in Eddington units. A similar conclusion is reached in a recent study focused on \gm-ray emitting blazars \citep[][]{2017MNRAS.469..255G}.

Efforts have been made to determine $M_{\rm BH}$ of NLSy1 galaxies by various non-conventional methods, such as spectropolarimetry and X-ray variability. \citet[][]{2016MNRAS.458L..69B} performed spectropolarimetry of a \gm-NLSy1 source PKS 2004$-$447 and reported a $M_{\rm BH}$ as high as $\sim$6$\times$10$^8$ \msun, which about 100 times larger than that estimated from the optical spectroscopy. As we discussed before, such a large $M_{\rm BH}$ is less likely to be reliable and also doubtful due to lack of the uncertainty measurement in the quoted value. Interestingly, \citet[][]{2018arXiv180508933P} performed a similar study on a sample of NLSy1 and BLSy1 sources. They reported the $M_{\rm BH}$ computed from the spectropolarimetry to be well in agreement with that measured from the reverberation technique. Similar conclusions are made from the X-ray variability study of a sample of NLSy1 and BLSy1 galaxies using the {\it XMM-Newton} data \citep[][]{2018MNRAS.480...96W}.

To summarize, it is possible that various phenomena reported in the literature, e.g., the flat BLR geometry, may be acting at different levels in \gm-NLSy1s galaxies. However, the overall effect of these scenarios on $M_{\rm BH}$ measurement could be minor. As discussed above, all of the available observations suggests \gm-NLSy1s to be `young' blazars and thus they could well-fit into the traditional blazar sequence. The accretion rate in Eddington units is found to be the driving factor to explain the physical properties of \gm-ray emitting blazars \citep[][P17]{2011MNRAS.414.2674G,2017MNRAS.469..255G} and our findings on \gm-NLSy1 sources are fully compatible with this postulation.

\section{Summary}\label{sec:summary}
We have performed a detailed broadband study of a sample of NLSy1 galaxies detected in the \gm-ray band. Our main results are summarized below.
\begin{enumerate}
\item The \gm-NLSy1s exhibit steep X-ray spectra below $\lesssim$2 keV which resemble the more general NLSy1 population and possibly originate from the X-ray corona. At hard X-ray energies, the spectra are relatively hard thus indicating the jet dominance similar to that observed from FSRQs.
\item We find tentative evidence for absorption in the soft X-ray spectrum of PKS~2004$-$447 and conclude it to be possibly due to the presence of `warm absorbers' along the line of sight.
\item There are weak hints for the detection of an Fe emission line in the X-ray spectrum of 1H~0323+342 as revealed by \suzaku, \xmm, and \nustar~observations. By fitting it with a relativistic disk line model, we note that the inner disk of this object is possibly truncated. Deeper X-ray observations and more importantly identifying more NLSy1s in \gm-rays are crucial to confirm the association of this feature with jetted AGNs.
\item The bulk Lorentz factor and the jet power derived for \gm-NLSy1 sources are found to be smaller than that known for blazars. This can explain the presence of small number of NLSy1 in \fermi-LAT catalogs.
\item In the $L_{\rm disk}$ versus $M_{\rm BH}$ plane, these objects form the tail end of the blazar population thus indicating their central engine to be less powerful but efficient in launching the relativistic jet.
\item The physical properties of \gm-ray emitting blazars are mostly understood on the basis of the accretion rate in Eddington units. Our study has extended this hypothesis to \gm-NLSy1s and confirms that this intriguing population also follow a similar pattern to that seen in blazars.
\end{enumerate}

\acknowledgments
We are thankful to the referee for a constructive criticism. We are grateful to the \swift~PI for approving our ToO requests and to the mission operation team for quickly scheduling the observations.
The \textit{Fermi} LAT Collaboration acknowledges generous ongoing support from a number of agencies and institutes that have supported both the development and the operation of the LAT as well as scientific data analysis. These include the National Aeronautics and Space Administration and the Department of Energy in the United States, the Commissariat \`a l'Energie Atomique and the Centre National de la Recherche Scientifique / Institut National de Physique
Nucl\'eaire et de Physique des Particules in France, the Agenzia Spaziale Italiana and the Istituto Nazionale di Fisica Nucleare in Italy, the Ministry of Education, Culture, Sports, Science and Technology (MEXT), High Energy Accelerator Research Organization (KEK) and Japan Aerospace Exploration Agency (JAXA) in Japan, and the K.~A.~Wallenberg Foundation, the Swedish Research Council and the Swedish National Space Board in Sweden. Additional support for science analysis during the operations phase is gratefully acknowledged from the Istituto Nazionale di Astrofisica in Italy and the Centre National d'\'Etudes Spatiales in France. This work performed in part under DOE Contract DE- AC02-76SF00515.

This work made use of data from the \nustar~mission, a project led by the California Institute of Technology, managed by the Jet Propulsion Laboratory, and funded by the National Aeronautics and Space Administration. We thank the \nustar~Operations, Software, and Calibration teams for support with the execution and analysis of these observations. This research has made use of the \nustar~Data Analysis Software (NuSTARDAS) jointly developed by the ASI Science Data Center (ASDC, Italy) and the California Institute of Technology (USA).

This research has made use of the XRT Data Analysis Software (XRTDAS). This work made use of data supplied by the UK Swift Science Data Centre at the University of Leicester. 

This research has made use of data obtained through the High Energy Astrophysics Science Archive Research Center Online Service, provided by the NASA/Goddard Space Flight Center.  This research has made use of the NASA/IPAC Extragalactic Database (NED), which is operated by the Jet Propulsion Laboratory, California Institute of Technology, under contract with the National Aeronautics and Space Administration. Part of this work is based on archival data, software or online services provided by the ASI Data Center (ASDC).

This work is based on observations obtained with XMM-Newton, an ESA science mission with instruments and contributions directly funded by ESA Member States and NASA. The scientific results reported in this article are based on data obtained from the Chandra Data Archive. This research has made use of software provided by the Chandra X-ray Center (CXC) in the application packages CIAO, ChIPS, and Sherpa.

This research has made use of the Palermo BAT Catalogue and database operated at INAF - IASF Palermo.

\software{SAS \citep[][]{2004ASPC..314..759G}, XSPEC \citep{Arnaud96}, CIAO \citep[][]{2006SPIE.6270E..1VF}, NuSTARDAS \citep[][]{2013ApJ...770..103H}}.

\bibliographystyle{aasjournal}
\bibliography{Master}

\cleardoublepage
\begin{deluxetable*}{lccccccc}
\tabletypesize{\small}
\tablecaption{Basic properties of \gm-ray detected NLSy1 galaxies studied in this work.\label{tab:basic_info}}
\tablewidth{0pt}
\tablehead{
\colhead{Name} & \multicolumn{2}{c}{Radio position (J2000)} & \colhead{$F_{\rm 1.4~GHz}$} & \colhead{$g'$} & \colhead{$z$}  & \colhead{$M_{\rm BH}$}  & \colhead{Ref.} \\
\cline{2-3}
\colhead{} & \colhead{hh mm ss.ss} & \colhead{dd mm ss.s} & \colhead{(mJy)} & \colhead{(mag)} & \colhead{} & \colhead{$M_{\sun}$}  & \colhead{}
}
\startdata
1H 0323+342	          & 03 24 41.1 & +34 10 46   & 613.5     & 15.7 & 0.60 &    7.30 & (1)   \\
SBS 0846+513          & 08 49 58.0 & +51 08 29   & 266.3   & 18.8 & 0.58 &    7.59 & (2)   \\
CGRaBS J0932+5306 & 09 32 41.1 & +53 06 33   & 481.6 & 18.9 & 0.60 &    7.66 & (2)   \\
GB6 J0937+5008      & 09 37 12.3 & +50 08 52   & 166.6   & 19.1 & 0.28 &    7.56 & (2)   \\
PMN J0948+0022     & 09 48 57.3 & +00 22 26    & 69.5   & 18.6 & 0.58 &    7.50 & (2) \\
TXS 0955+326	      & 09 58 20.9 & +32 24 02    & 1247.1 & 16.1 & 0.53 &    7.72 & (2)   \\
FBQS J1102+2239    & 11 02 22.9  & +22 39 19 & 2.5      & 19.2 & 0.45 &   8.06 & (3)  \\
CGRaBS J1222+0413 & 12 22 22.5 & +04 13 16    & 800.3   & 17.0 & 0.97 &    8.30 & (4)  \\
SDSS J1246+0238   & 12 46 34.7 & +02 38 09 & 35.7 & 18.4 & 0.36 & 7.78 & (3) \\
TXS 1419+391   	      & 14 21 06.0 & +38 55 23    & 85.7     & 18.6 & 0.49 &    7.36 &  (2)  \\
PKS 1502+036	      & 15 05 06.5 & +03 26 31    & 394.8   & 18.6 & 0.41 &    7.26 & (2)   \\
TXS 1518+423	      & 15 20 39.6 & +42 11 09    & 138.4   & 19.3 & 0.48      &    7.80 &  (2)  \\
RGB J1644+263       & 16 44 42.5 & +26 19 13    & 128.4   & 18.0 & 0.14  &    7.21 &  (2)  \\
PKS 2004$-$447      & 20 07 55.2 & $-$44 34 44 & 471.0 & 19.3 & 0.24 &    6.70 &  (5)  \\
TXS 2116$-$077     & 21 18 52.9  & $-$07 32 28 & 96.1       & 16.5  & 0.26 &    7.21 & (2)   \\
PMN J2118+0013     & 21 18 17.4 & +00 13 17    & 147.9   & 19.3 & 0.46 &    7.47 &  (2)  \\
\enddata
\tablecomments{Radio coordinates and flux values are from NRAO VLA Sky Survey \citep[][]{1998AJ....115.1693C} and $g'$ magnitudes are from \citet[][]{2017ApJS..229...39R}, except for PKS 2004$-$447. We collect basic information of this object from \citet[][]{2006MNRAS.370..245G}. $M_{\rm BH}$ is the logarithmic black hole mass derived from the single epoch optical spectroscopy, in units of solar mass. The references for the black hole mass are as follows: (1): \citet[][]{2007ApJ...658L..13Z}; (2): \citet[][]{2017ApJS..229...39R}; (3): \citep[][]{F15}; (4) \citet[][]{2015MNRAS.454L..16Y}; (5): \citet[][]{2001ApJ...558..578O}.}

\end{deluxetable*}

\begin{table*}
\begin{center}
\caption{Details of the X-ray observations used for the X-ray spectroscopy sample. For brevity, we report the FPMA exposure times only for \nustar.}
\label{table:xray_obs}
\begin{tabular}{l l c r l}
\hline
Source & Instrument & Date & Length & Obs Id\\
\hline
1H 0323+342 & \emph{Suzaku} & 2009-07-26 & 168 ks & 704034010\\
& \emph{Suzaku} & 2013-03-01 & 203 ks & 707015010\\
& \emph{NuSTAR} & 2014-03-15 & 102 ks & 60061360002\\
& \emph{XMM-Newton} & 2015-08-23 & 64 ks & 0764670101\\
PMN J0948+0022 & \emph{XMM-Newton} & 2011-05-28 & 29 ks & 0673730101\\
& \emph{XMM-Newton} & 2016-11-04 & 65 ks & 0790860101\\
& \emph{NuSTAR} & 2016-11-04 & 193 ks & 60201052002\\
FBQS J1102+2239 & \emph{XMM-Newton} & 2012-06-11 & 3 ks & 0690090301\\
CGRaBS J1222+0413 & \emph{XMM-Newton} & 2006-07-12 & 2 ks & 0401790601\\
& \emph{NuSTAR} & 2017-06-27 & 32 ks & 60301018002\\
SDSS J1246+0238 & \emph{XMM-Newton} & 2012-12-14 & 13 ks & 0690090201\\
PKS 1502+036 & \emph{XMM-Newton} & 2012-08-07 & 14 ks & 0690090101\\
& \emph{NuSTAR} & 2017-02-12 & 115 ks & 60201044002\\
RGB J1644+263 & \emph{XMM-Newton} & 2017-03-03 & 45 ks & 0783230101\\
PKS 2004$-$447 & \emph{XMM-Newton} & 2004-04-11 & 17 ks & 0200360201\\
& \emph{XMM-Newton} & 2012-05-01 & 18 ks & 0694530101\\
& \emph{XMM-Newton} & 2012-10-18 & 28 ks & 0694530201\\
& \emph{XMM-Newton} & 2016-05-05 & 32 ks & 0790630101\\
TXS 2116$-$077 & \emph{XMM-Newton} & 2016-05-10 & 19 ks & 0784090201\\
& \emph{XMM-Newton} & 2016-10-27 & 16 ks & 0784090301\\
\hline
\end{tabular}
\end{center}
\end{table*}

\clearpage
\pagebreak
\begin{deluxetable*}{lccccccc}
\tabletypesize{\small}
\tablecaption{Fit parameters for the basic fits to the X-ray spectra. \label{table:xrayfitparams}}
\tablewidth{0pt}
\tablehead{
\colhead{Source Name} & \colhead{Spectrum} & \colhead{$\Gamma_1$} & \colhead{$\Gamma_2$} & \colhead{kT (keV)}  & \colhead{$\chi^2$/dof}  & \colhead{$F_\mathrm{0.5-10}$} & \colhead{$F_\mathrm{10-50}$}}
\startdata
1H 0323+342       & NuSTAR 2014       & 1.80 $\pm$ 0.01          &  --                      & --                      & 372/333   &  --   & 12.8  \\
                  &                   & 1.86 $\pm$ 0.04          & 0.3 $\pm$ 1.5            & --                      & 362/331   &  --   & 12.8  \\
                  &                   & 1.79 $\pm$ 0.02          &  --                      &0.5 $\pm$ 0.5            & 371/331   &  --   & 12.9  \\
                  & Suzaku 2009       & 2.008 $\pm$ 0.006        &  --                      & --                      & 918/222   &  --   &  --   \\
                  &                   & 1.53 $\pm$ 0.06          & 2.84 $\pm$ 0.09          & --                      & 347/220   &  --   &  --   \\
                  &                   & 1.86 $\pm$ 0.01          &  --                      &0.145 $\pm$ 0.003        & 312/220   &  --   &  --   \\
                  & Suzaku 2013       & 1.969 $\pm$ 0.005        &  --                      & --                      & 943/256   &  --   &  --   \\
                  &                   & 1.73 $\pm$ 0.03          & 3.3 $\pm$ 0.1            & --                      & 423/254   &  --   &  --   \\
                  &                   & 1.863 $\pm$ 0.008        &  --                      &0.128 $\pm$ 0.003        & 369/254   &  --   &  --   \\
                  & XMM 2015          & 2.041 $\pm$ 0.003        &  --                      & --                      & 2078/168  & 11.9  &  --   \\
                  &                   & 2.30 $\pm$ 0.02          & 0.92 $\pm$ 0.06          & --                      & 432/166   & 13.0  &  --   \\
                  &                   & 1.802 $\pm$ 0.007        &  --                      &0.138 $\pm$ 0.001        & 439/166   & 12.8  &  --   \\
PMN J0948+0022    & XMM 2011          & 1.86 $\pm$ 0.01          &  --                      & --                      & 601/128   & 3.29  &  --   \\
                  &                   & 1.35 $\pm$ 0.06          & 2.9 $\pm$ 0.1            & --                      & 103/126   & 3.72  &  --   \\
                  &                   & 1.59 $\pm$ 0.02          &  --                      &0.103 $\pm$ 0.003        & 111/126   & 3.66  &  --   \\
                  & XMM/NuSTAR 2016   & 1.607 $\pm$ 0.006        &  --                      & --                      & 936/354   & 2.08  & 2.08  \\
                  &                   & 1.35 $\pm$ 0.03          & 2.7 $\pm$ 0.1            & --                      & 349/352   & 2.05  & 2.05  \\
                  &                   & 1.482 $\pm$ 0.009        &  --                      &0.109 $\pm$ 0.004        & 406/352   & 2.08  & 2.08  \\
FBQS J1102+2239   & XMM 2012          & 2.3 $\pm$ 0.2            &  --                      & --                      & 0/1       & 0.133 &  --   \\
CGRaBS J1222+0413 & NuSTAR 2017       & 1.37 $\pm$ 0.05          &  --                      & --                      & 40/40     &  --   & 1.98  \\
                  &                   & 10.0 $\pm$ 11.0          & 1.36 $\pm$ 0.05          & --                      & 40/38     &  --   & 1.98  \\
                  &                   & 1.37 $\pm$ 0.05          &  --                      &0.04 $\pm$ 0.04          & 40/38     &  --   & 1.99  \\
                  & XMM 2006          & 1.50 $\pm$ 0.04          &  --                      & --                      & 37/28     & 2.73  &  --   \\
                  &                   & 0.8 $\pm$ 0.9            & 1.9 $\pm$ 0.3            & --                      & 29/26     & 2.98  &  --   \\
                  &                   & 1.3 $\pm$ 0.1            &  --                      &0.16 $\pm$ 0.04          & 31/26     & 2.95  &  --   \\
SDSS J1246+0238   & XMM 2012          & 1.9 $\pm$ 0.1            &  --                      & --                      & 6/4       & 0.0963&  --   \\
                  &                   & 2.1 $\pm$ 0.1            & -2.9 $\pm$ -3.0          & --                      & 2/2       & 0.0963&  --   \\
                  &                   & 1.5 $\pm$ 0.3            &  --                      &0.12 $\pm$ 0.06          & 3/2       & 0.108 &  --   \\
PKS 1502+036      & NuSTAR 2017       & 1.11 $\pm$ 0.07          &  --                      & --                      & 23/22     &  --   & 0.372 \\
                  &                   & 10.0 $\pm$ 6.3           & 1.05 $\pm$ 0.06          & --                      & 22/20     &  --   & 0.372 \\
                  &                   & 1.07 $\pm$ 0.08          &  --                      &0.057 $\pm$ 0.05         & 20/20     &  --   & 16.9  \\
                  & XMM 2012          & 1.87 $\pm$ 0.05          &  --                      & --                      & 59/28     & 0.294 &  --   \\
                  &                   & 2.3 $\pm$ 0.2            & 0.7 $\pm$ 1.2            & --                      & 40/26     & 0.348 &  --   \\
                  &                   & 1.6 $\pm$ 0.1            &  --                      &0.093 $\pm$ 0.03         & 44/26     & 0.324 &  --   \\
RGB J1644+263     & XMM 2017          & 1.780 $\pm$ 0.008        &  --                      & --                      & 208/134   & 2.74  &  --   \\
                  &                   & 2.3 $\pm$ 0.3            & 1.5 $\pm$ 0.2            & --                      & 158/132   & 2.74  &  --   \\
                  &                   & 1.70 $\pm$ 0.01          &  --                      &0.113 $\pm$ 0.008        & 150/132   & 2.82  &  --   \\
PKS 2004$-$447      & XMM 2004          & 1.57 $\pm$ 0.02          &  --                      & --                      & 89/74     & 1.37  &  --   \\
                  &                   & 7.0 $\pm$ 2.2            & 1.53 $\pm$ 0.03          & --                      & 72/72     & 1.37  &  --   \\
                  &                   & 1.53 $\pm$ 0.02          &  --                      &0.045 $\pm$ 0.01         & 72/72     & 1.39  &  --   \\
                  & XMM 2012A         & 1.66 $\pm$ 0.03          &  --                      & --                      & 30/37     & 0.431 &  --   \\
                  &                   & 2.6 $\pm$ 2.6            & 1.6 $\pm$ 1.6            & --                      & 30/35     & 0.431 &  --   \\
                  &                   & 1.64 $\pm$ 0.07          &  --                      &0.2 $\pm$ 0.2            & 30/35     & 0.433 &  --   \\
                  & XMM 2012B         & 1.66 $\pm$ 0.02          &  --                      & --                      & 94/67     & 0.665 &  --   \\
                  &                   & 9.8 $\pm$ 4.3            & 1.65 $\pm$ 0.02          & --                      & 93/65     & 0.665 &  --   \\
                  &                   & 1.62 $\pm$ 0.04          &  --                      &0.29 $\pm$ 0.04          & 90/65     & 0.662 &  --   \\
                  & XMM 2016          & 1.62 $\pm$ 0.02          &  --                      & --                      & 78/83     & 0.888 &  --   \\
                  &                   & 6.5 $\pm$ 2.5            & 1.59 $\pm$ 0.02          & --                      & 74/81     & 0.888 &  --   \\
                  &                   & 1.59 $\pm$ 0.02          &  --                      &0.048 $\pm$ 0.02         & 74/81     & 0.895 &  --   \\
TXS 2116$-$077      & XMM 2016A         & 1.65 $\pm$ 0.02          &  --                      & --                      & 55/50     & 0.647 &  --   \\
                  &                   & 1.7 $\pm$ 1.7            & 1.7 $\pm$ 1.7            & --                      & 55/48     & 0.647 &  --   \\
                  &                   & 1.58 $\pm$ 0.05          &  --                      &0.31 $\pm$ 0.04          & 42/48     & 0.636 &  --   \\
                  & XMM 2016B         & 1.50 $\pm$ 0.04          &  --                      & --                      & 14/19     & 0.27  &  --   \\
                  &                   & 1.5 $\pm$ 1.5            & 1.5 $\pm$ 1.5            & --                      & 14/17     & 0.27  &  --   \\
                  &                   & 1.49 $\pm$ 0.07          &  --                      &0.5 $\pm$ 0.2            & 11/17     & 0.261 &  --   \\
\enddata
\tablecomments{We fit three models to each spectrum: a powerlaw, two powerlaws, and a powerlaw plus black body. Fluxes are in units of 10$^{-12}$~erg~cm$^{-2}$~s$^{-1}$. We report 0.5--10~keV fluxes for \xmm\ and 10--50~keV fluxes for \nustar. Note that only the powerlaw model was fit to FBQS J1102+2239, as the spectrum did not have enough bins to constrain more complex models.}
\end{deluxetable*}

\begin{table*}[h!]
\centering
\caption{Best-fit parameters for the relativistic line model fit to the X-ray spectra of 1H~0323+342.}
\label{tab:relline}
\begin{tabular}{l c l}
\hline
Parameter & Value & Description\\
\hline
$E$			&	$6.79_{-0.06}^{+0.05}$	&	Energy (keV)	\\
$i$			&	$<7$		&	Inclination (degrees) \\
$R_\mathrm{in}$	&	$9.7_{-0.8}^{+1.1}$		&	Inner radius ($R_\mathrm{G}$) \\
$\chi^2$/dof	&	1208/948 & Fit statistic\\
\hline
\end{tabular}
\end{table*}

\begin{table*}[t!]
\begin{center}
\caption{Summary of the soft X-ray analysis done for the broadband SED modeling.\label{tab:x-ray_sed}}
\begin{tabular}{lcccccccc}
\tableline\tableline
 Name                  & Exp.    &      $N_{\rm H}$             & $\Gamma_{\rm X}/\Gamma_{\rm X,1}$   &   $\Gamma_{\rm X,2}$       & $E_{\rm b}$    & $F_{\rm X}$                                             & Stat. & Fit\\
                           & (ksec) &(10$^{20}$ cm$^{-2}$)  &                                                                 &                                             & (keV)              & (10$^{-12}$ erg cm$^{-2}$ s$^{-1}$)  & $\chi^2$ or C-stat./dof  & \\ 
\tableline
1H 0323+342            & 149.5   & 12.7    & 2.08$^{+0.03}_{-0.03}$   & 1.85$^{+0.02}_{-0.02}$   & 1.40$^{+0.15}_{-0.14}$     & 24.86$^{+0.28}_{-0.24}$  & 716.13/538 & chi\\
SBS 0846+513          & 50.13   & 2.91    & 1.55$^{+0.09}_{-0.09}$   & ...                                      & ...                                        & 0.73$^{+0.07}_{-0.06}$    & 41.67/30     & chi\\
CGRaBS J0932+5306 & 5.89     & 1.49    & 1.63$^{+0.14}_{-0.15}$   & ...                                      & ...                                        & 0.46$^{+0.05}_{-0.05}$  & 154.63/162   & c-stat\\
GB6 J0937+5008      & 4.90     & 1.33    & 1.46$^{+0.14}_{-0.15}$   & ...                                      & ...                                        & 3.10$^{+0.43}_{-0.39}$  & 167.48/187   & c-stat\\
PMN J0948+0022      & 84.53   & 5.08    & 1.78$^{+0.11}_{-0.09}$   & 1.51$^{+0.05}_{-0.06}$   & 1.17$^{+0.48}_{-0.24}$     & 4.02$^{+0.20}_{-0.14}$  & 192.99/235   & chi\\
TXS 0955+326          & 1.95     & 1.60    & 1.69$^{+0.57}_{-0.56}$   & ...                                      & ...                                        & 0.60$^{+0.48}_{-0.23}$  & 9.24/19         & c-stat\\
FBQS J1102+2239    & 9.05    & 1.12    & 2.00$^{+0.14}_{-0.14}$   & ...                                      & ...                                        & 0.38$^{+0.04}_{-0.04}$  & 229.13/160     & c-stat \\
CGRaBS J1222+0413 & 27.26   & 1.63    & 1.28$^{+0.06}_{-0.06}$   & ...                                      & ...                                        & 3.80$^{+0.22}_{-0.24}$  & 104.44/72     & chi \\
SDSS J1246+0238 & 24.25 & 1.96 & 1.57$^{+0.30}_{-0.29}$   & ...                                      & ...                                        & 0.17$^{+0.06}_{-0.04}$  & 99.34/73     & c-stat \\
TXS 1419+391          &  2.98    & 1.01    & 1.79$^{+0.44}_{-0.43}$   & ...                                      & ...                                        & 0.57$^{+0.34}_{-0.17}$  & 30.52/28       & c-stat\\
PKS 1502+036          & 43.45   & 3.93    & 1.63$^{+0.15}_{-0.15}$   & ...                                      & ...                                        & 0.40$^{+0.07}_{-0.04}$  & 23.95/31       & chi \\
TXS 1518+423          & 3.08     & 1.89    & 1.70$^{+0.68}_{-0.69}$   & ...                                      & ...                                        & 0.39$^{+0.37}_{-0.16}$  & 19.81/18       & c-stat\\
RGB J1644+263        & 16.82   & 5.14    & 1.80$^{+0.11}_{-0.11}$   & ...                                      & ...                                        & 1.78$^{+0.19}_{-0.15}$  & 54.30/55       & chi \\
PKS 2004$-$447      & 74.38   & 3.17    & 1.49$^{+0.08}_{-0.08}$   & ...                                      & ...                                         & 0.61$^{+0.05}_{-0.05}$  & 35.74/37       & chi \\
TXS 2116$-$077      & 16.23   & 6.99    & 1.50$^{+0.08}_{-0.07}$   & ...                                      & ...                                         & 0.35$^{+0.03}_{-0.02}$  & 64.35/69       & chi \\
PMN J2118+0013     & 14.33   & 5.82    & 1.94$^{+0.28}_{-0.27}$   & ...                                      & ...                                        & 0.35$^{+0.06}_{-0.06}$  & 66.33/79       & c-stat\\
 \tableline
\end{tabular}
\end{center}
\tablecomments{The spectral parameter $\Gamma_{\rm X}$/$\Gamma_{\rm X,1}$ represents the best-fit photon index for a power-law model or photon index before the break ($E_{\rm b}$, in keV) for the broken power-law model, respectively. On the other hand, $\Gamma_{\rm X,2}$ refers to the photon index after the break. $F_{\rm X}$ is the integrated X-ray flux in the energy range of 0.3$-$10 keV. Note that the spectral parameters associated with CGRaBS J0932+5306, FBQS J1102+2239 are derived from the {\it Chandra}  data analysis, whereas, {\it XMM-Newton} data are considered for TXS 2116$-$077.}
\end{table*}

\begin{table*}[t!]
\begin{center}
\caption{Summary of the hard X-ray analysis done for the broadband SED modeling.\label{tab:hard-x-ray_sed}}
\begin{tabular}{lcccccccc}
\tableline\tableline
 Name                  & $\Gamma_{\rm X}$    & $F_{\rm X}$                                                & Stat./dof                        & Instrument\\
                           &                                   & (10$^{-12}$ erg cm$^{-2}$ s$^{-1}$)  & $\chi^2$  &                 \\ 
\tableline
1H 0323+342            & 1.62$^{+0.31}_{-0.29}$  & 27.31$^{+2.44}_{-3.41}$  & 3.87/6       & BAT\\
PMN J0948+0022      & 1.05$^{+0.33}_{-0.35}$  & 20.05$^{+1.30}_{-1.29}$  & 4.09/6       & BAT\\
CGRaBS J1222+0413 & 1.45$^{+0.21}_{-0.20}$  & 36.22$^{+3.30}_{-3.56}$  & 1.93/6       & BAT \\
PKS 1502+036          & 1.15$^{+0.10}_{-0.09}$  & 2.34$^{+0.30}_{-0.34}$  & 95.35/101   & \nustar\\
PKS 2004$-$447      & 1.69$^{+0.06}_{-0.06}$   & 2.90$^{+0.26}_{-0.22}$  & 256.95/217 & \nustar \\
 \tableline
\end{tabular}
\end{center}
\tablecomments{The parameters are derived by fitting a simple power-law model in the energy range of 14$-$195 keV for BAT and 3$-$79 keV for \nustar~spectra.}
\end{table*}

\begin{table*}
\caption{The physical parameters derived from the modeling of the \gm-NLSy1 galaxies.\label{tab:sed_param}}
\begin{center}
\begin{tabular}{lcccccccccccccccc}
\hline
Name & $\theta_{\rm v}$ & $M_{\rm BH}$ & $L_{\rm disk}$ & $L_{\rm cor}$ & $R_{\rm diss}$ & $R_{\rm BLR}$ & $\delta_{\rm b}$ & $\Gamma_{\rm b}$ & $B$ & $p$ & $q$ & $\gamma_{\rm min}$ & $\gamma_{\rm b}$ & $\gamma_{\rm max}$ & $U_{\rm e}$ & CD \\
~[1] & [2] & [3]  & [4] & [5] & [6]  & [7] & [8] & [9] & [10] & [11] & [12] & [13] & [14] & [15]& [16]&[17]\\ 
\hline
1H 0323+342            & 3.0  &  7.30  &  45.30  &  44.30  &  0.023  &  0.046	    &    13.6  &  8	  &  1.05  &  1.8    &    4.9  &  1  &  89	&    2000  &   0.073 &  43.27\\
SBS 0846+513          & 2.0  &  7.59  &  44.43  &  43.43  &  0.023  &  0.017	    &    19.1  &  11  &  1.70  &  1.6    &    3.8  &  1  &  590	&    3000  &   0.210 &  1.64\\
CGRaBS J0932+5306 & 3.0  &  8.00  &  45.70  &  44.70  &  0.033  &  0.072	&    14.7  &  9	  &  1.30  &  1.9    &    3.7  &  1  &  77	&    3000  &   0.119 &  31.31\\
GB6 J0937+5008      & 3.0  &  7.56  &  43.71  &  42.71  &  0.014  &  0.007	    &    15.4  &  10  &  0.80  &  1.7    &    3.6  &  1  &  860	&    4000  &   1.410 &  0.61\\
PMN J0948+0022      & 3.0  &  8.18  &  45.70  &  45.20  &  0.050  &  0.072	&    15.7  &  10  &  2.70  &  1.6    &    3.9  &  1  &  63	&    1500  &   0.232 &  9.87\\
TXS 0955+326          & 3.0  &  8.70  &  46.23  &  44.71  &  0.043  &  0.133	    &    12.3  &  7	  &  1.90  &  1.8    &    4.2  &   1  &  84	&    3000  &   0.193 &  9.75\\
FBQS J1102+2239    & 3.0  &  7.78  &  44.00  &  43.48  &  0.108  &  0.010	    &    19.0  &  17	  &  0.50  &  1.6    &    4.4  &   20  &  467&    5000  &   0.012 &  3.72\\
CGRaBS J1222+0413 & 3.0  &  8.85  &  46.18  &  45.35  &  0.094  &  0.125	&    16.5  &  11  &  2.20  &  1.7    &    4.4  &  1  &  62	&    2000  &   0.139 &  21.63\\
SDSS J1246+0238   & 3.0  &  8.48  &  44.48  &  43.38  &  0.037  &  0.029 	&    17.8  &  13  &  1.30  &  1.8    &    5.3  &  5  &  251	&    1500  &   0.030 &  5.99\\
TXS 1419+391          & 3.0  &  8.48  &  45.34  &  44.82  &  0.031  &  0.048	&    13.6  &  8	  &  2.30  &  1.7    &    4.2  &  1  &  41	&    3000  &   0.321 &  11.46\\
PKS 1502+036          & 3.0  &  7.60  &  44.78  &  43.78  &  0.126  &  0.025	&    17.2  &  12  &  0.25  &  1.6    &    4.3  &  2  &  968	&    10000 &   0.011 &  11.53\\
TXS 1518+423          & 3.0  &  7.85  &  44.70  &  44.00  &  0.100  &  0.023	&    17.8  &  13  &  0.45  &  2.1    &    4.8  &  3  &  1629    &   9000   &  0.011  & 2.66\\
RGB J1644+263        & 3.0  &  7.70  &  44.48  &  43.86  &  0.017  &  0.018	&    14.7  &  9	  &  1.30  &  2.1    &    4.5  &  1  &  103	&    2000  &   0.089 &  32.73\\
PKS 2004$-$447      & 3.0  &  6.70  &  43.00  &  42.00  &  0.019  &  0.003	&    17.2  &  12  &  0.35  &  1.8    &    4.0  &  7  &  1213    &   6500   &  0.489  & 3.68\\
TXS 2116$-$077      & 3.0  &  7.20  &  44.00  &  43.00  &  0.038  &  0.010	&    17.2  &  12  &  0.30  &  1.9    &    4.8  &  4  &  1144    &   10000  &  0.067  & 5.17\\
PMN J2118+0013     & 3.0  &  7.77  &  44.94  &  44.42  &  0.035  &  0.030	&    14.7  &  9	  &  1.90  &  1.6    &    4.5  &  1  &  193	&    2500  &   0.070 &  3.56\\
\hline
\end{tabular}
\end{center}
\tablecomments{The column information are as follows: Col.[1] name of the source; Col.[2]: viewing angle, in degrees; Col.[3]: log scale $M_{\rm BH}$, in solar mass units; Col.[4]: log scale $L_{\rm disk}$, in \lum; Col.[5]: log scale corona luminosity, in \lum; Col.[6]: distance of the emission region from the central black hole, in parsec; Col.[7]: size of the BLR, in parsec; Col.[8] and [9]:  Doppler and the bulk Lorentz factors, respectively; Col.[10]: magnetic field, in Gauss; Col.[11] and [12]: low-and high-energy slopes of the electron energy distribution, respectively; Col.[13], [14], and [15]: the minimum, break, and the maximum Lorentz factors of the radiating electrons; Col.[16]: log scale electron energy density, in erg cm$^{-3}$; and Col.[17]: Compton dominance.}
\end{table*}

\begin{table*}
\caption{Various jet powers computed from the SED modeling.\label{tab:jet_param}}
\begin{center}
\begin{tabular}{lccccc}
\hline
Name &  $P_{\rm ele}$ & $P_{\rm mag}$ & $P_{\rm rad}$ & $P_{\rm kin}$ & $P_{\rm jet}$\\
~[1] & [2] & [3]  & [4] & [5] & [6]  \\ 
\hline
1H 0323+342	           & 43.34  &  43.42  &  43.94  &  45.82  &  45.82\\
SBS 0846+513	       & 44.06  &  44.10  &  44.96  &  46.05  &  46.05\\
CGRaBS J0932+5306	& 43.99  &  44.04  &  44.96  &  46.54  &  46.54\\
GB6 J0937+5008	   & 44.35  &  42.91  &  44.51  &  46.41  &  46.41\\
PMN J0948+0021	   & 44.72  &  45.12  &  46.05  &  47.11  &  47.11\\
TXS 0955+326	       & 44.19  &  44.37  &  45.01  &  46.68  &  46.68\\
FBQS J1102+2239     & 44.57  &  44.78  &  45.52 &  45.86  &  45.91\\
CGRaBS J1222+0413	& 45.12  &  45.56  &  46.68  &  47.59  &  47.59\\
SDSS J1246+0238    & 43.81   &  44.45  & 44.68   & 45.67   & 45.70 \\
TXS 1419+391	       & 44.24  &  44.35  &  44.83  &  46.77  &  46.77\\
PKS 1502+036	       & 44.36  &  44.01  &  45.26  &  46.08  &  46.09\\
TXS 1518+423	       & 44.24  &  44.39  &  44.76  &  46.32  &  46.32\\
RGB J1644+263	       & 43.25  &  43.43  &  43.74  &  45.91  &  45.91\\
PKS 2004$-$447       & 44.37  &  42.67  &  44.37  &  45.91  &  45.92\\
TXS 2116$-$077	       & 44.10  &  43.13  &  44.15  &  45.92  &  45.93\\
PMN J2118+0013     & 43.80  &  44.41  &  44.66  &  45.99  &  46.00\\
\hline
\end{tabular}
\end{center}
\tablecomments{Col.[1]: name  of the source; Col.[2], [3], [4], and [5]: the jet power in electrons, magnetic field, radiation, and in protons, respectively. Col.[6] represents the total jet power with $P_{\rm jet}$ = $P_{\rm ele}$ + $P_{\rm mag}$ + $P_{\rm kin}$. All quantities are evaluated assuming a two-sided jet.}
\end{table*}

\clearpage
\appendix
\section{X-ray Spectral Results}
\begin{figure*}
\centering
\includegraphics[width=0.32\linewidth]{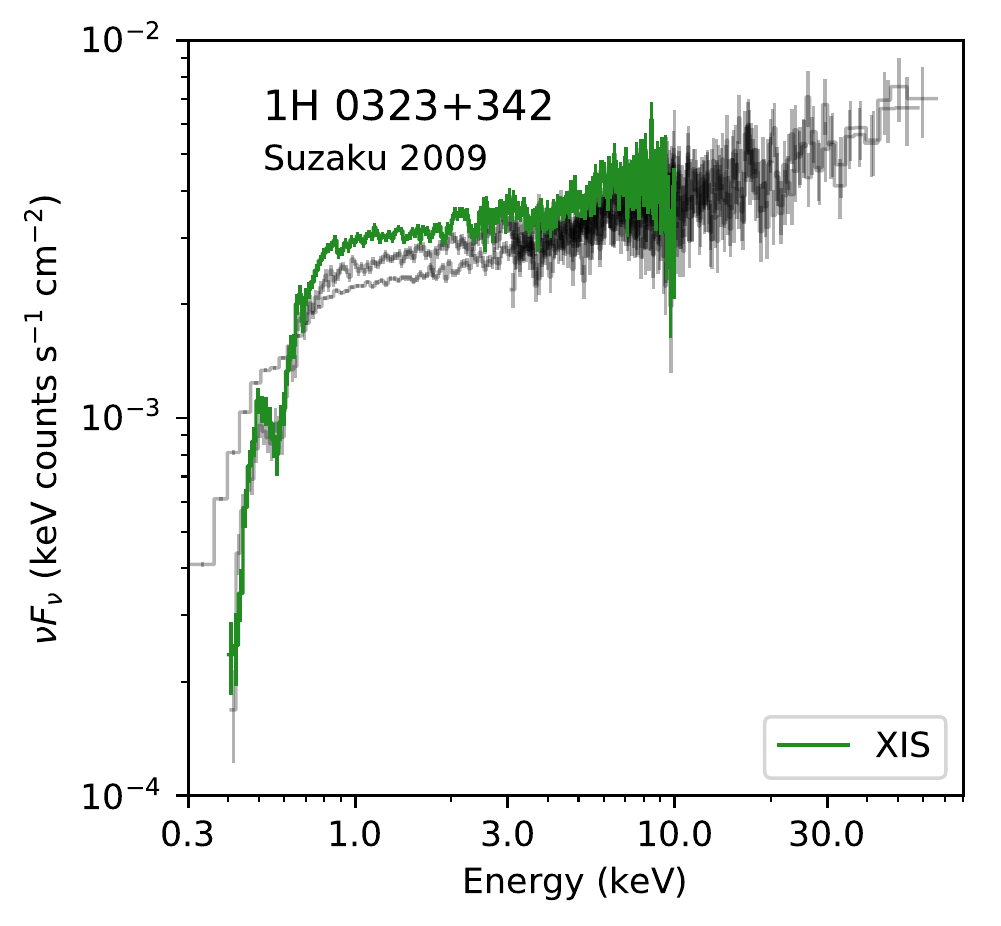}
\includegraphics[width=0.32\linewidth]{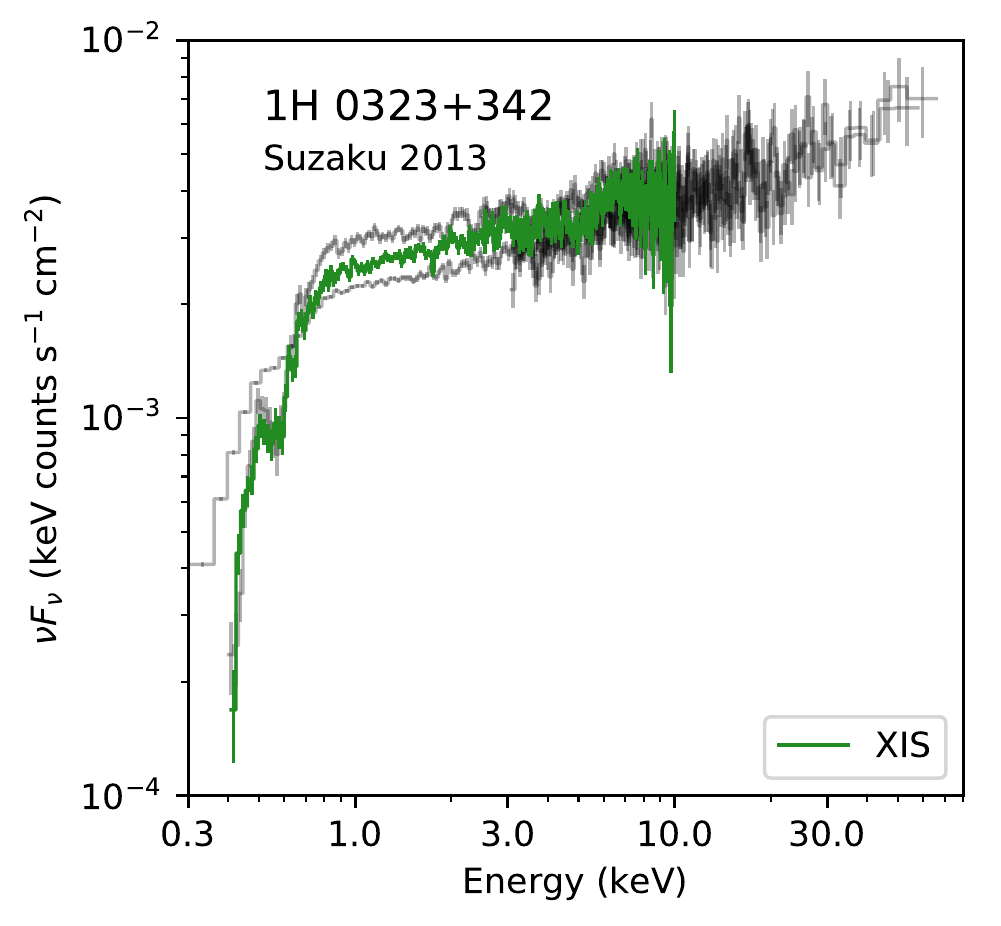}
\includegraphics[width=0.32\linewidth]{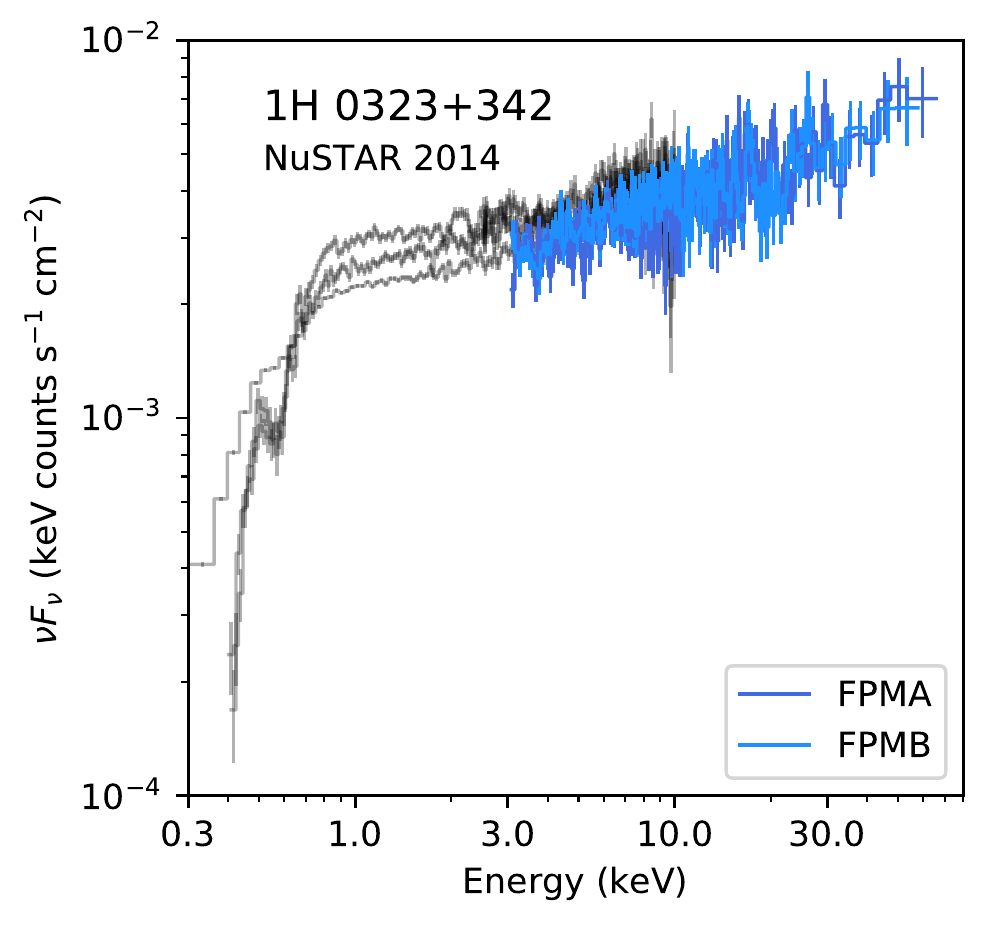}
\includegraphics[width=0.32\linewidth]{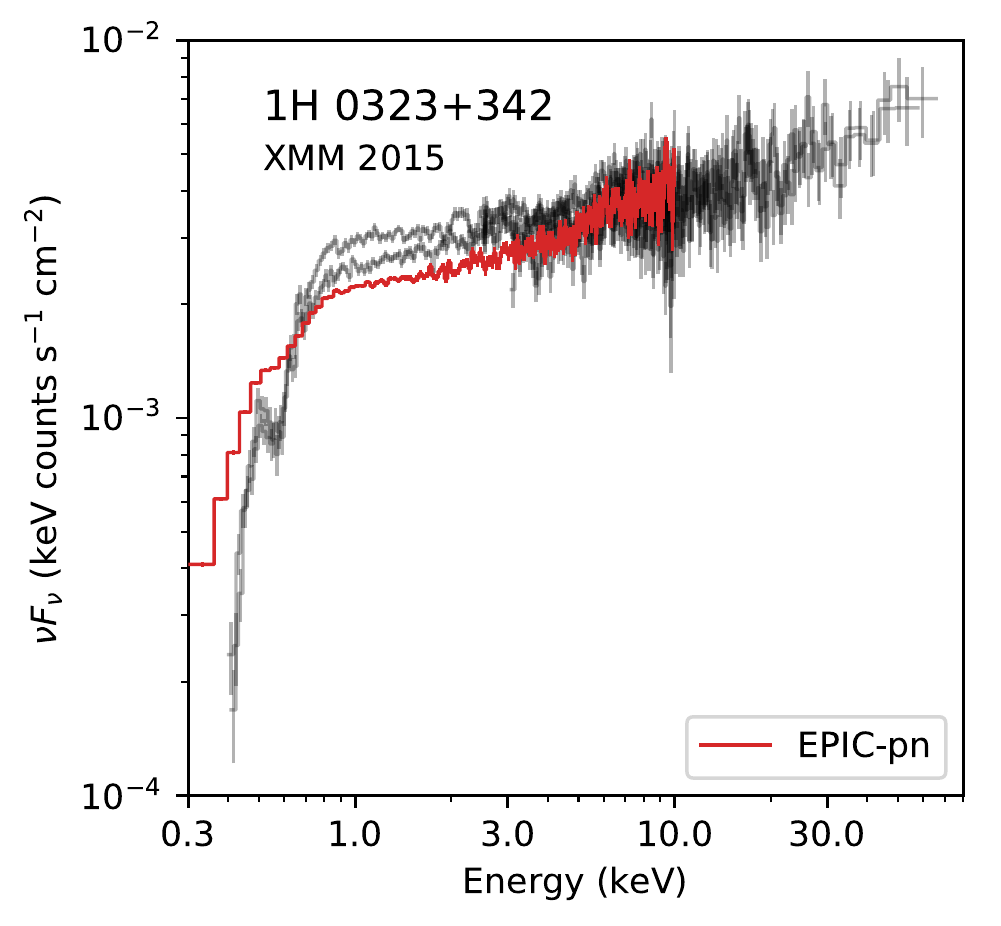}
\includegraphics[width=0.32\linewidth]{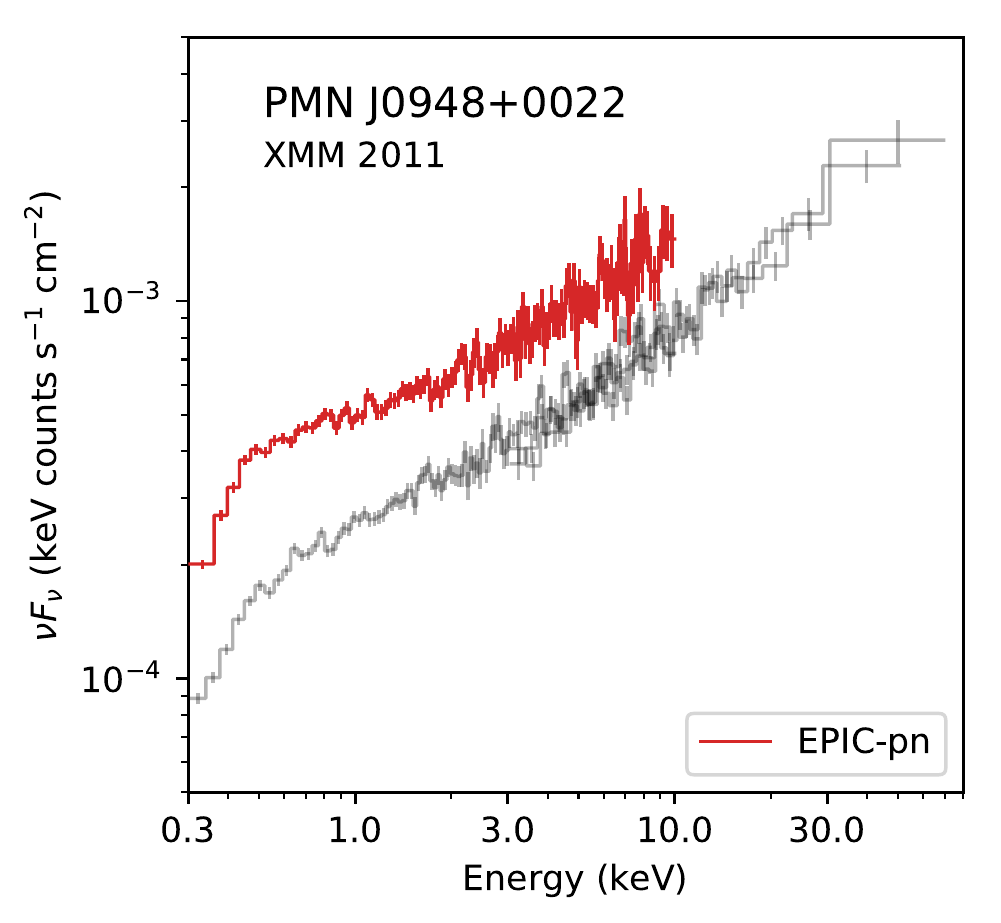}
\includegraphics[width=0.32\linewidth]{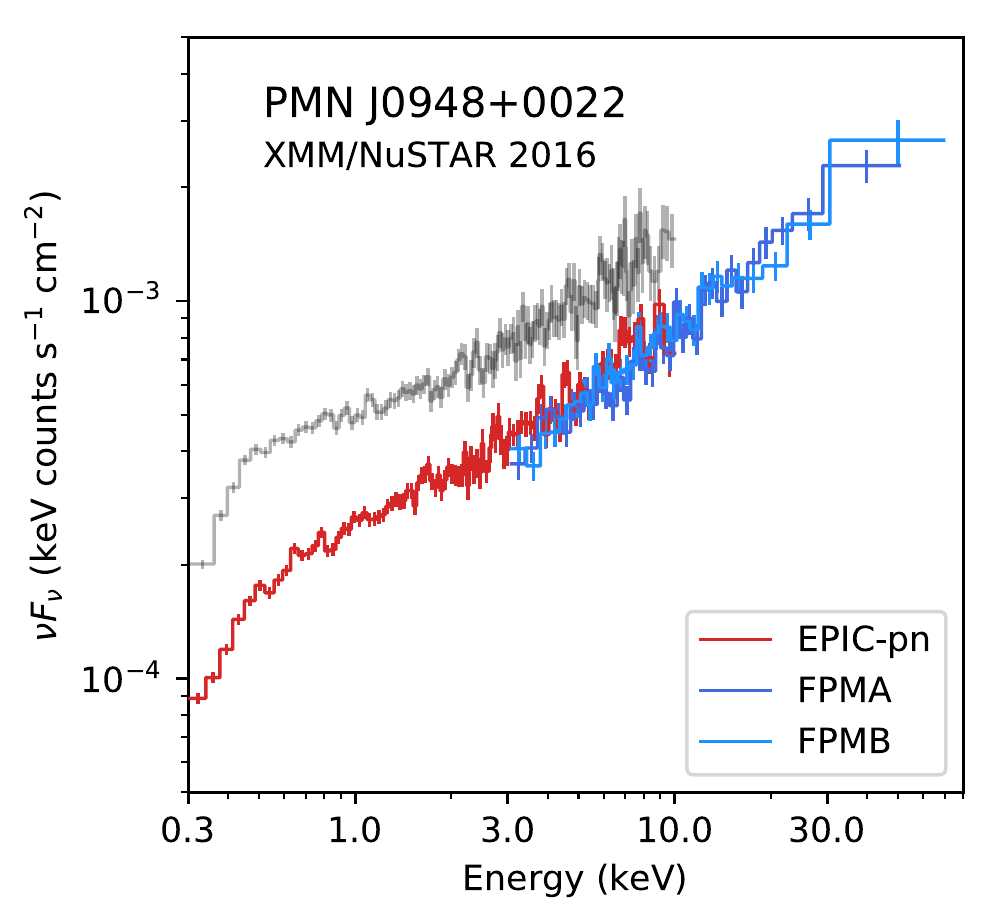}
\includegraphics[width=0.32\linewidth]{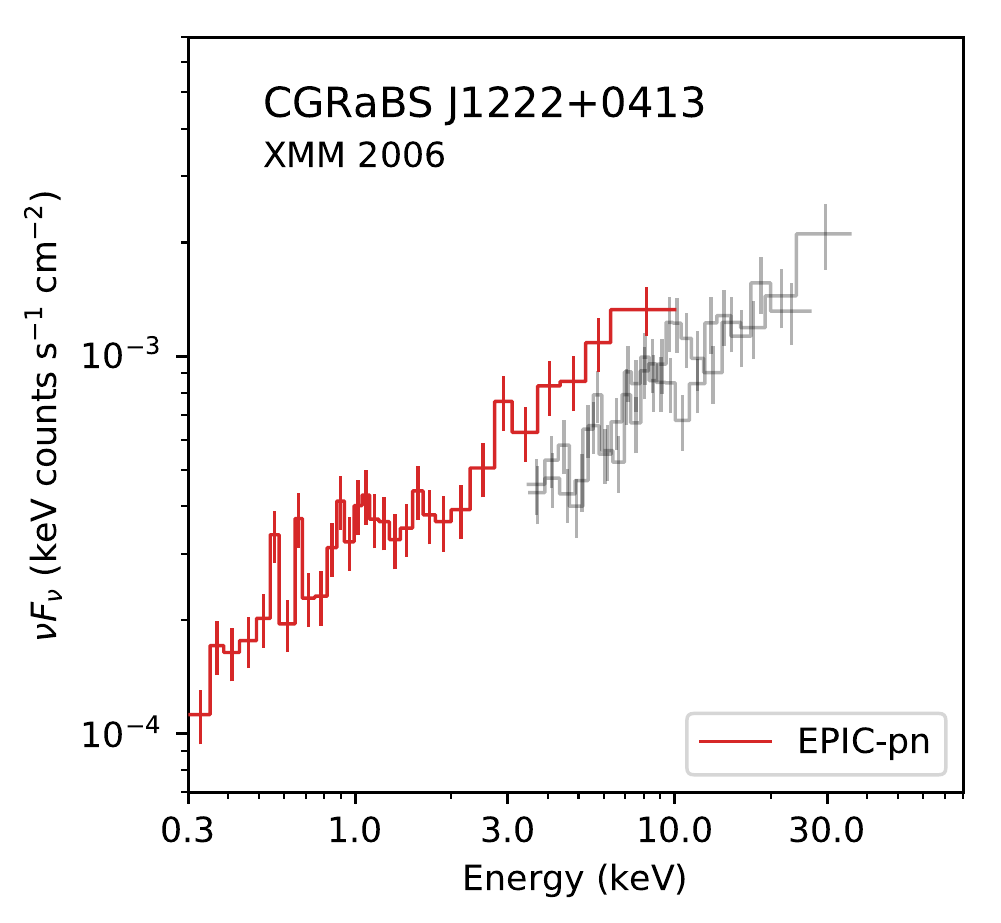}
\includegraphics[width=0.32\linewidth]{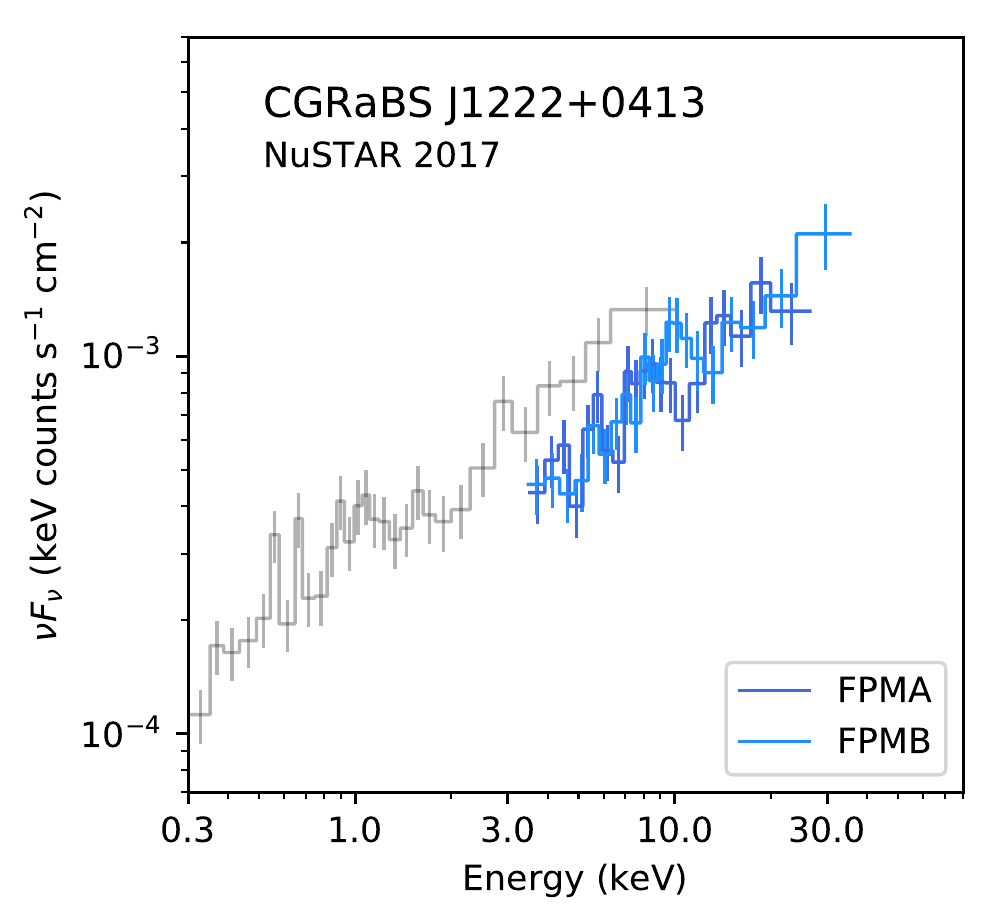}
\includegraphics[width=0.32\linewidth]{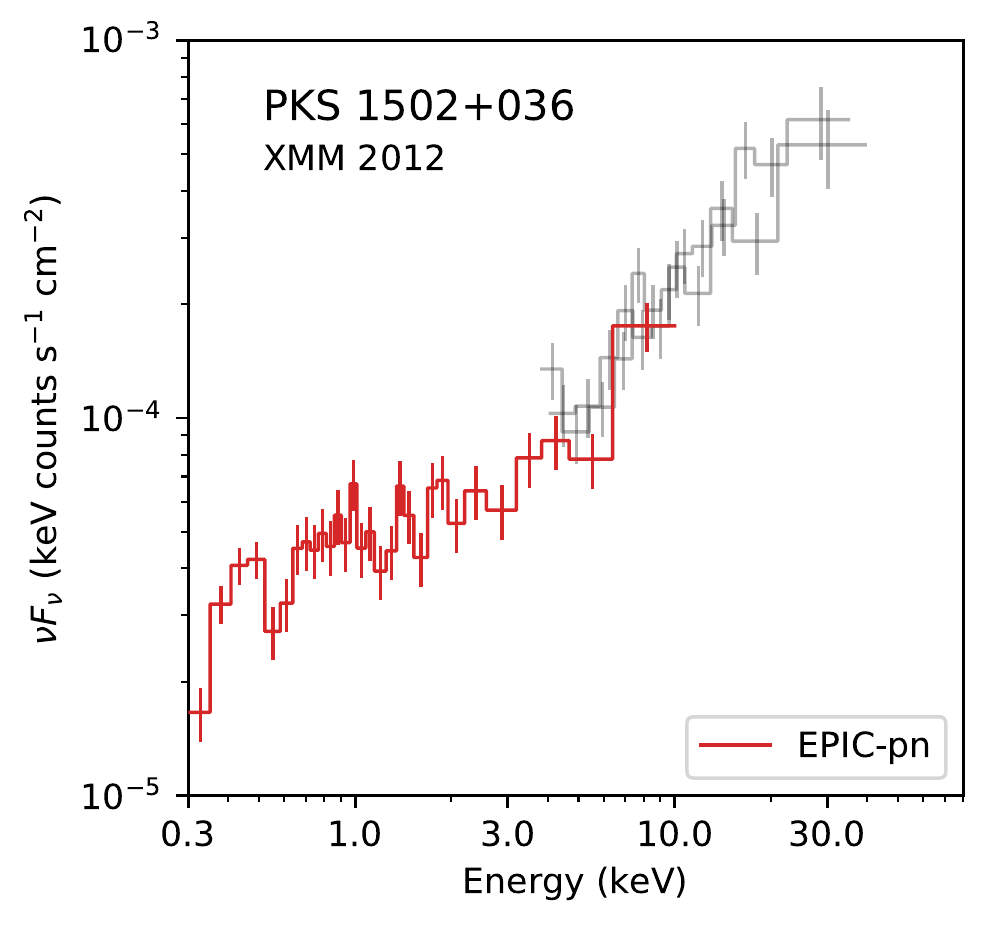}
\includegraphics[width=0.32\linewidth]{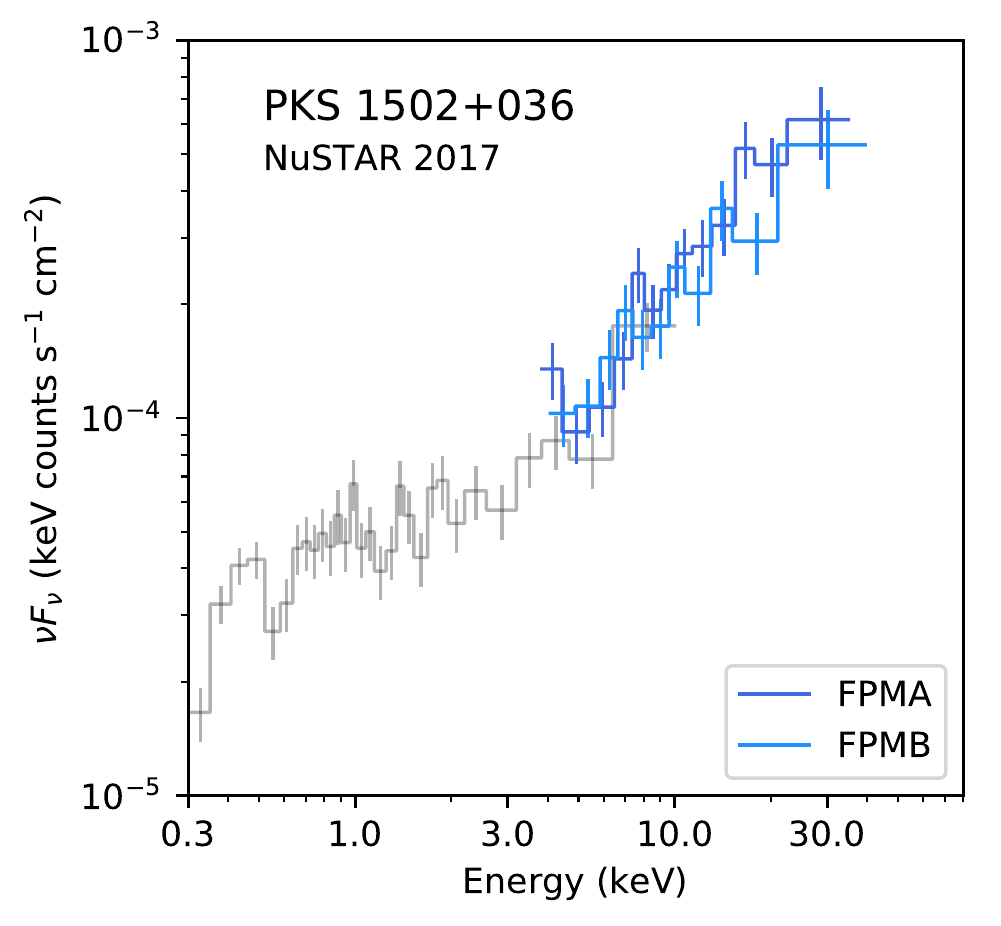}
\includegraphics[width=0.32\linewidth]{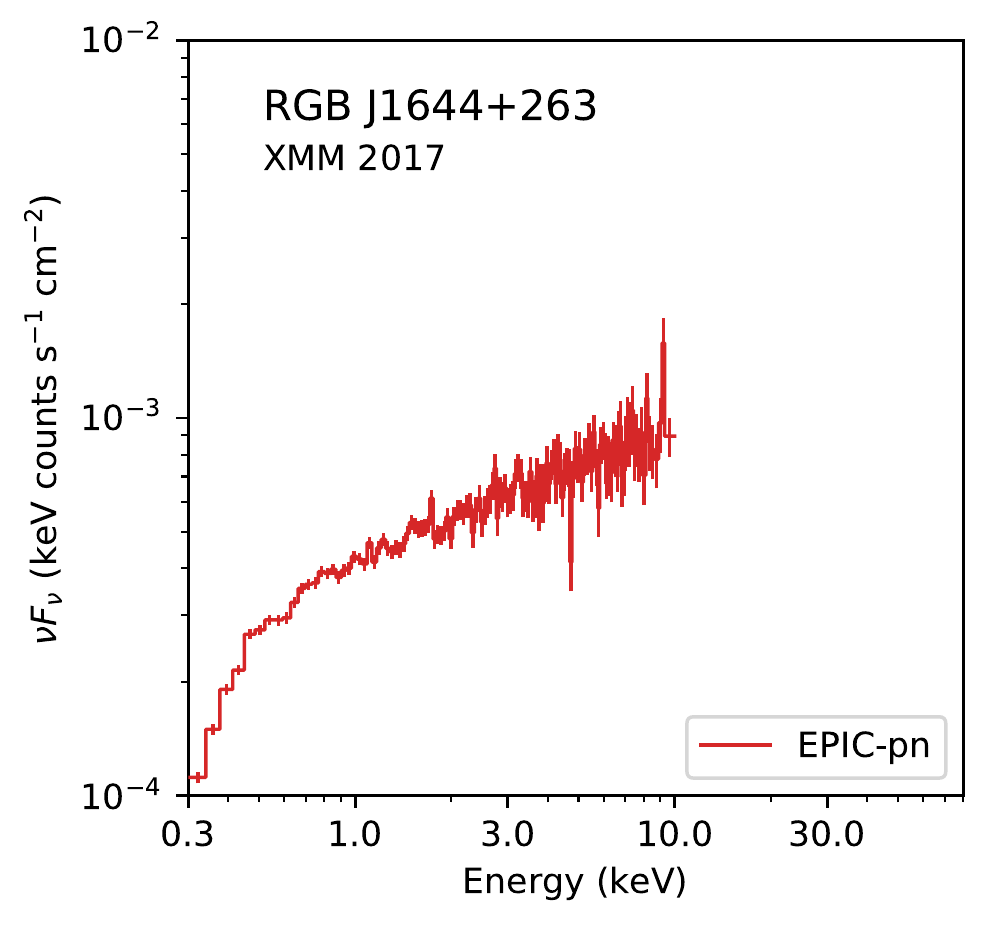}
\includegraphics[width=0.32\linewidth]{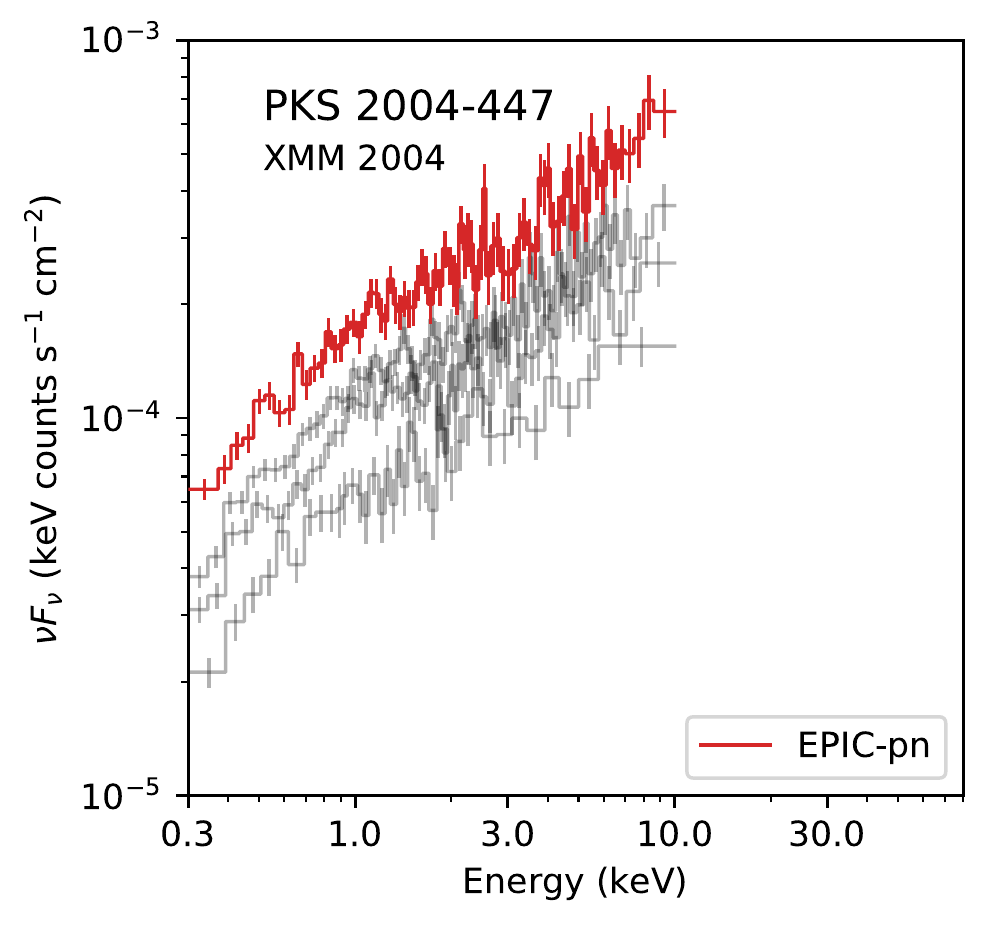}
\caption{Spectra of the sources in the X-ray sample. In each figure, the spectra from different observations are plotted as grey lines.}
\label{fig_xrayspectra_1}
\end{figure*}

\begin{figure*}
\centering
\includegraphics[width=0.32\linewidth]{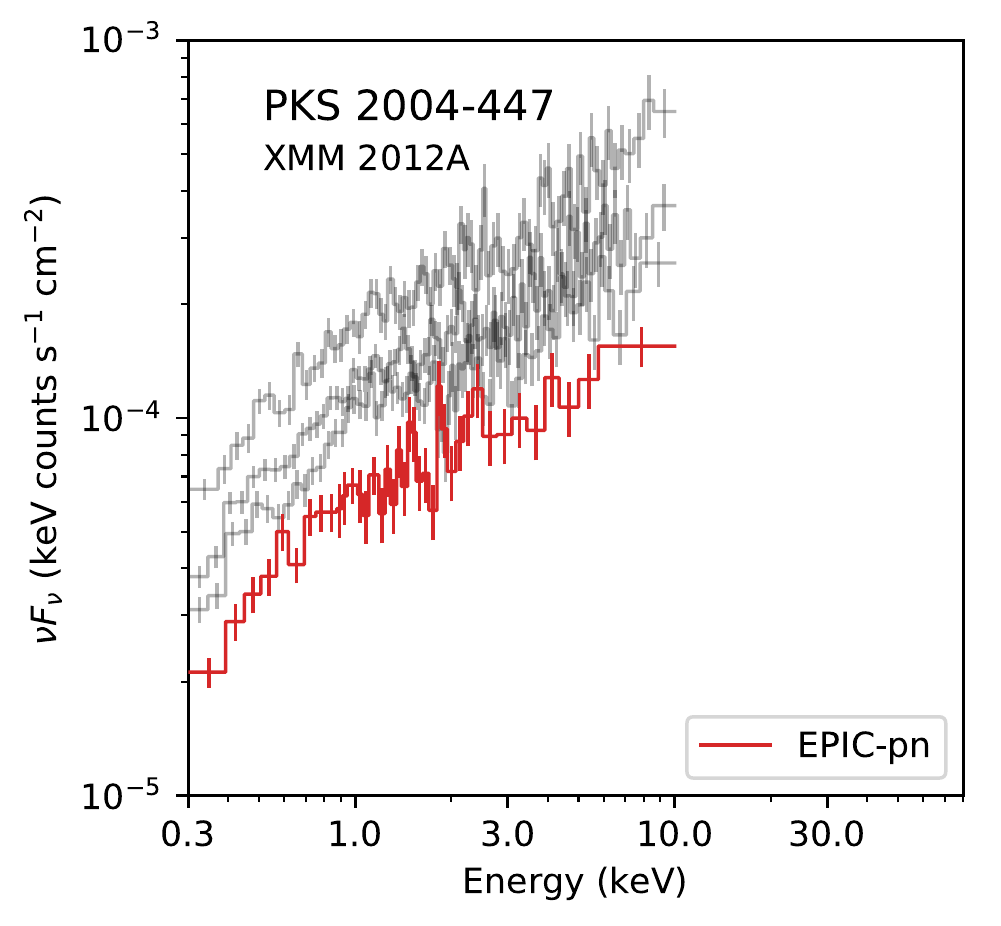}
\includegraphics[width=0.32\linewidth]{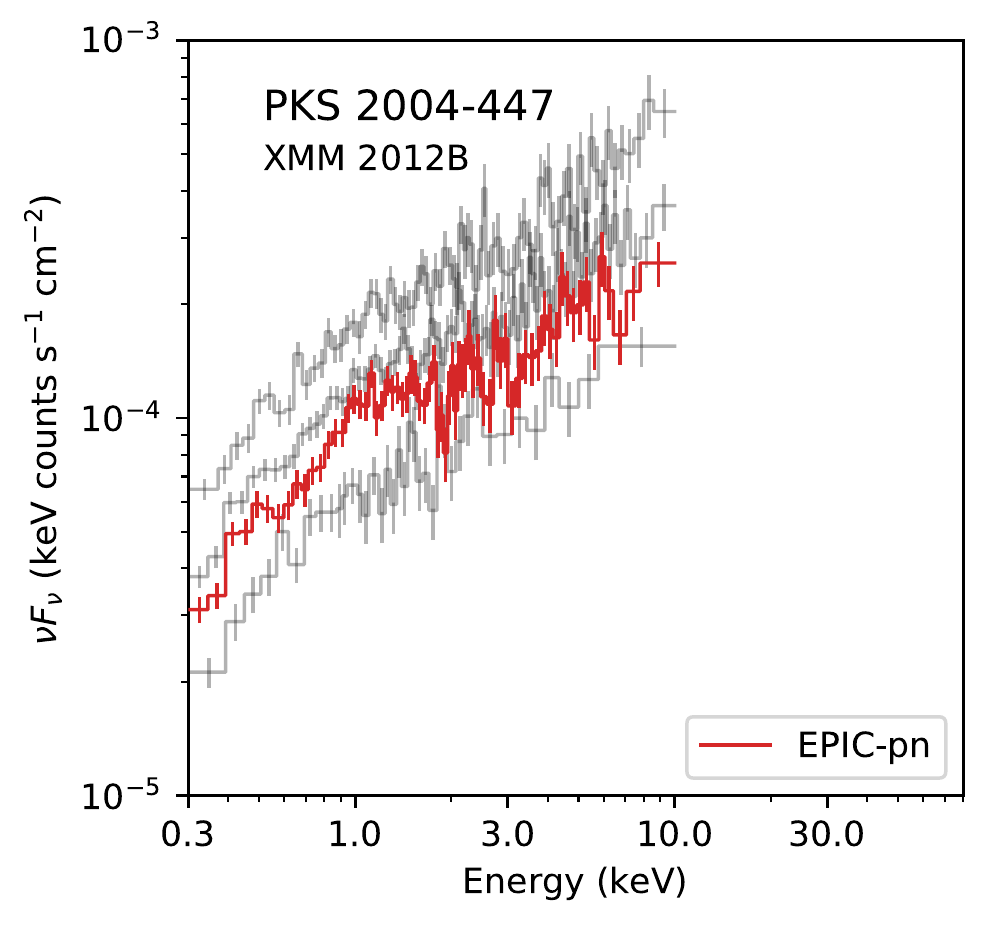}
\includegraphics[width=0.32\linewidth]{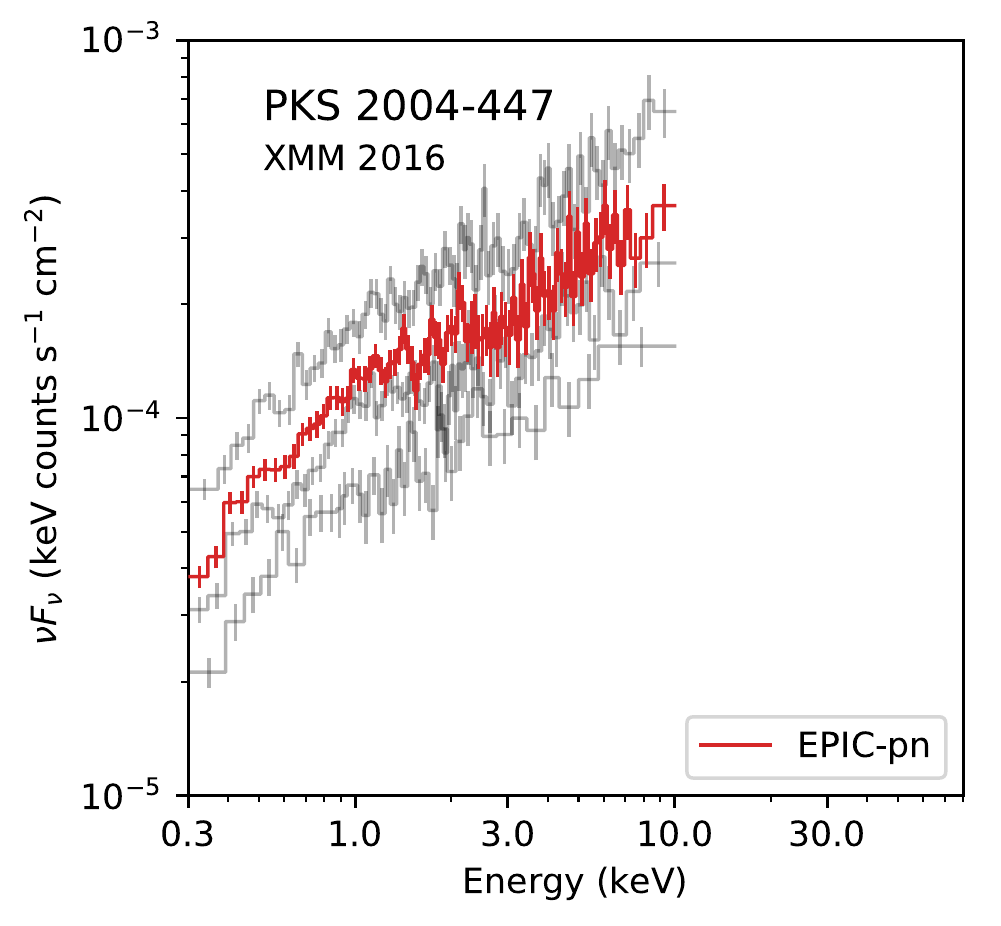}
\includegraphics[width=0.32\linewidth]{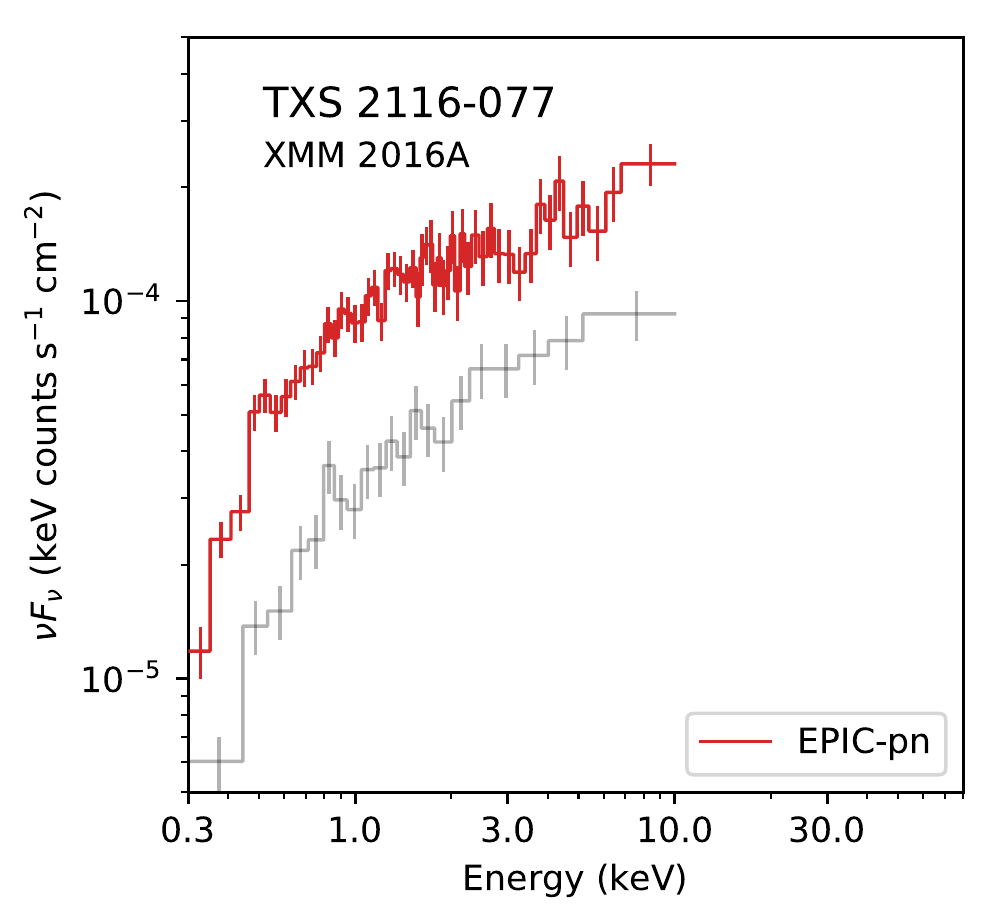}
\includegraphics[width=0.32\linewidth]{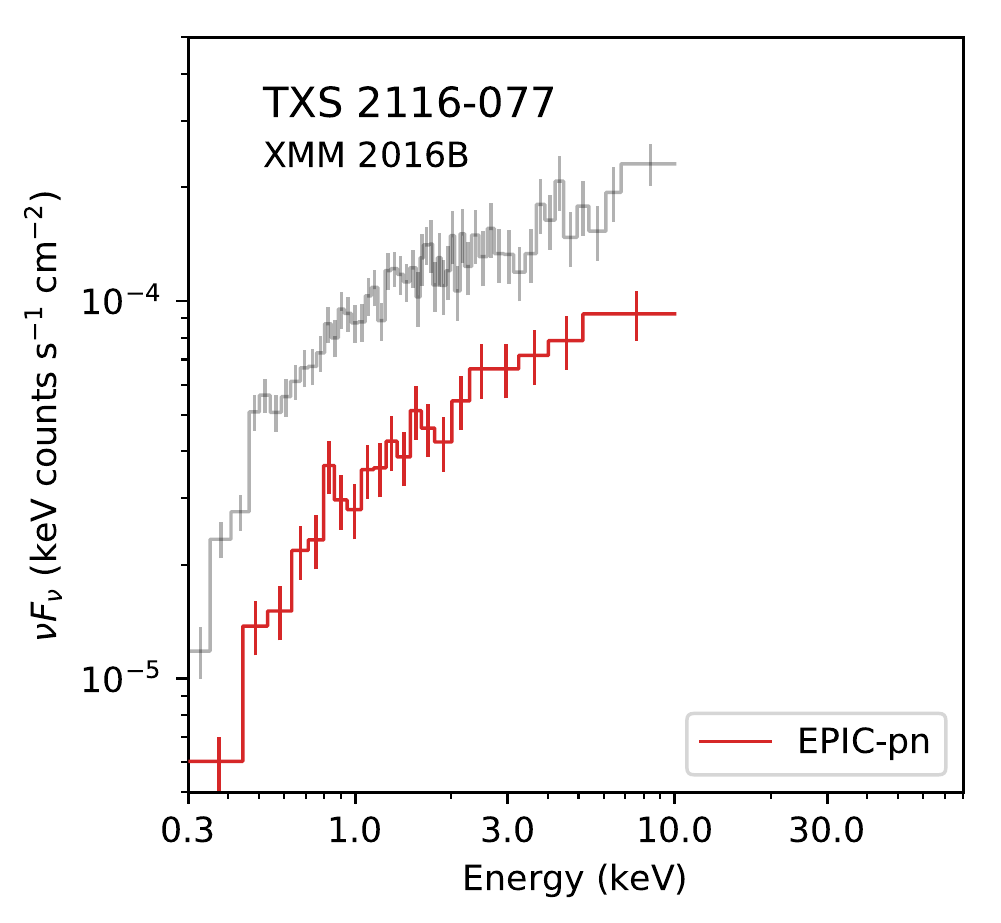}
\includegraphics[width=0.32\linewidth]{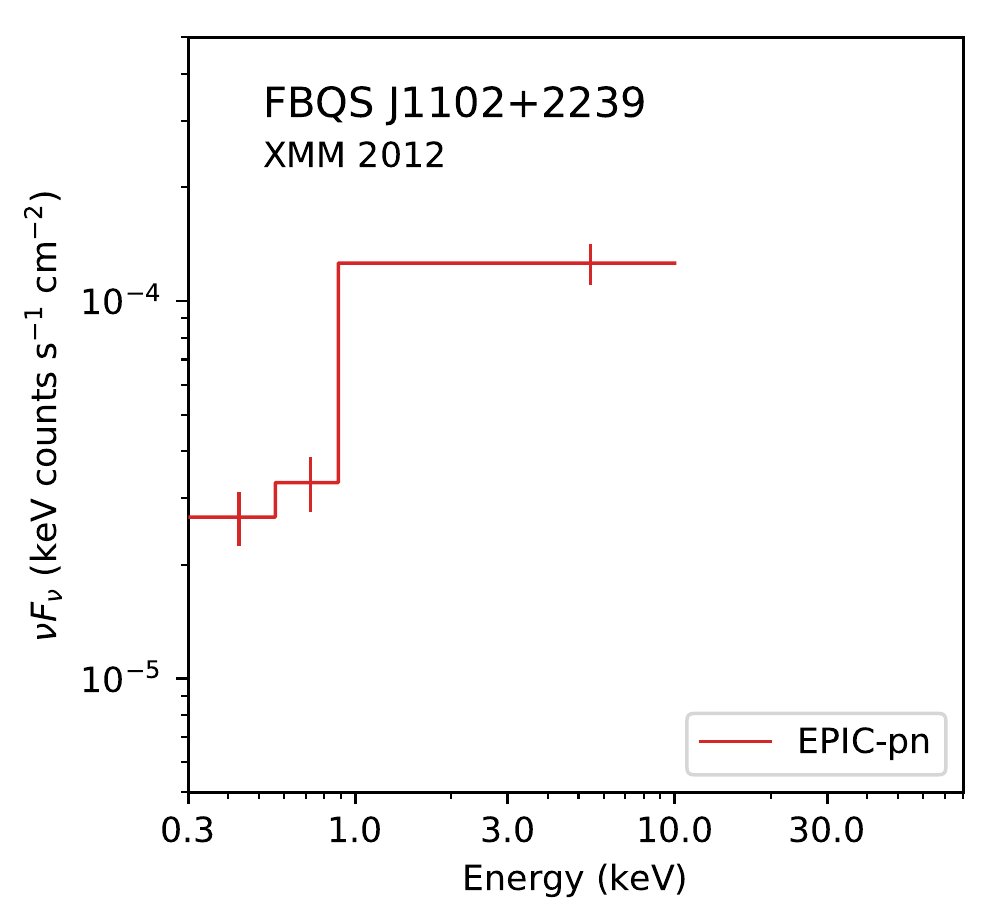}
\includegraphics[width=0.32\linewidth]{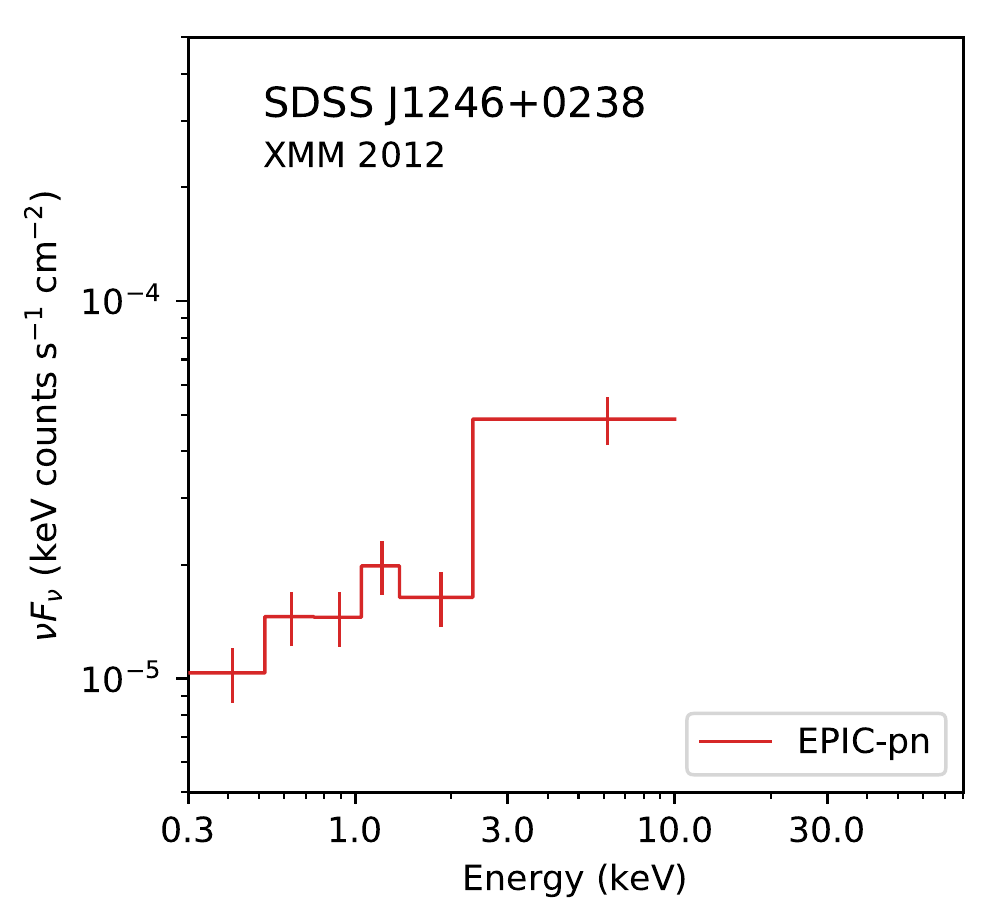}
\caption{Continued from Figure~\ref{fig_xrayspectra_1}}
\end{figure*}

\begin{figure*}
\centering
\includegraphics[width=0.32\linewidth]{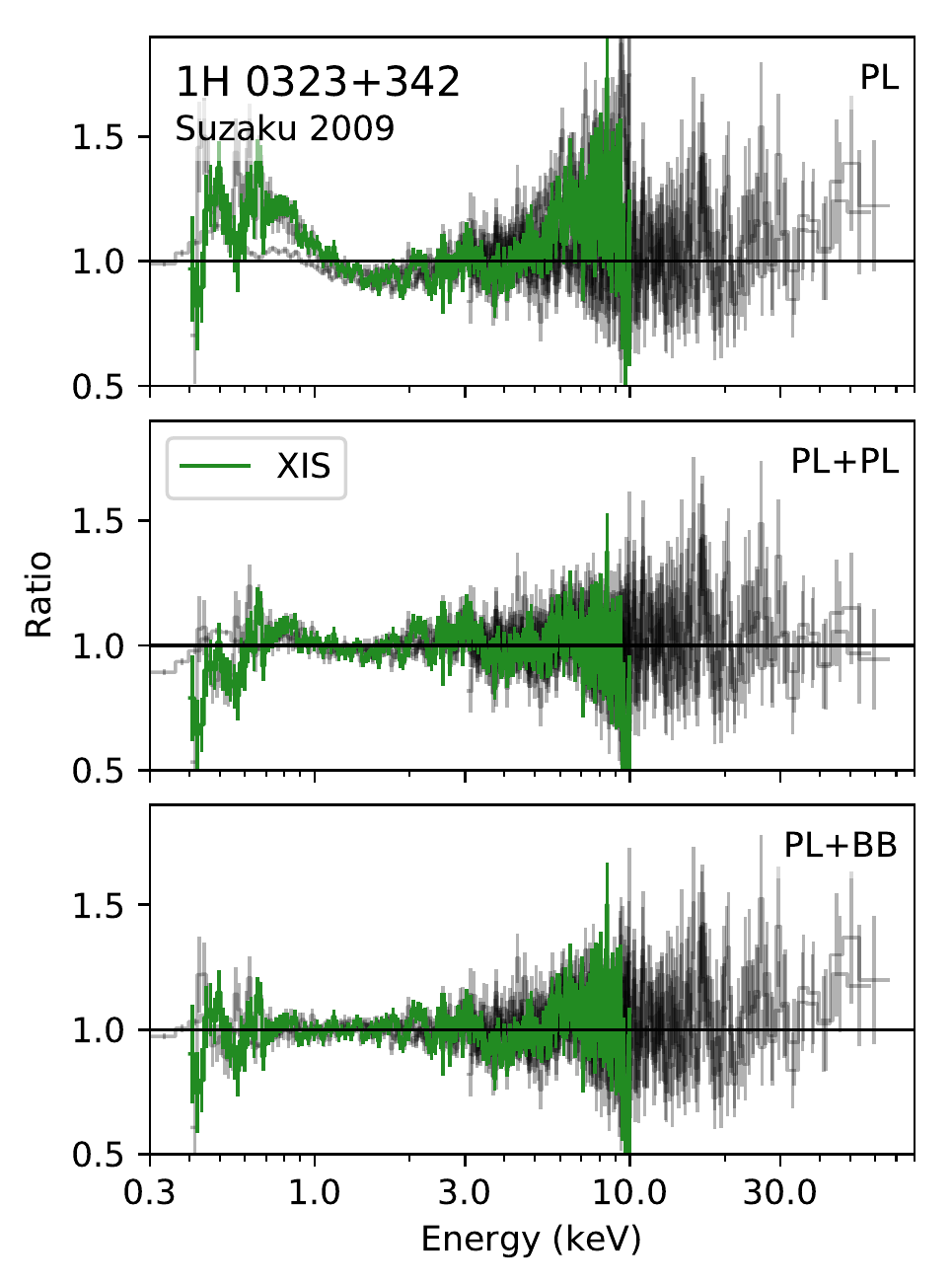}
\includegraphics[width=0.32\linewidth]{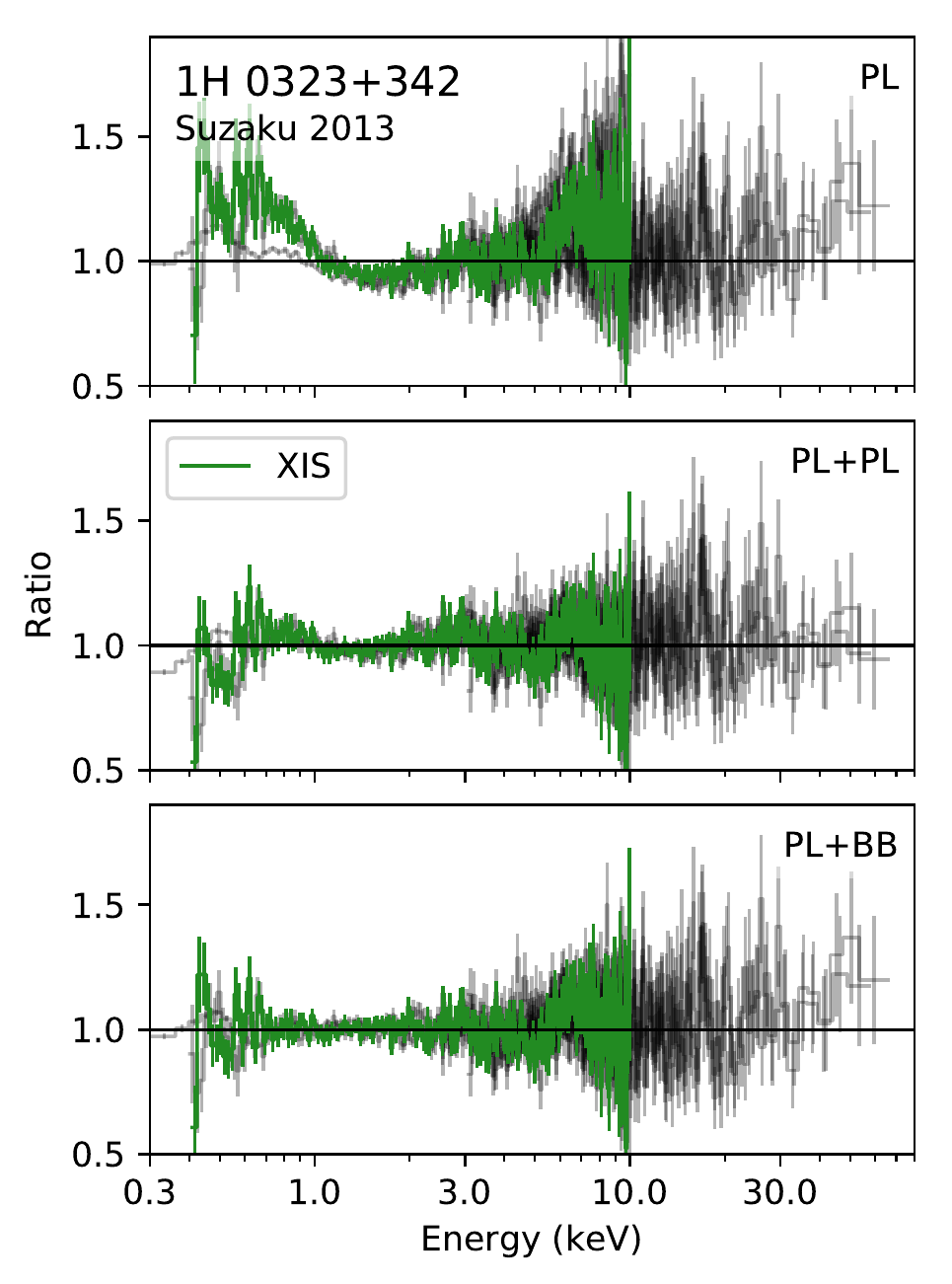}
\includegraphics[width=0.32\linewidth]{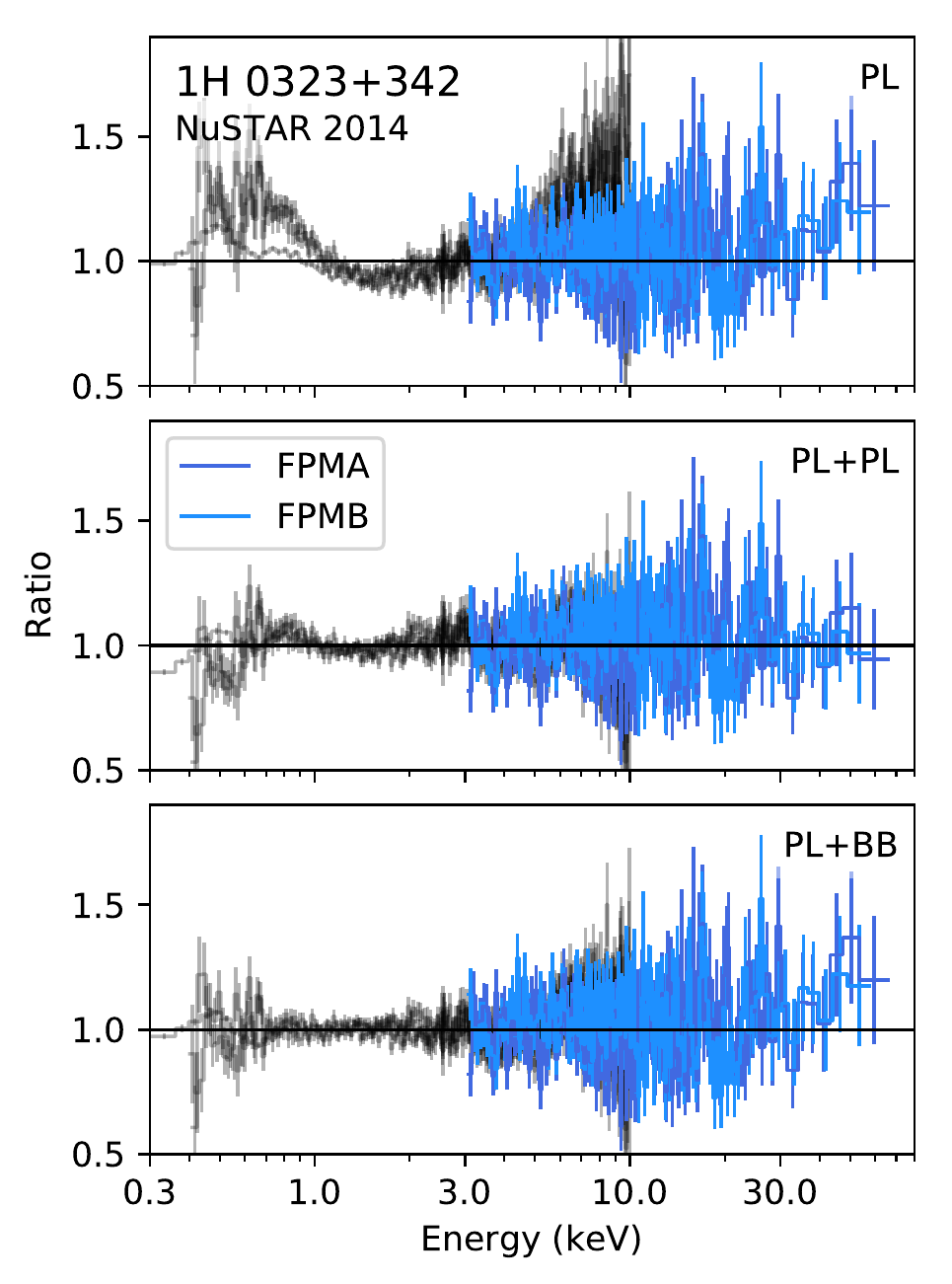}
\includegraphics[width=0.32\linewidth]{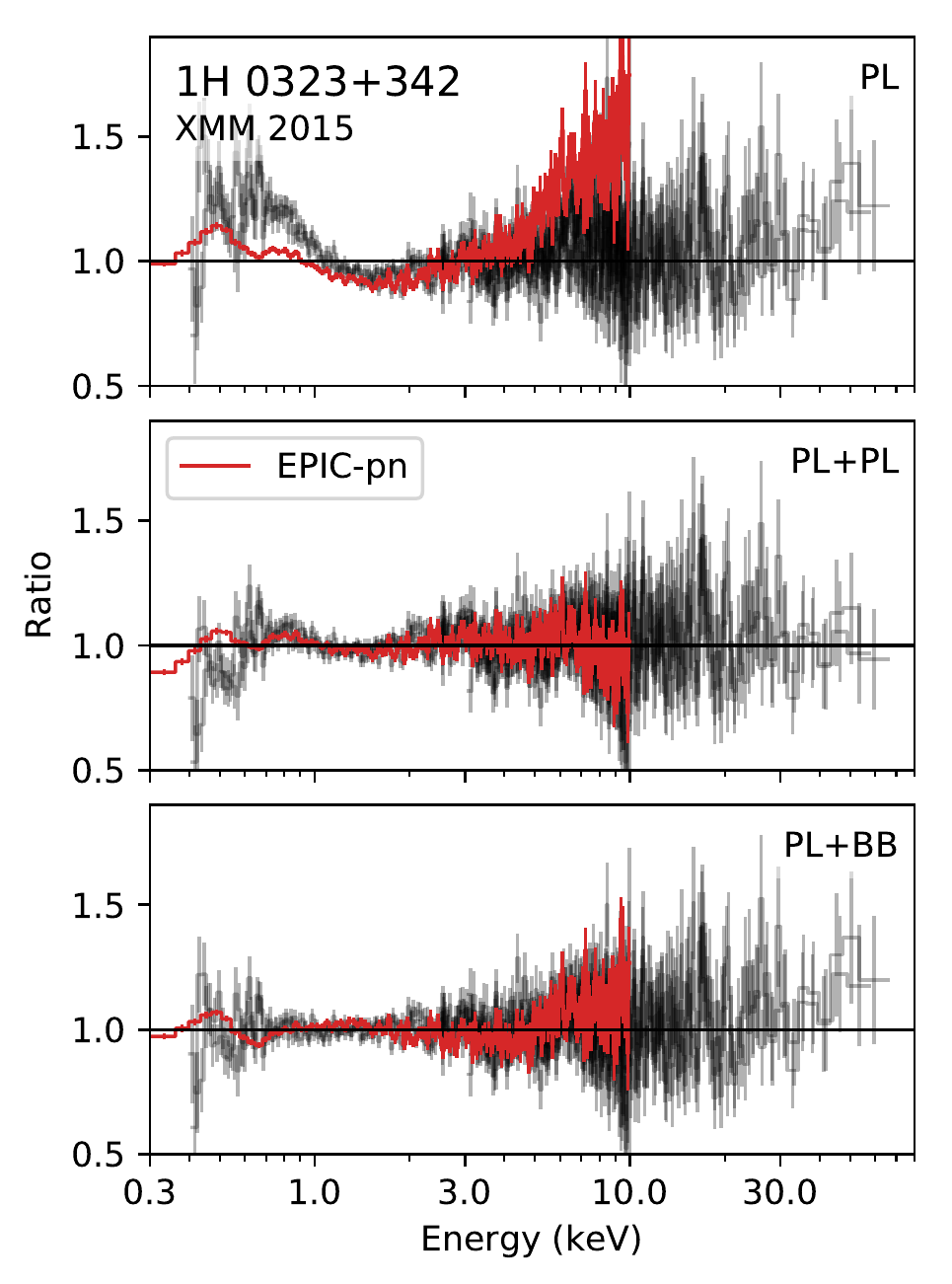}
\includegraphics[width=0.32\linewidth]{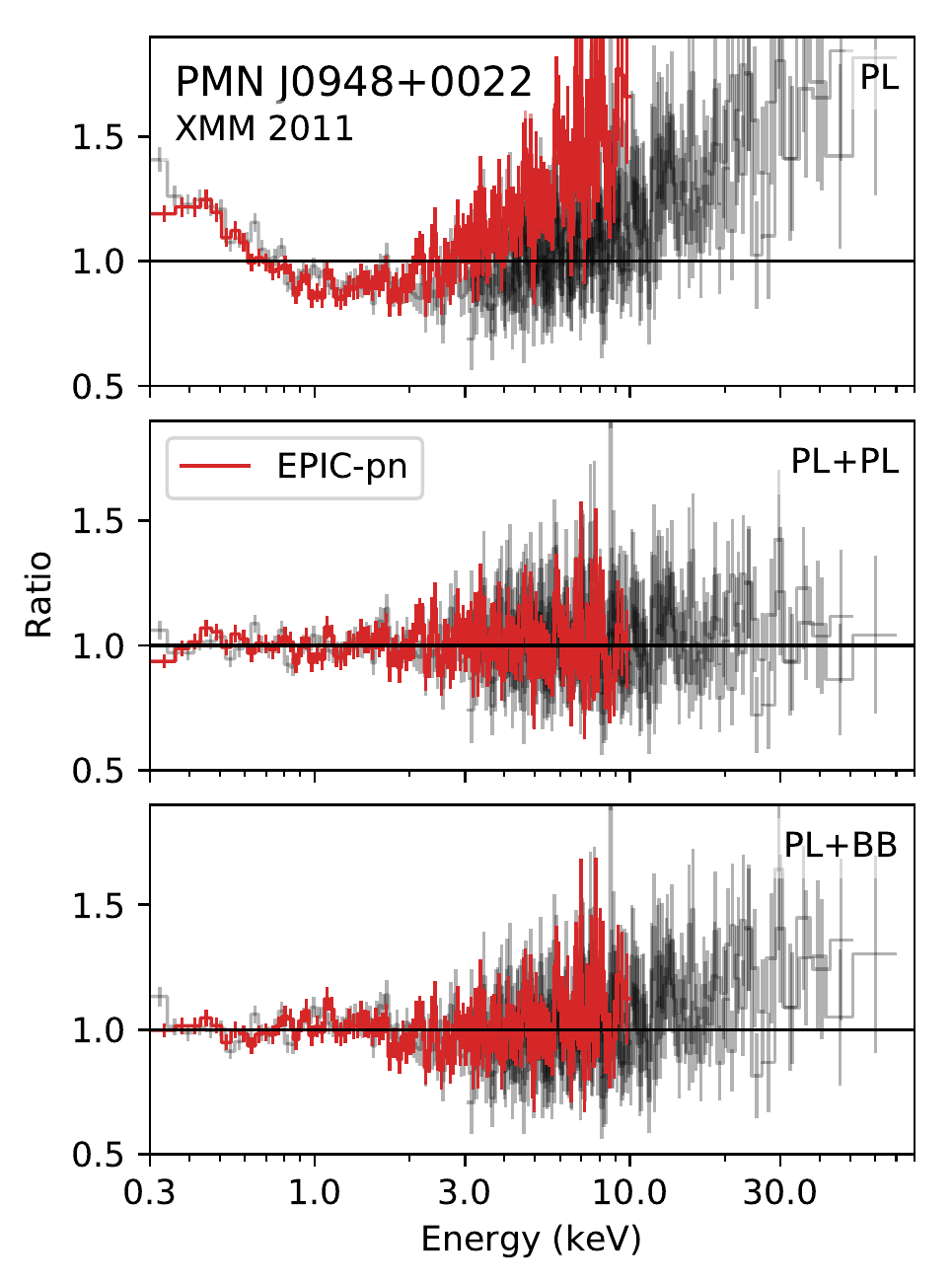}
\includegraphics[width=0.32\linewidth]{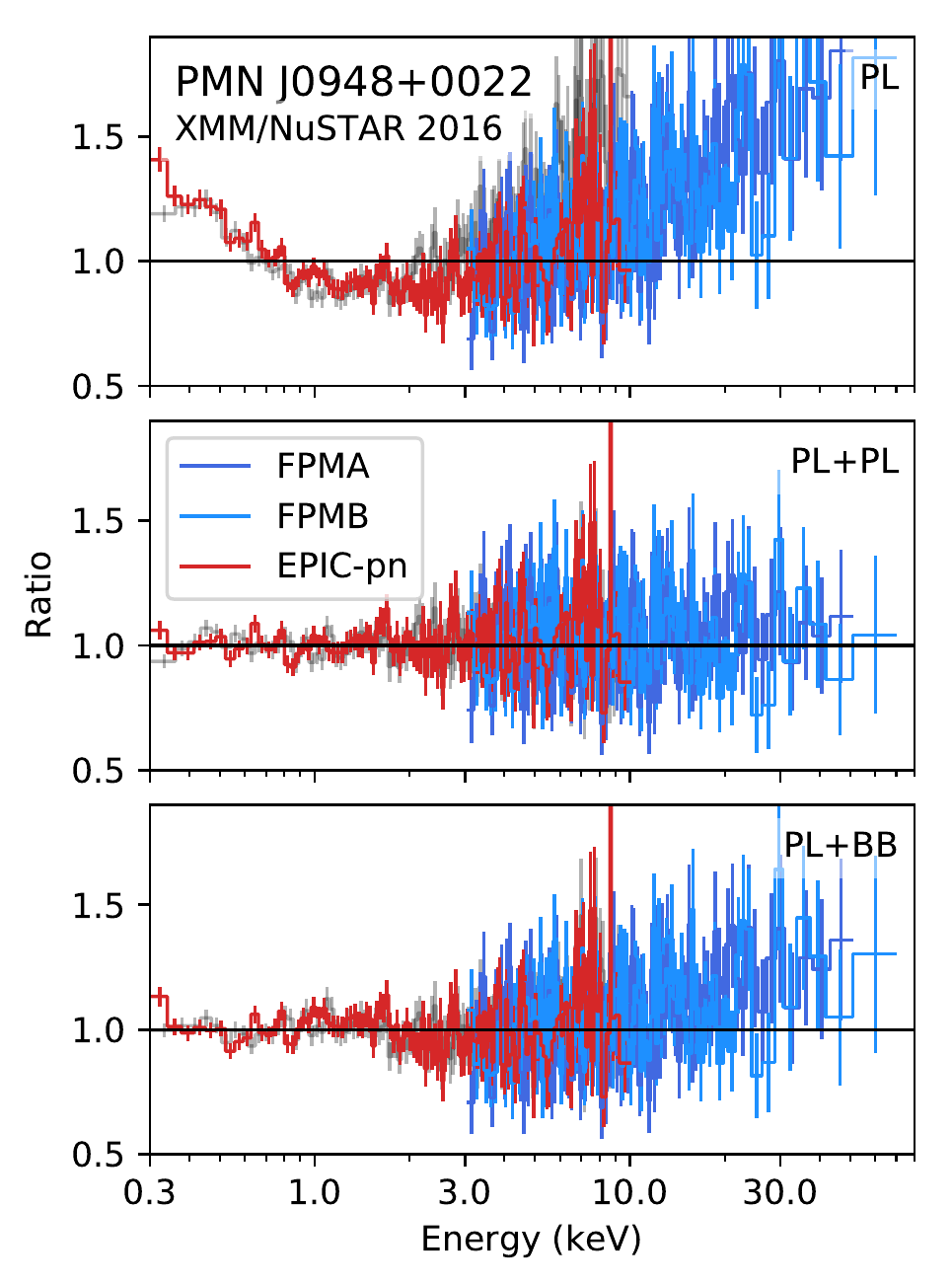}
\includegraphics[width=0.32\linewidth]{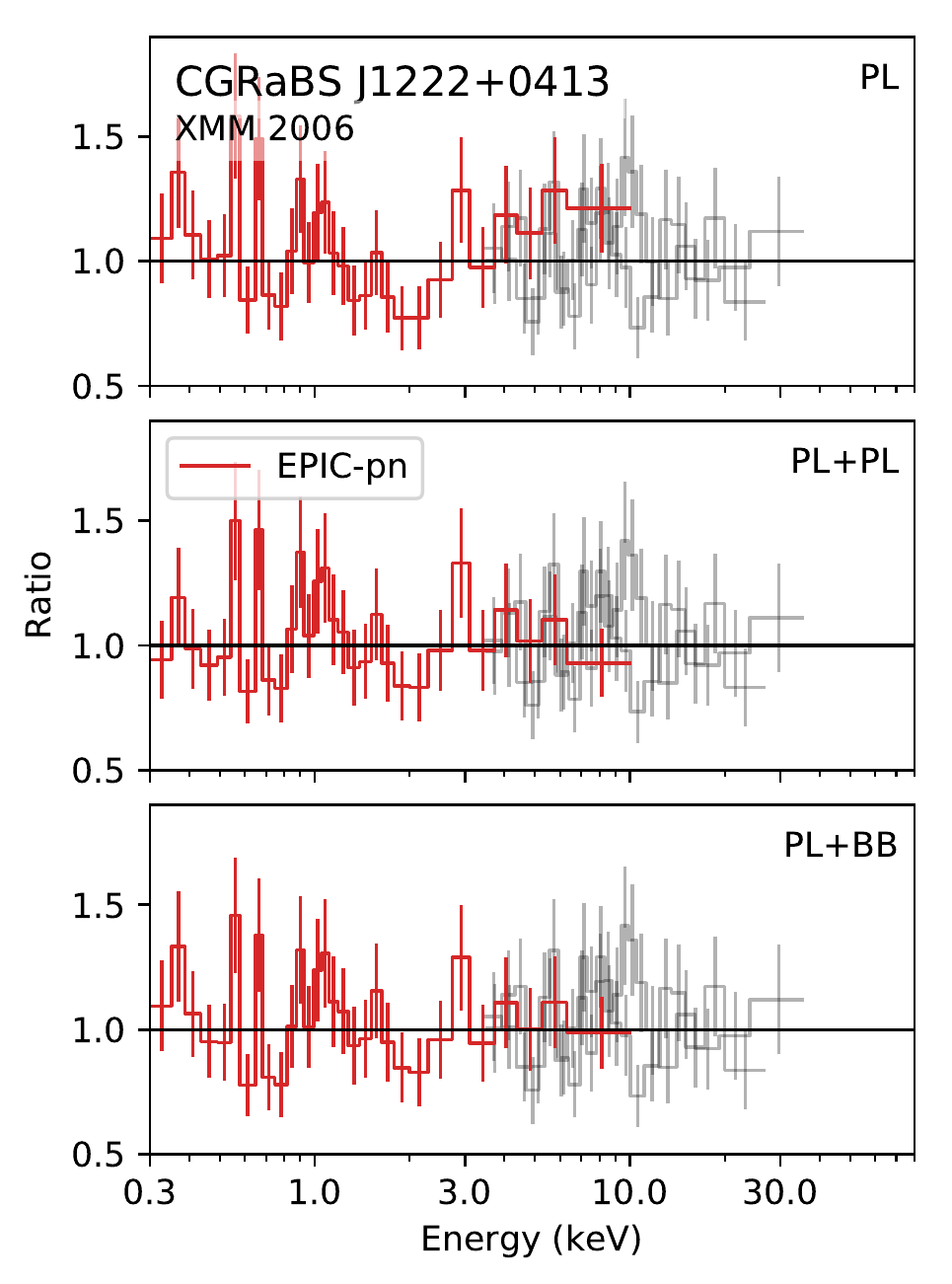}
\includegraphics[width=0.32\linewidth]{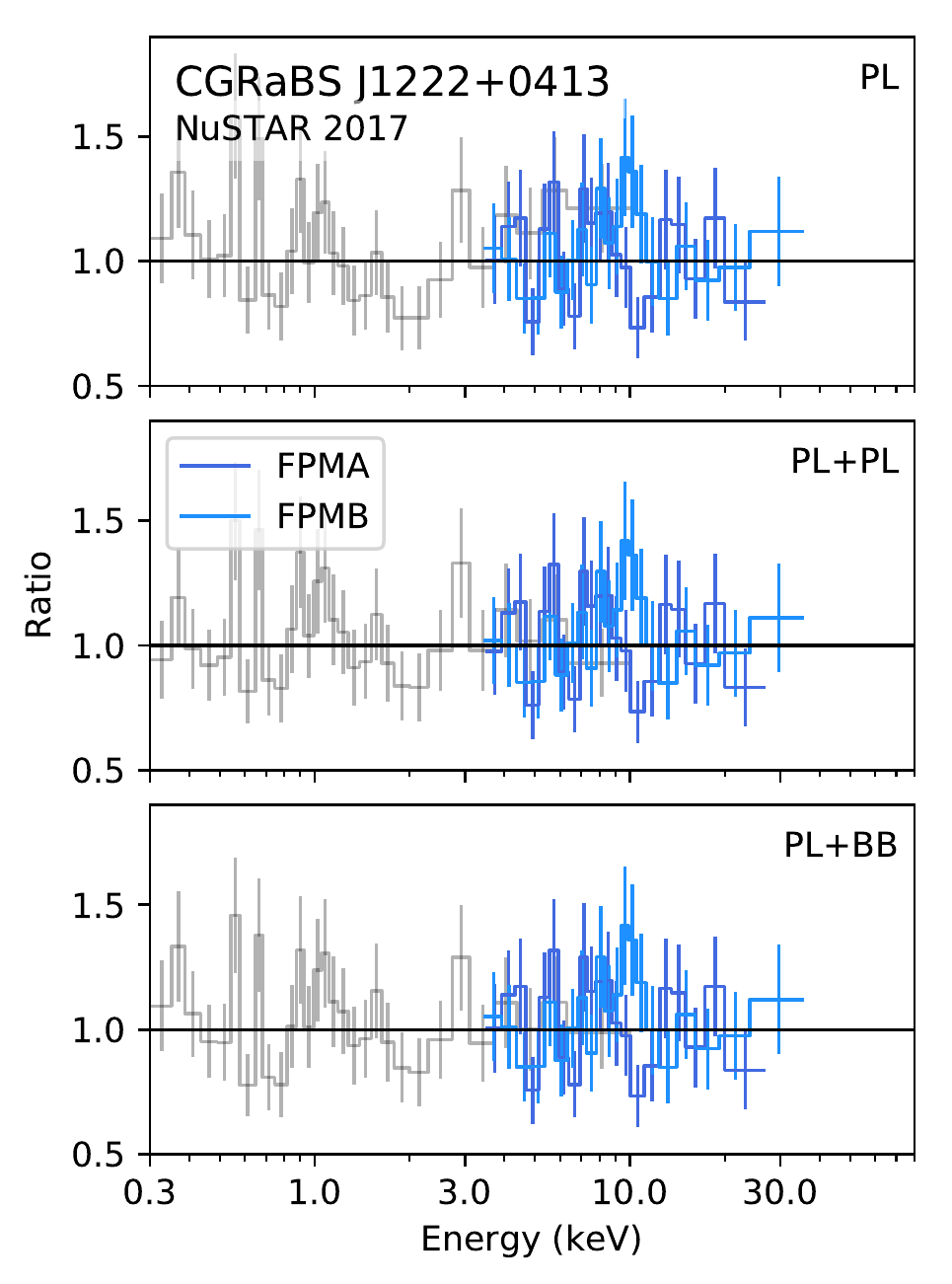}
\includegraphics[width=0.32\linewidth]{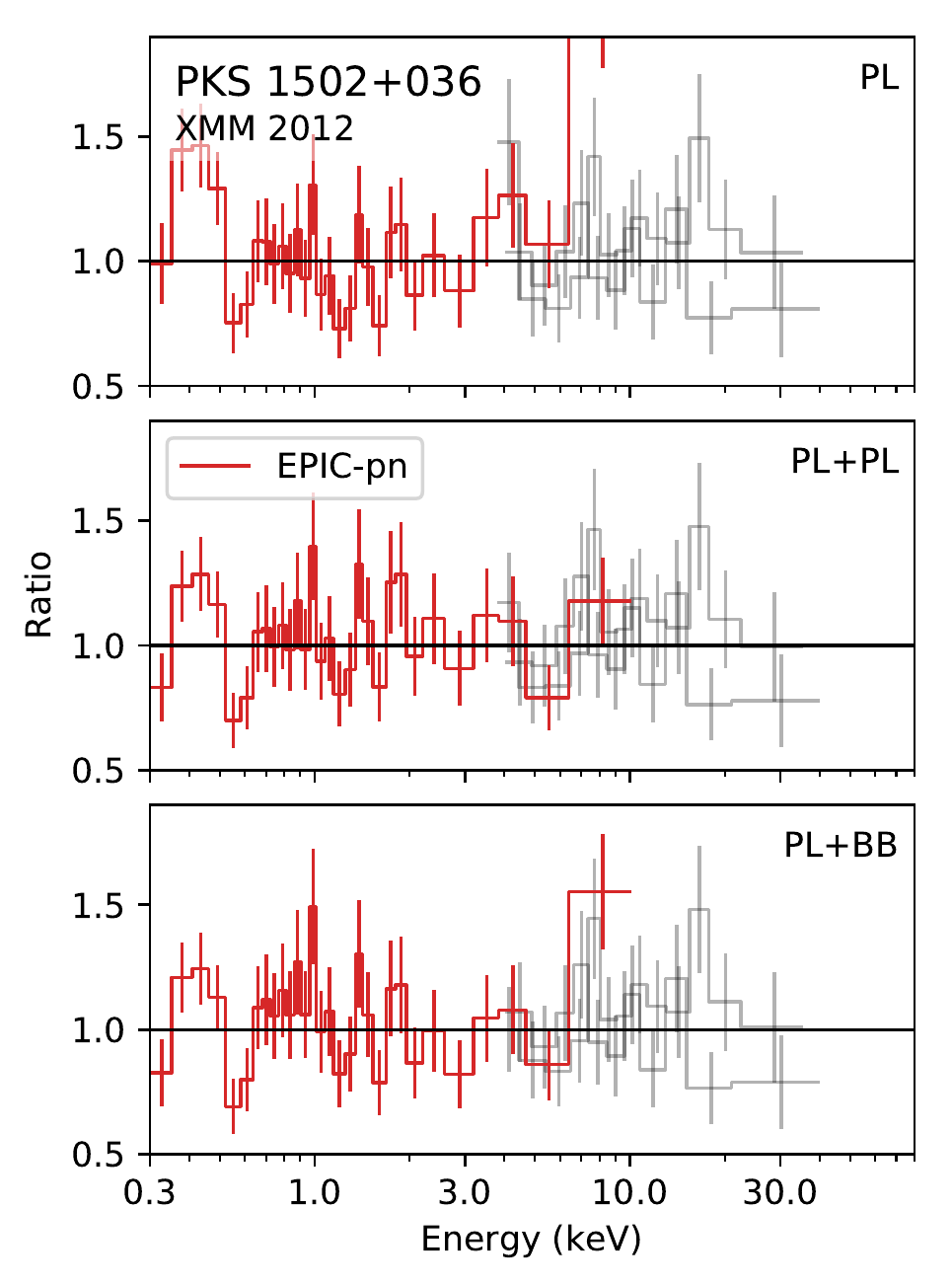}
\caption{Residual spectra for the basic fits to the X-ray data. We plot each dataset separately, but plot the residuals for other observations in grey for comparison.}
\label{fig_xray_residuals}
\end{figure*}

\begin{figure*}
\centering
\includegraphics[width=0.32\linewidth]{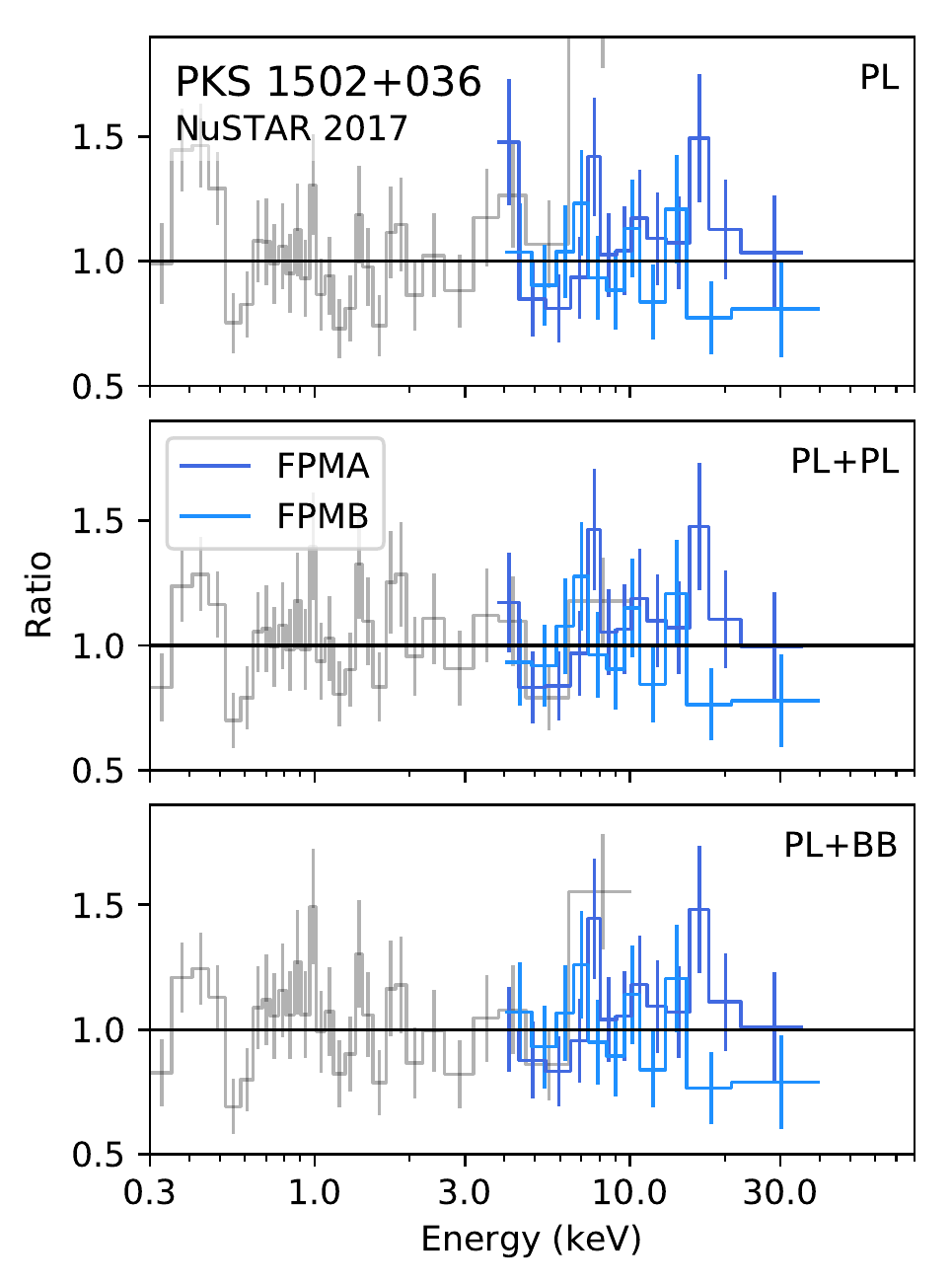}
\includegraphics[width=0.32\linewidth]{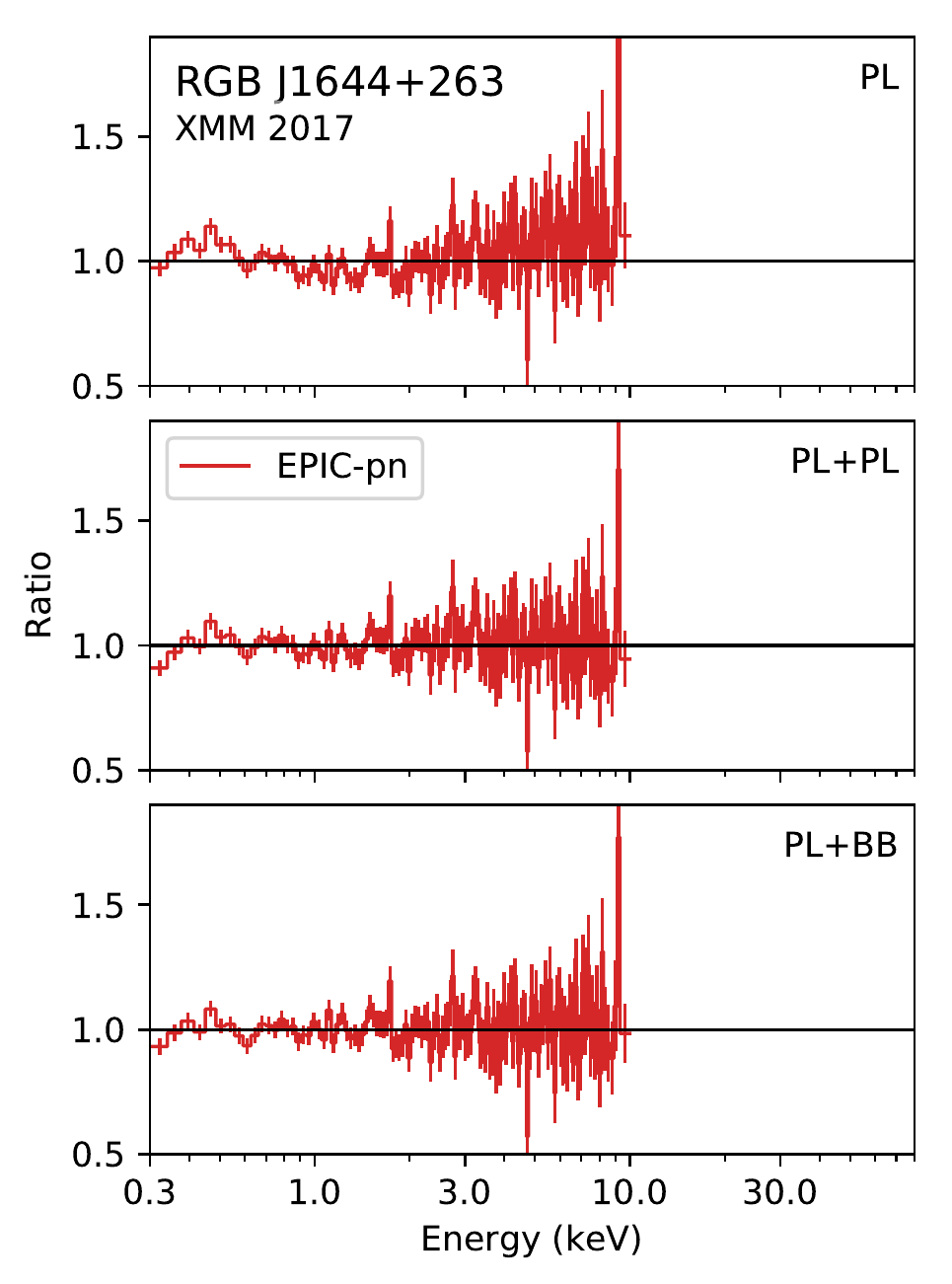}
\includegraphics[width=0.32\linewidth]{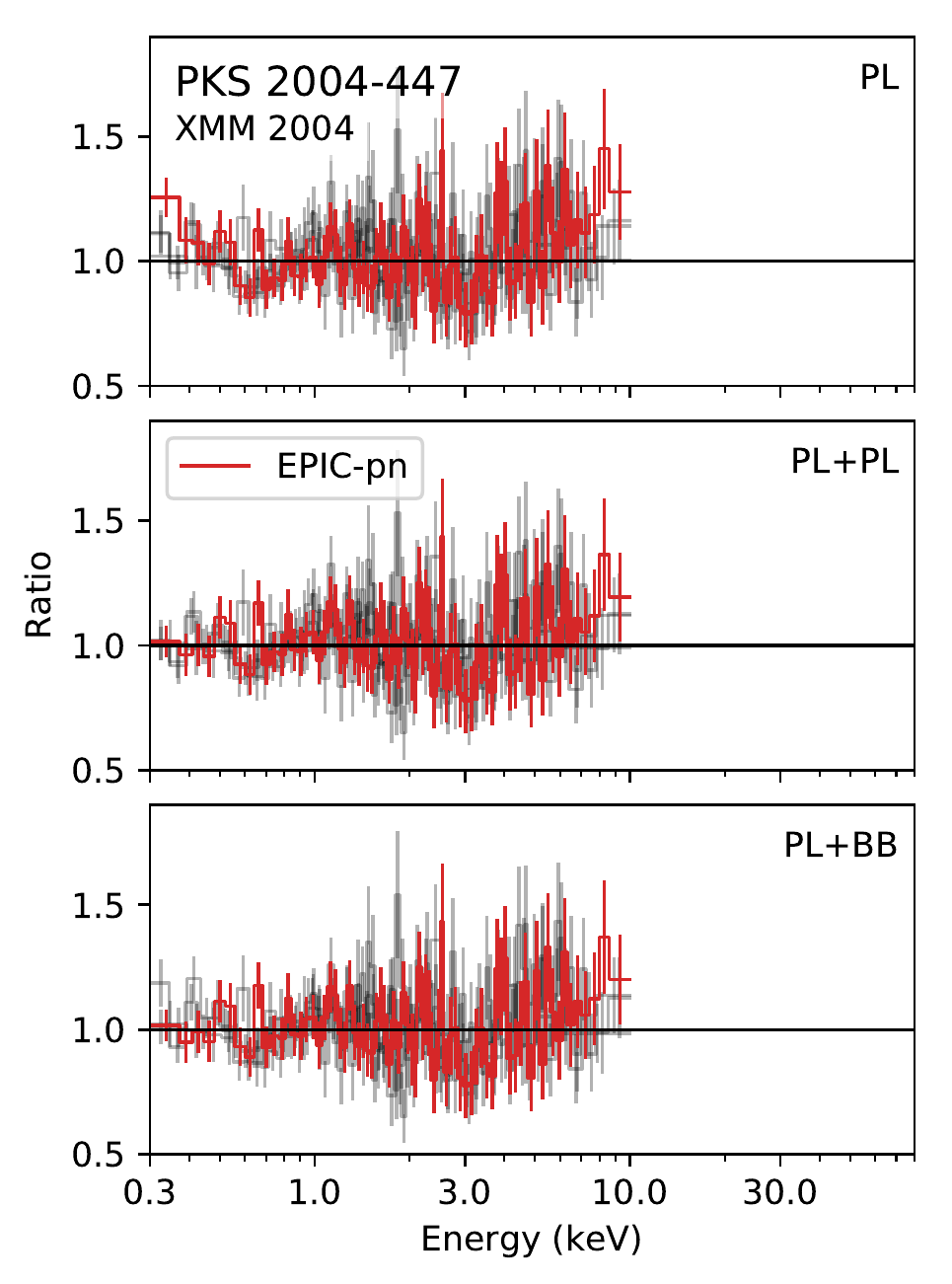}
\includegraphics[width=0.32\linewidth]{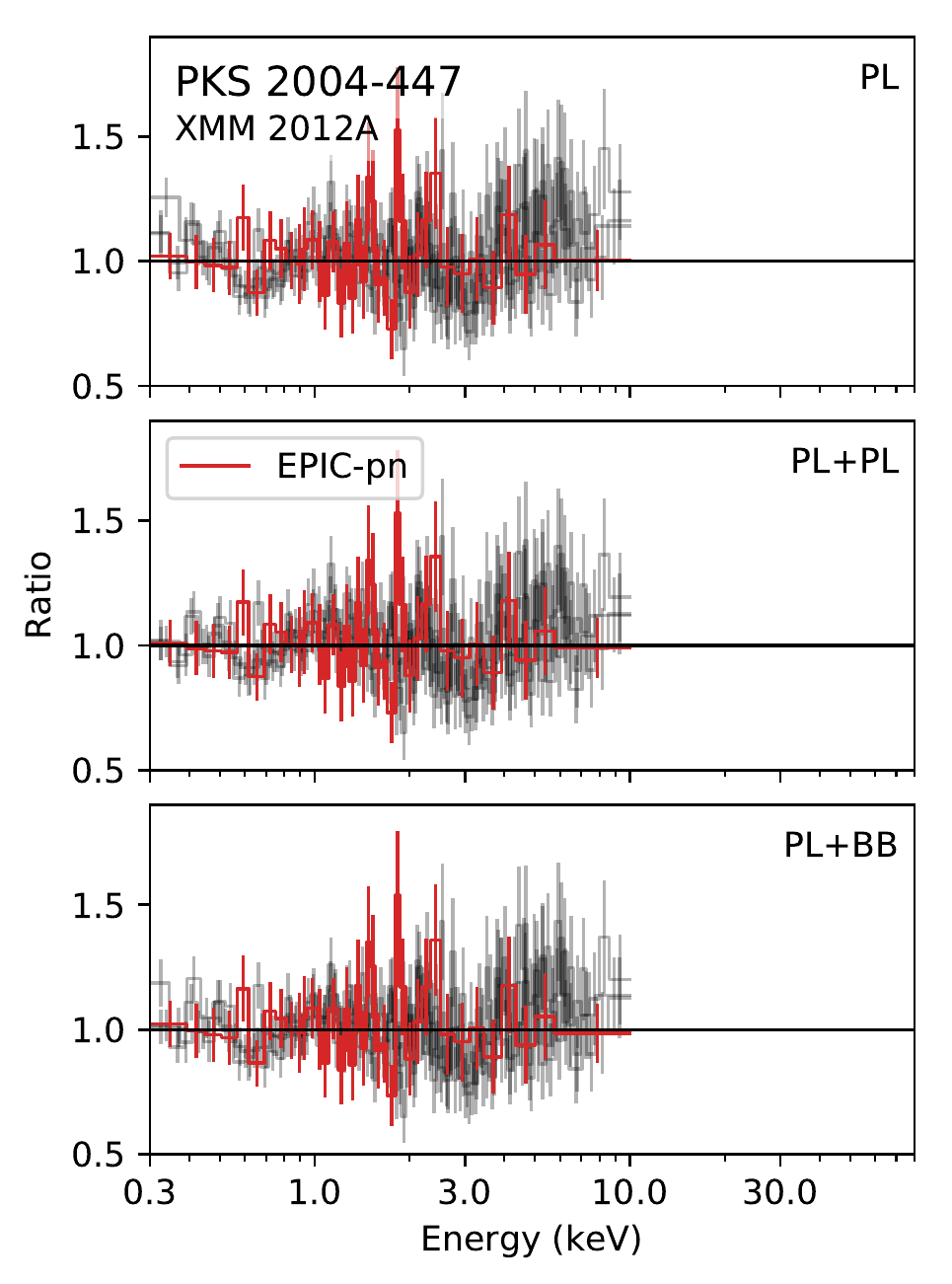}
\includegraphics[width=0.32\linewidth]{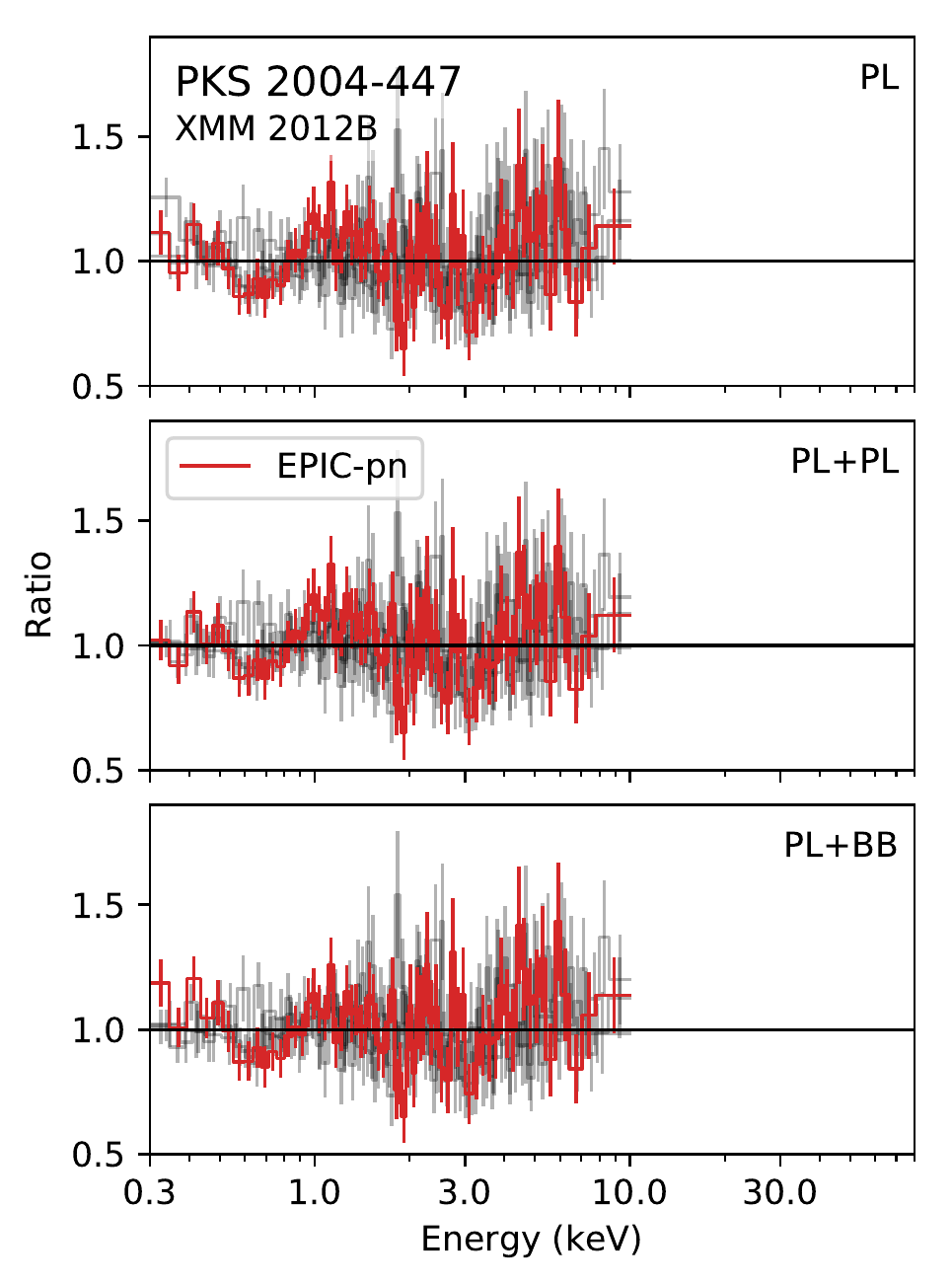}
\includegraphics[width=0.32\linewidth]{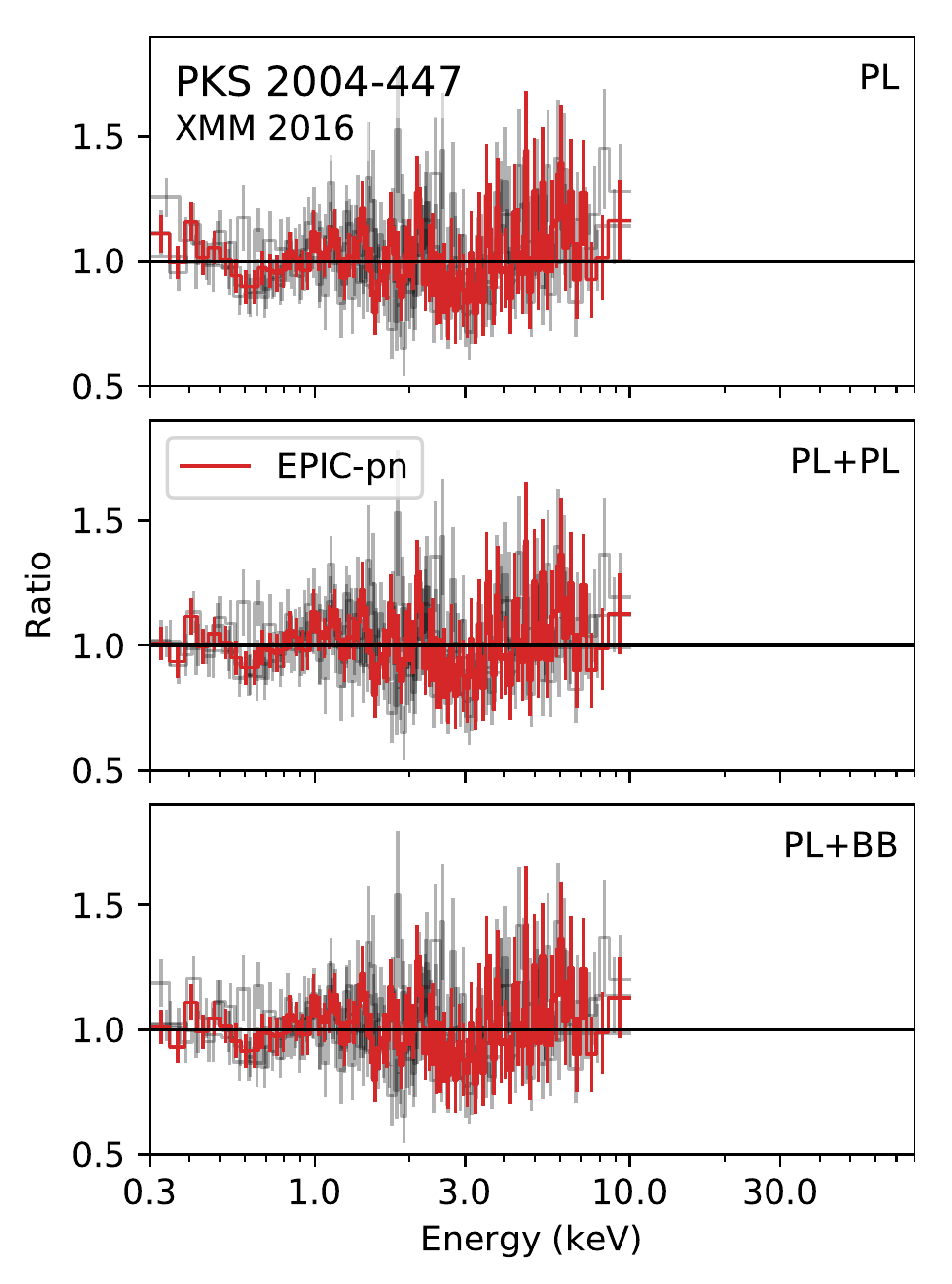}
\includegraphics[width=0.32\linewidth]{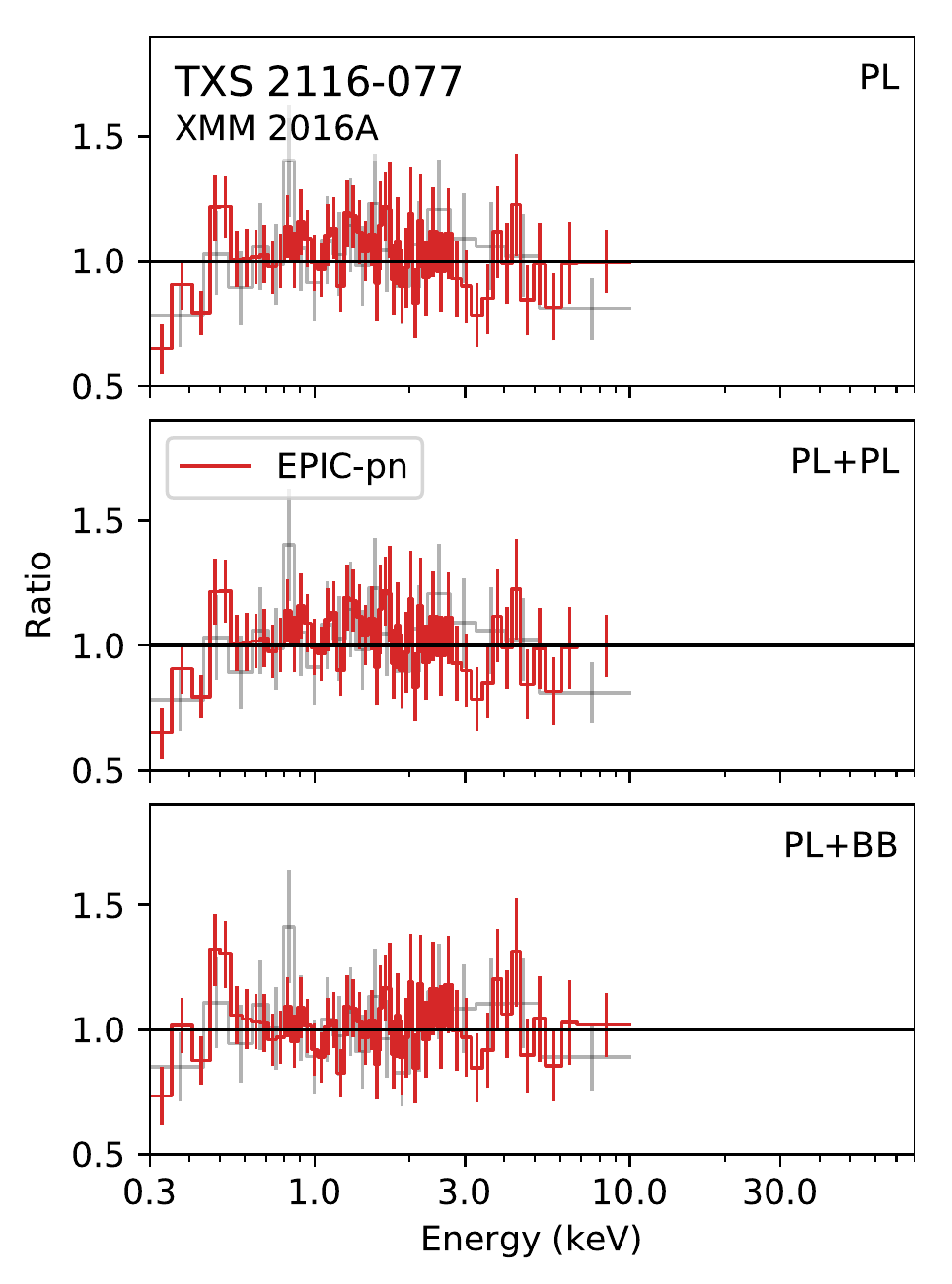}
\includegraphics[width=0.32\linewidth]{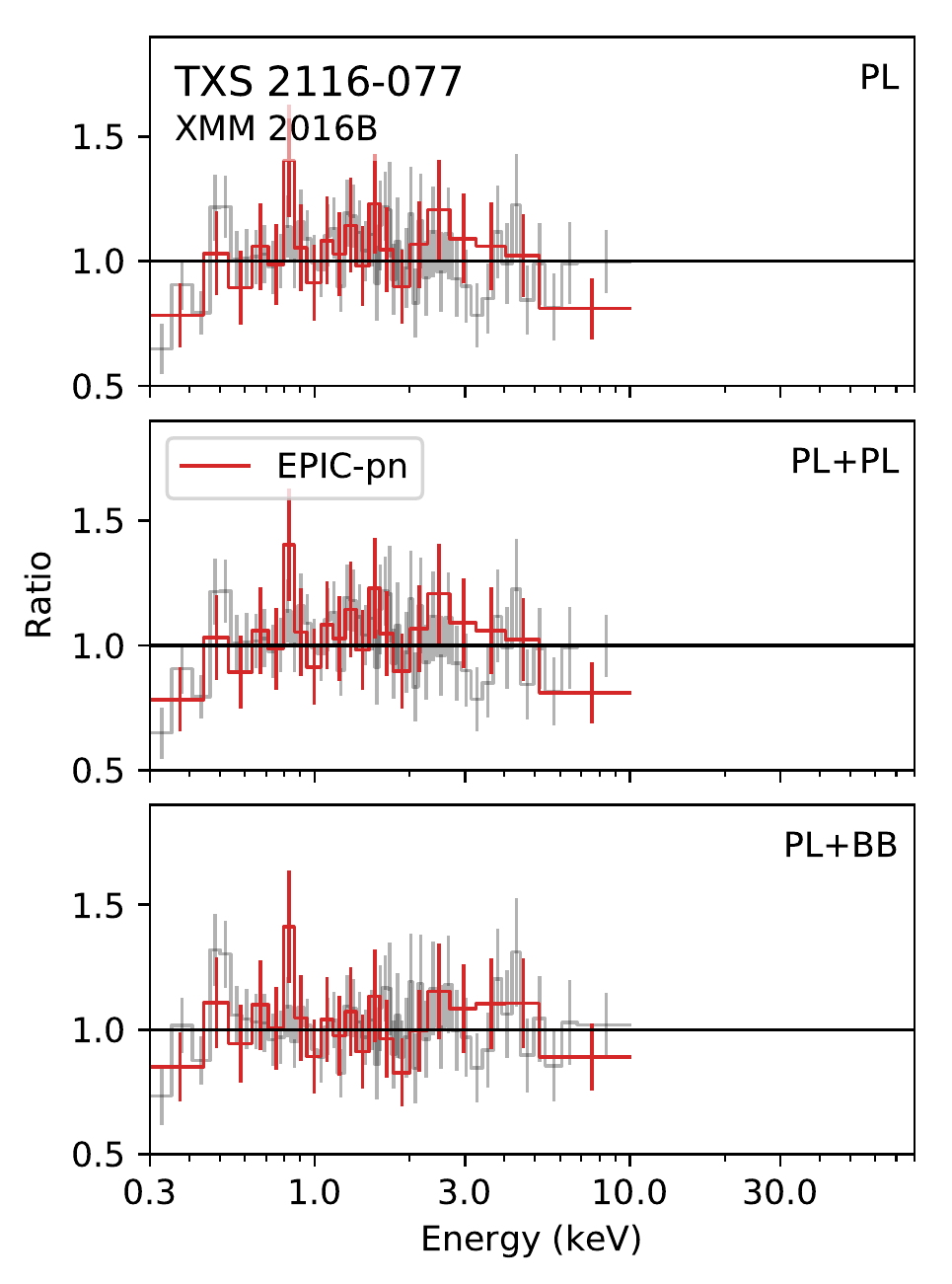}
\includegraphics[width=0.32\linewidth]{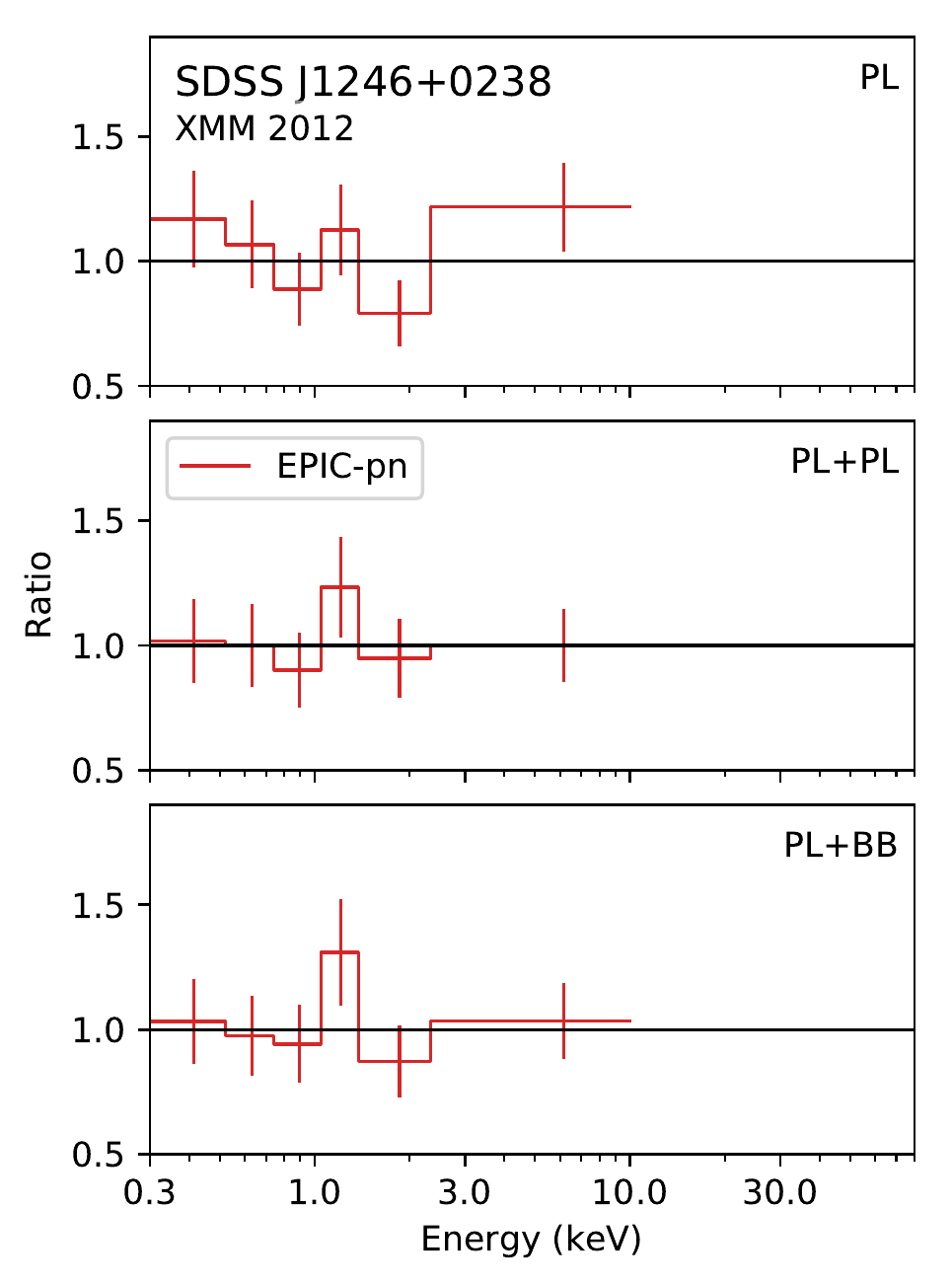}
\caption{Continued from Figure~\ref{fig_xray_residuals}}
\end{figure*}

\begin{figure*}
\centering
\includegraphics[width=0.32\linewidth]{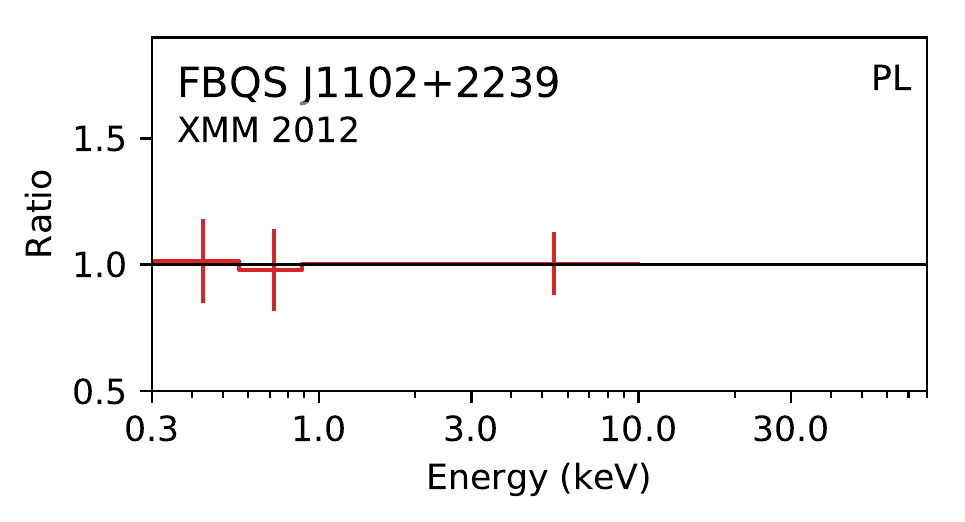}
\caption{Continued from Figure~\ref{fig_xray_residuals}}
\end{figure*}

\section{Modeled SED of all \gm-NLSy1 Galaxies}
\begin{figure*}[t]
\hbox{
\includegraphics[scale=0.48]{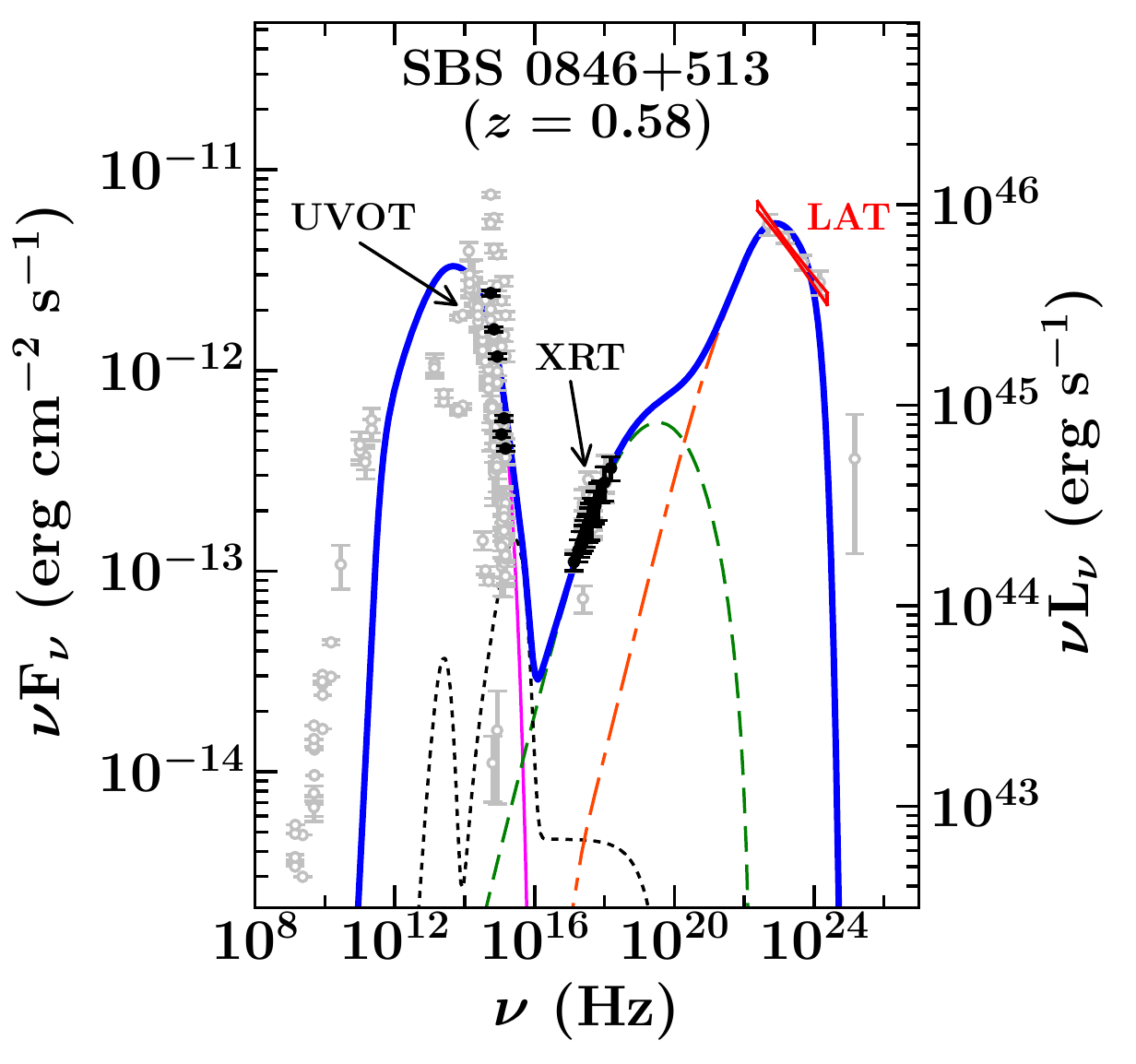}
\includegraphics[scale=0.48]{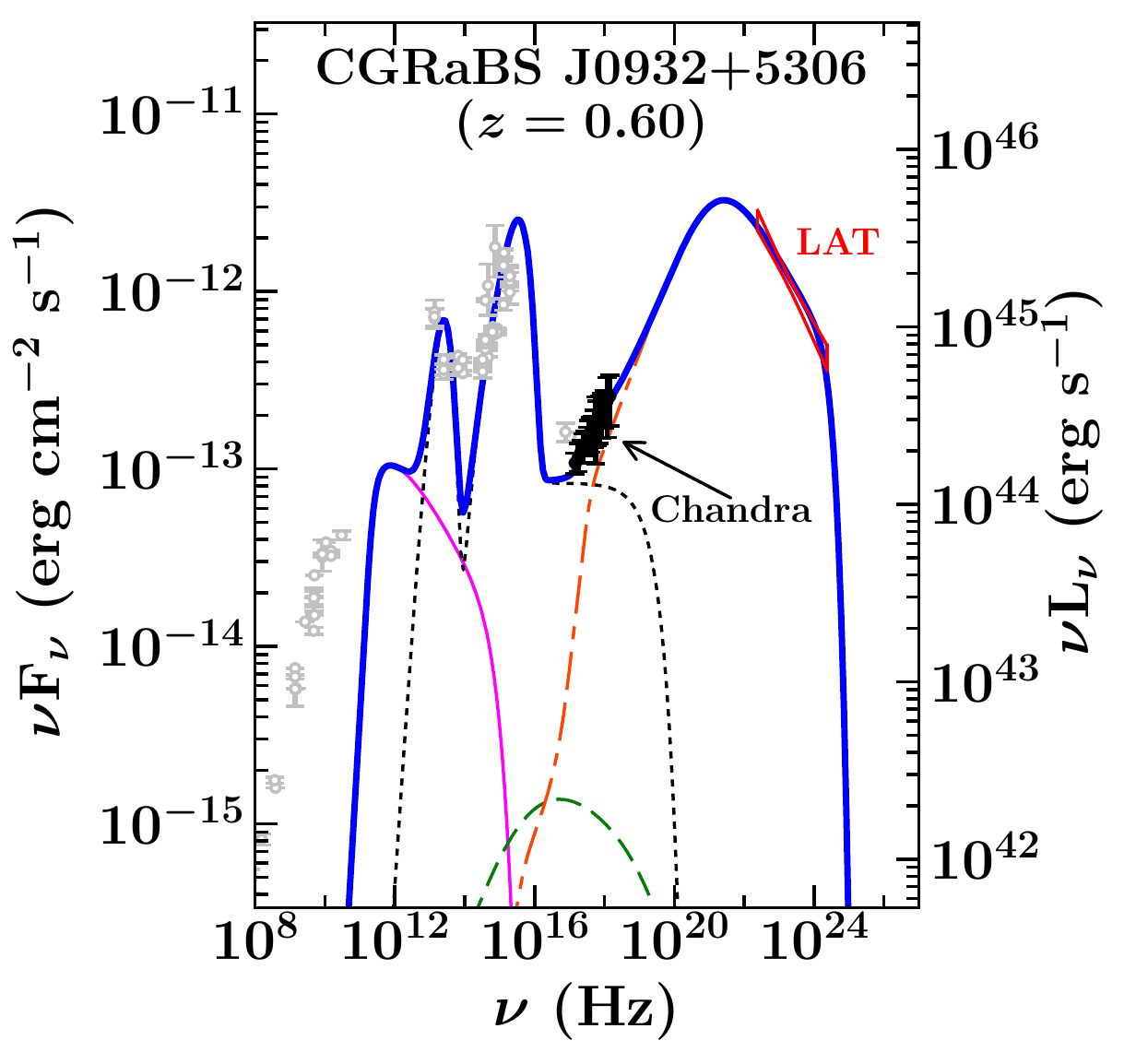}
\includegraphics[scale=0.48]{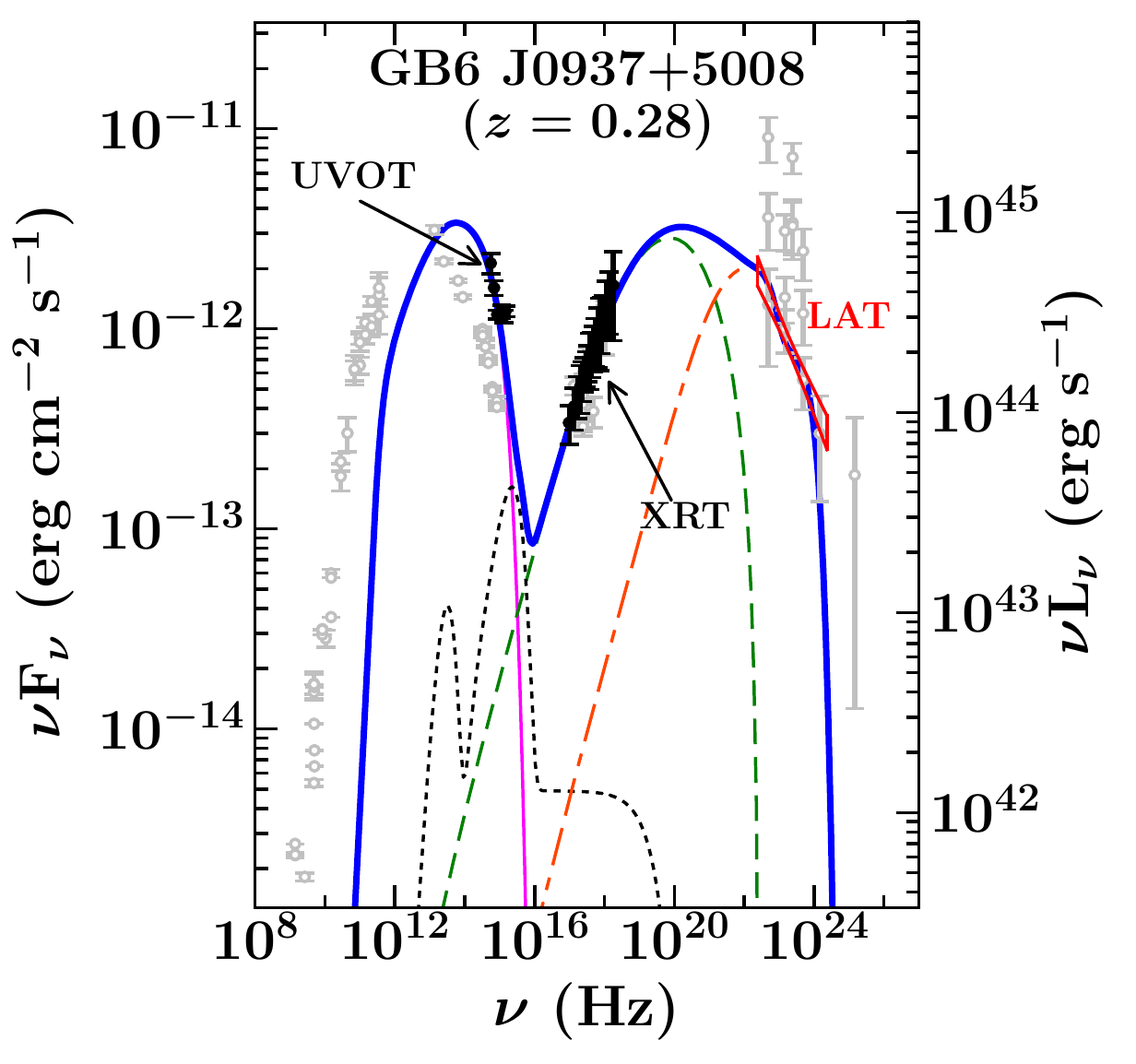}
}
\hbox{
\includegraphics[scale=0.48]{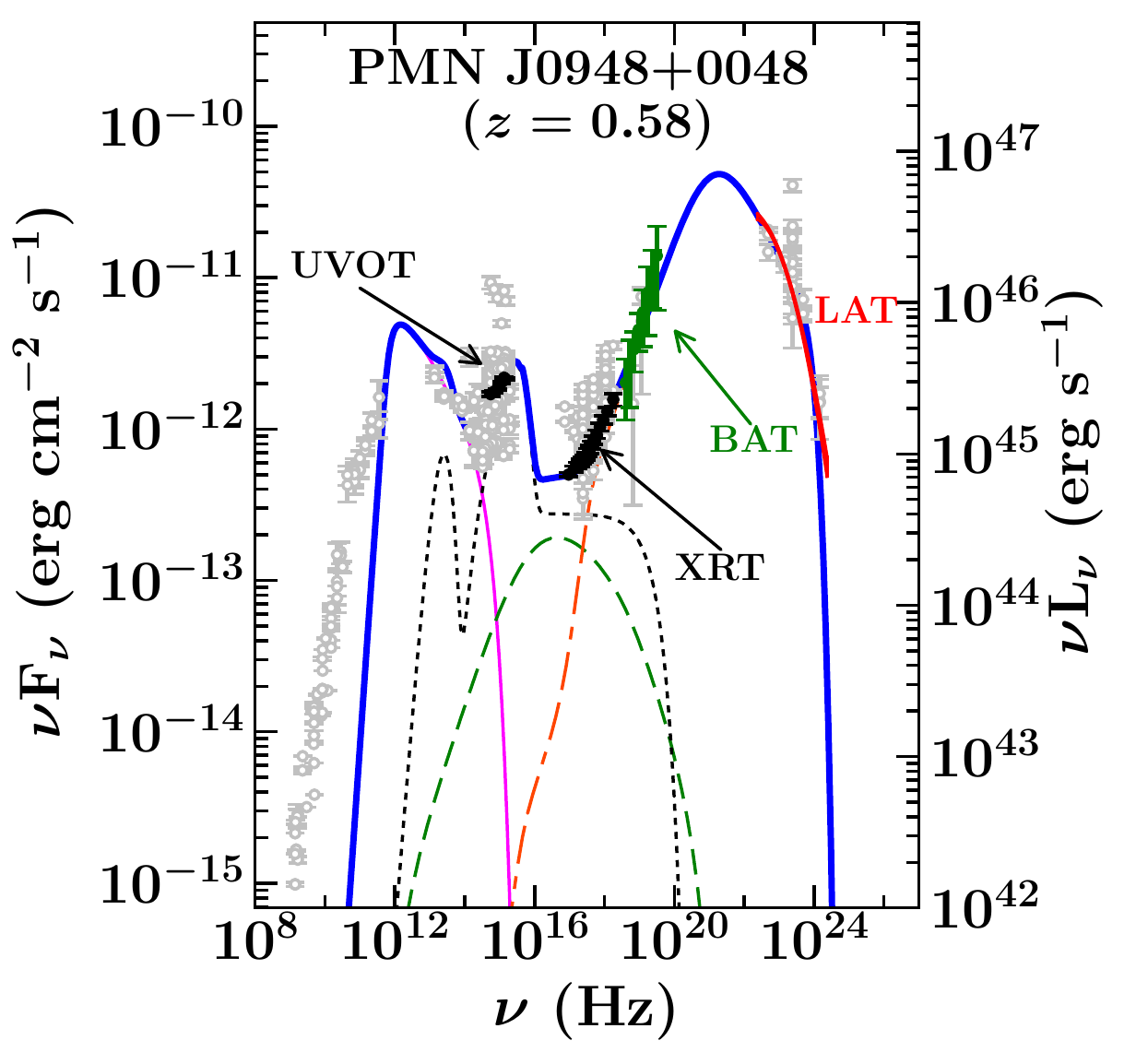}
\includegraphics[scale=0.48]{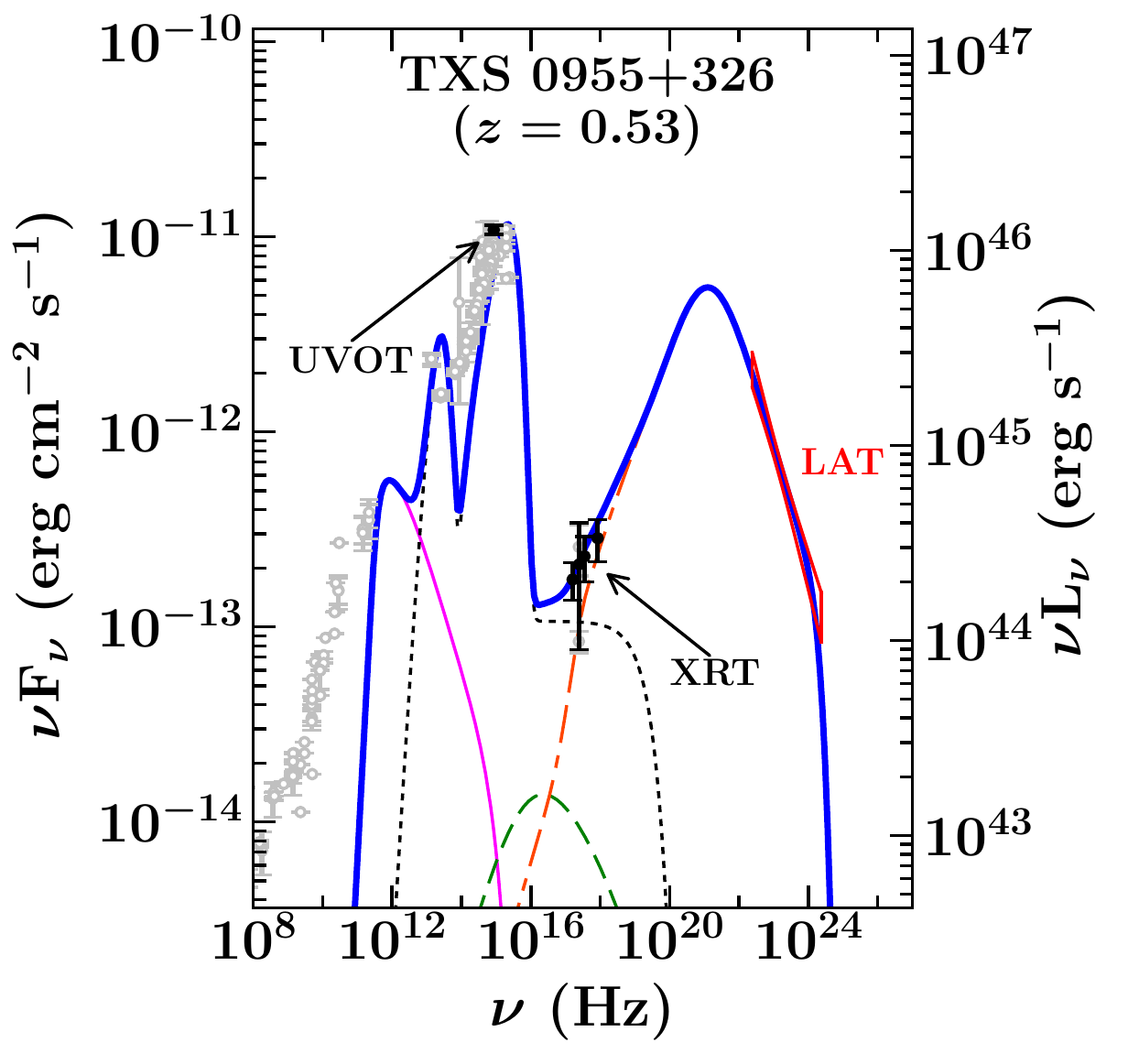}
\includegraphics[scale=0.48]{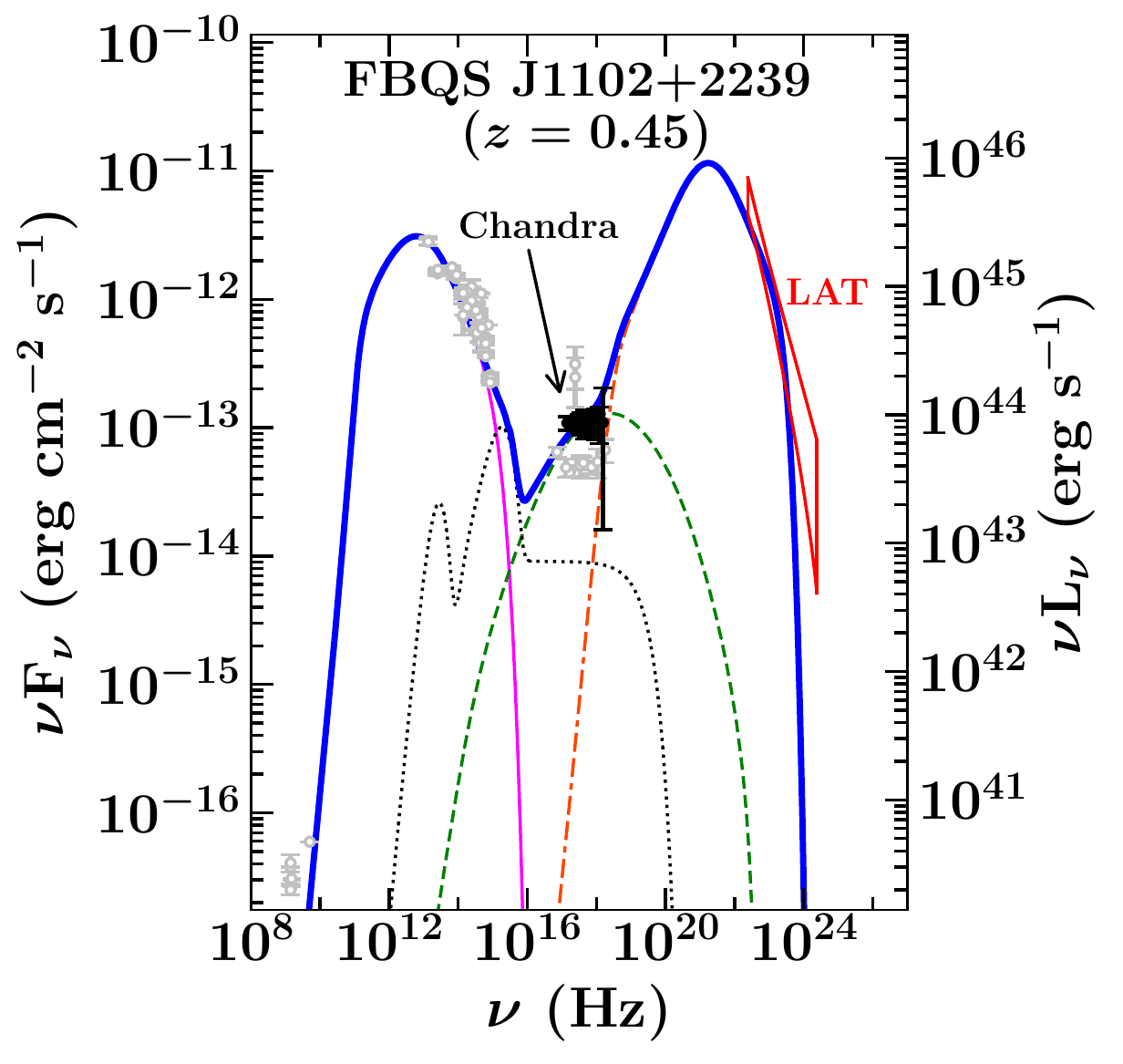}
}
\hbox{
\includegraphics[scale=0.48]{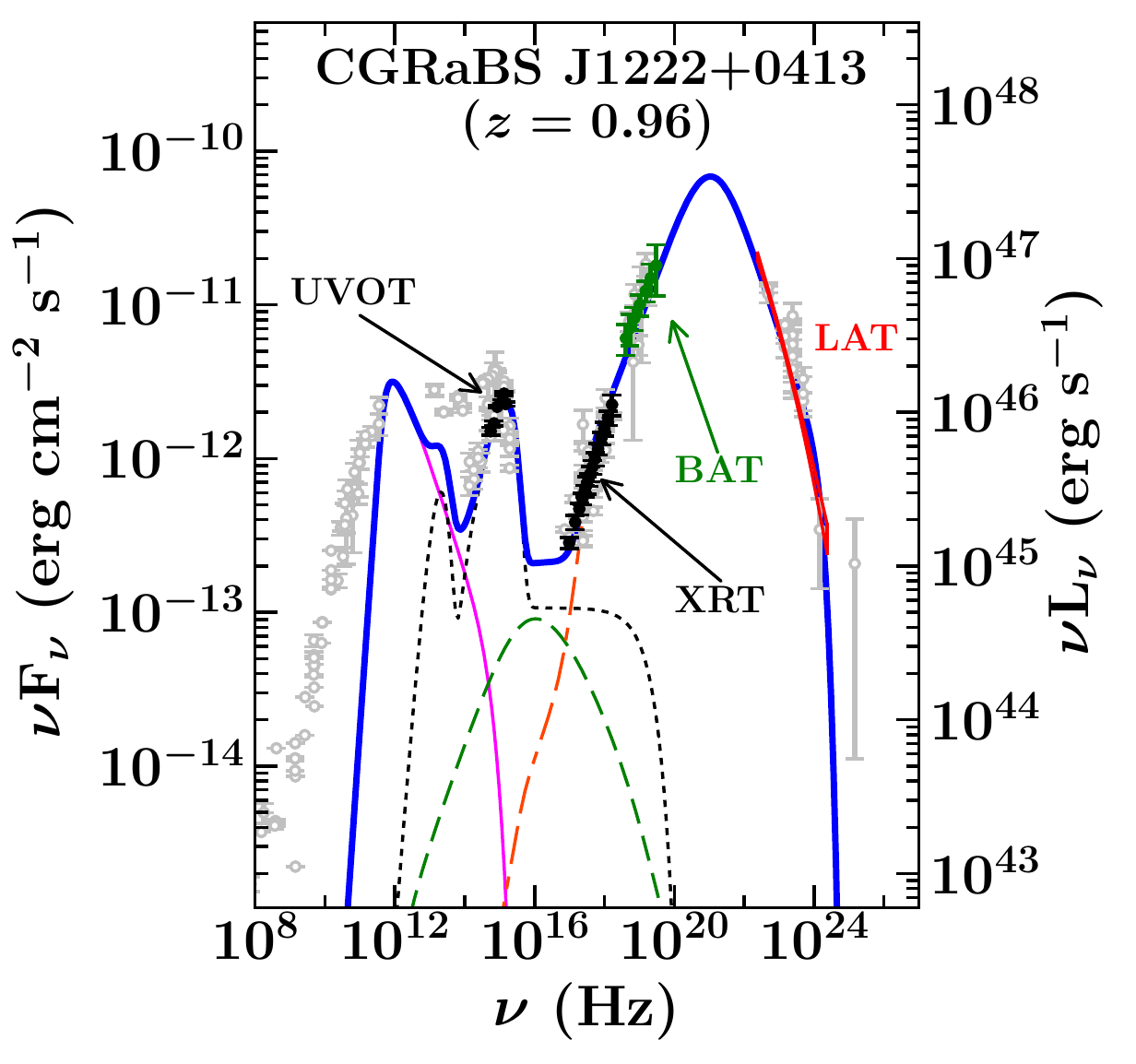}
\includegraphics[scale=0.48]{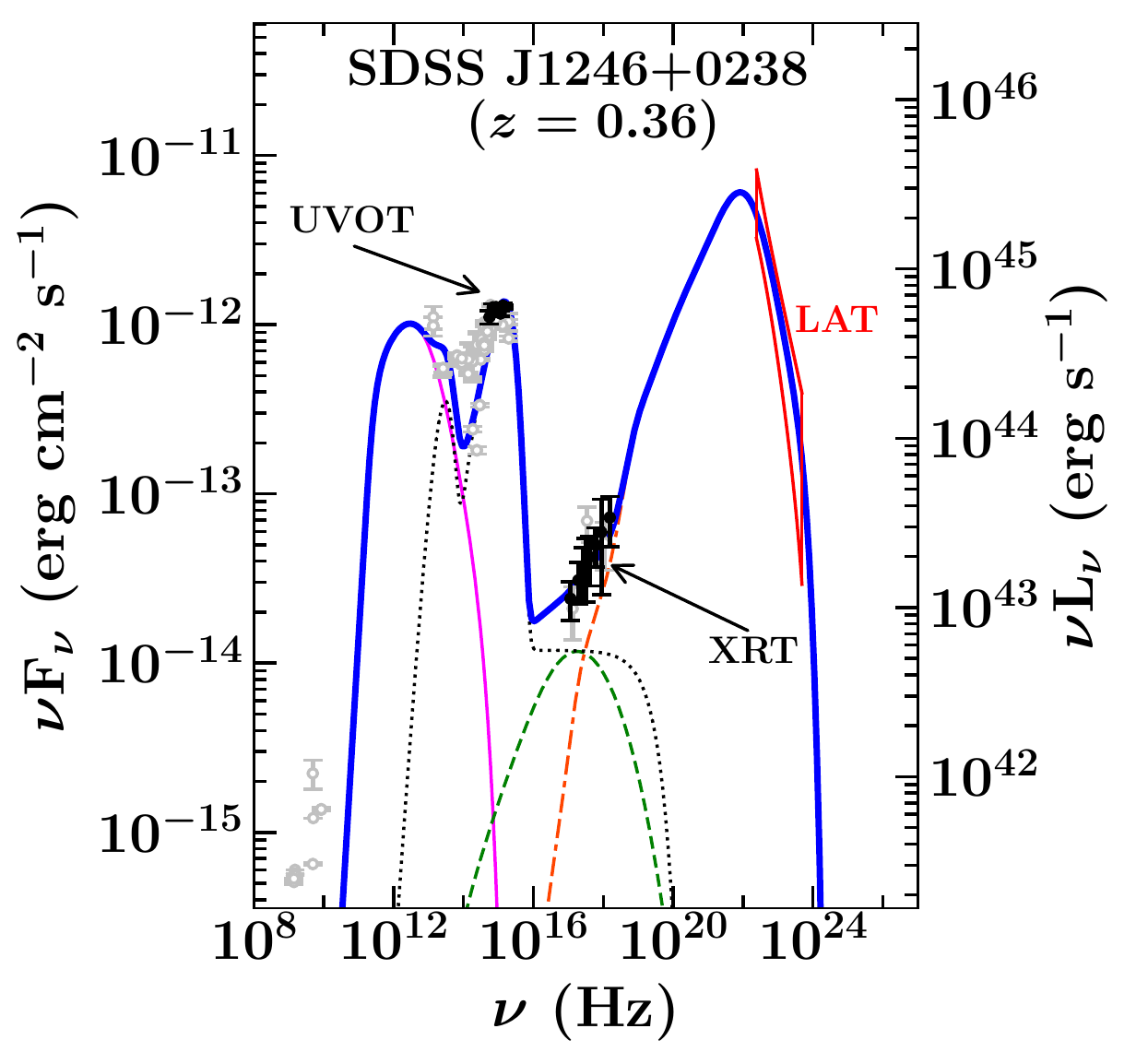}
\includegraphics[scale=0.48]{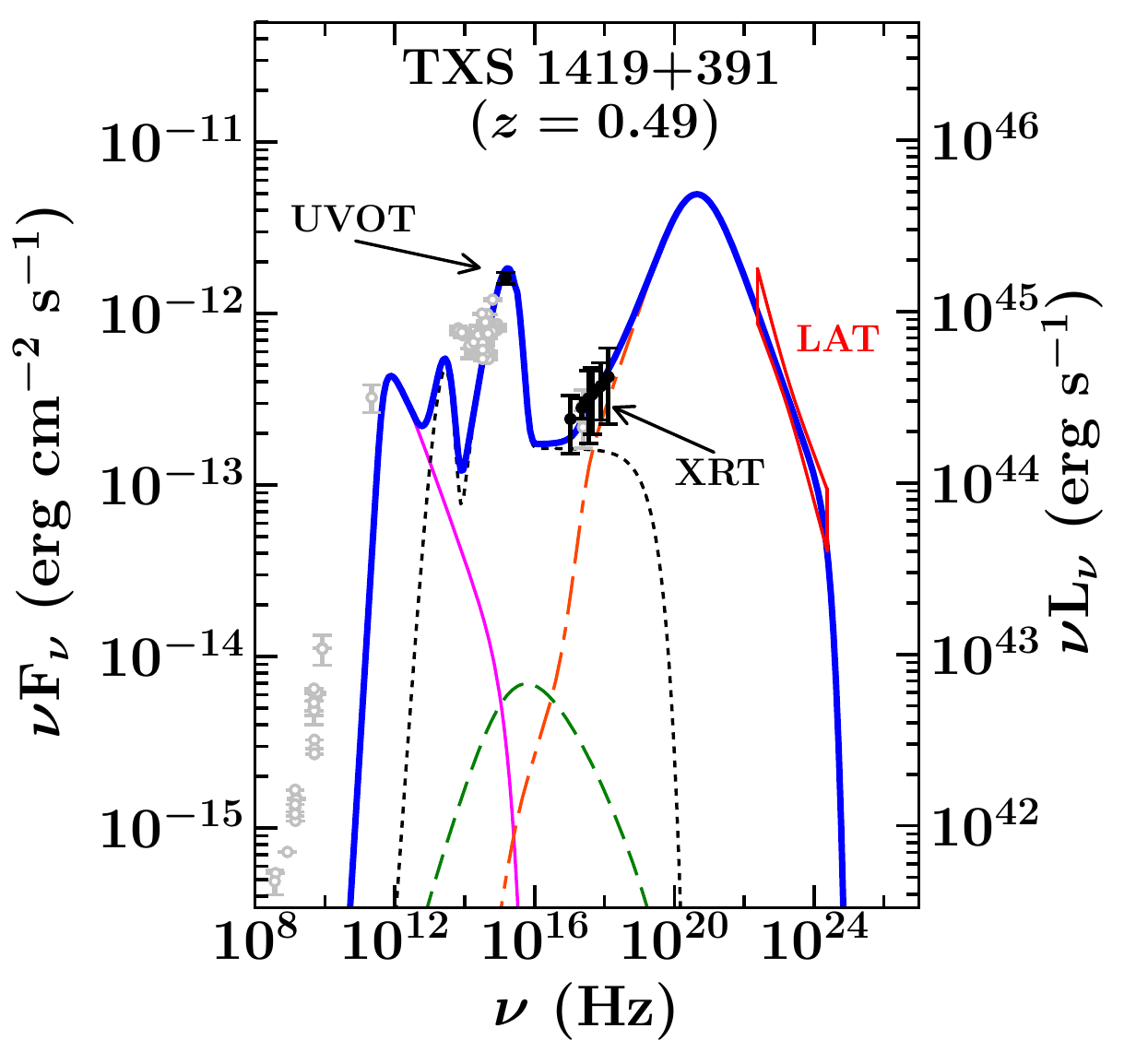}
}
\caption{The modeled SED of \gm-NLSy1 galaxies. Other information are same as in Figure \ref{fig:SED}.\label{fig:app_SED1}}
\end{figure*}

\begin{figure*}[t]
\hbox{
\includegraphics[scale=0.48]{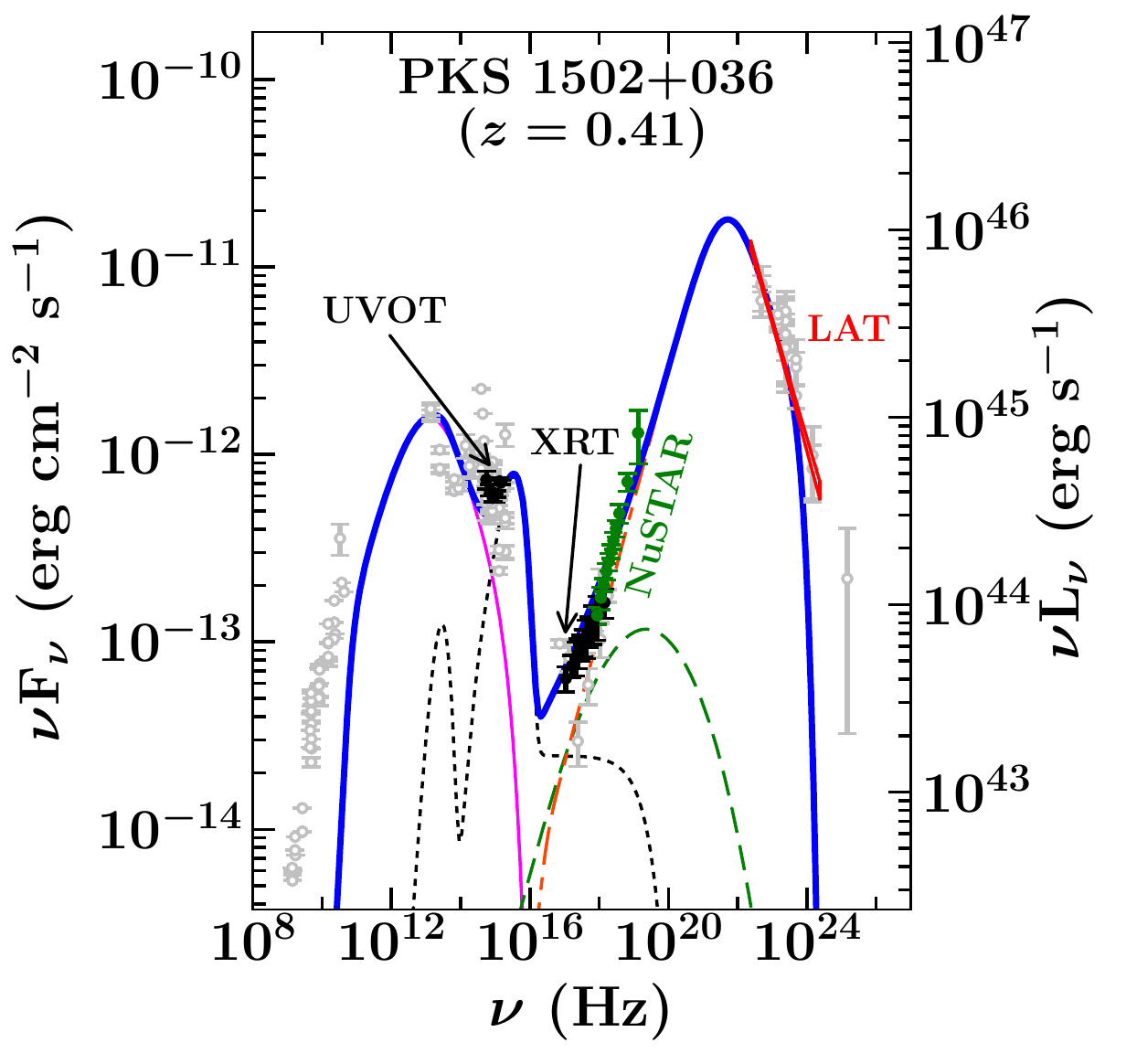}
\includegraphics[scale=0.48]{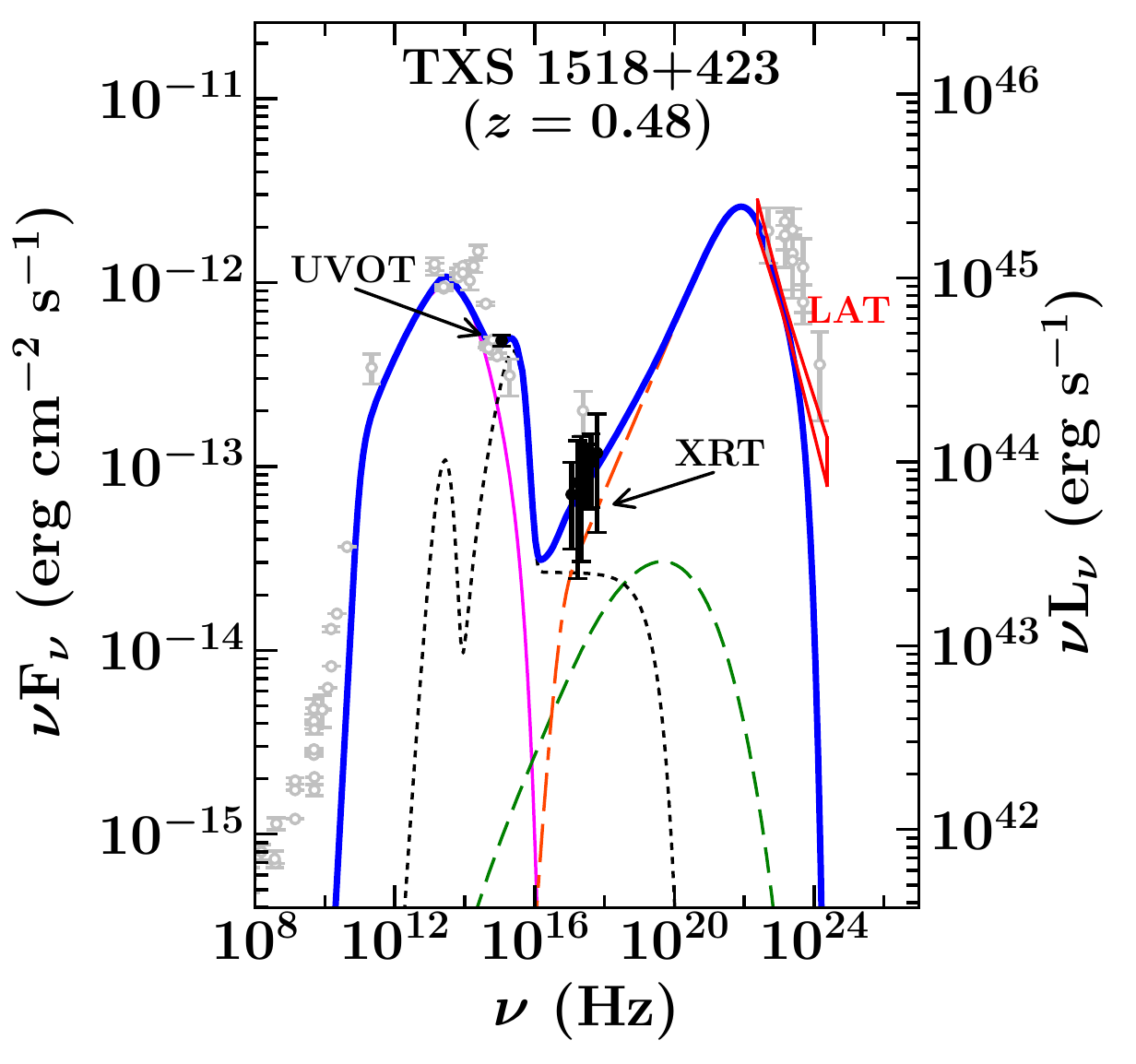}
\includegraphics[scale=0.48]{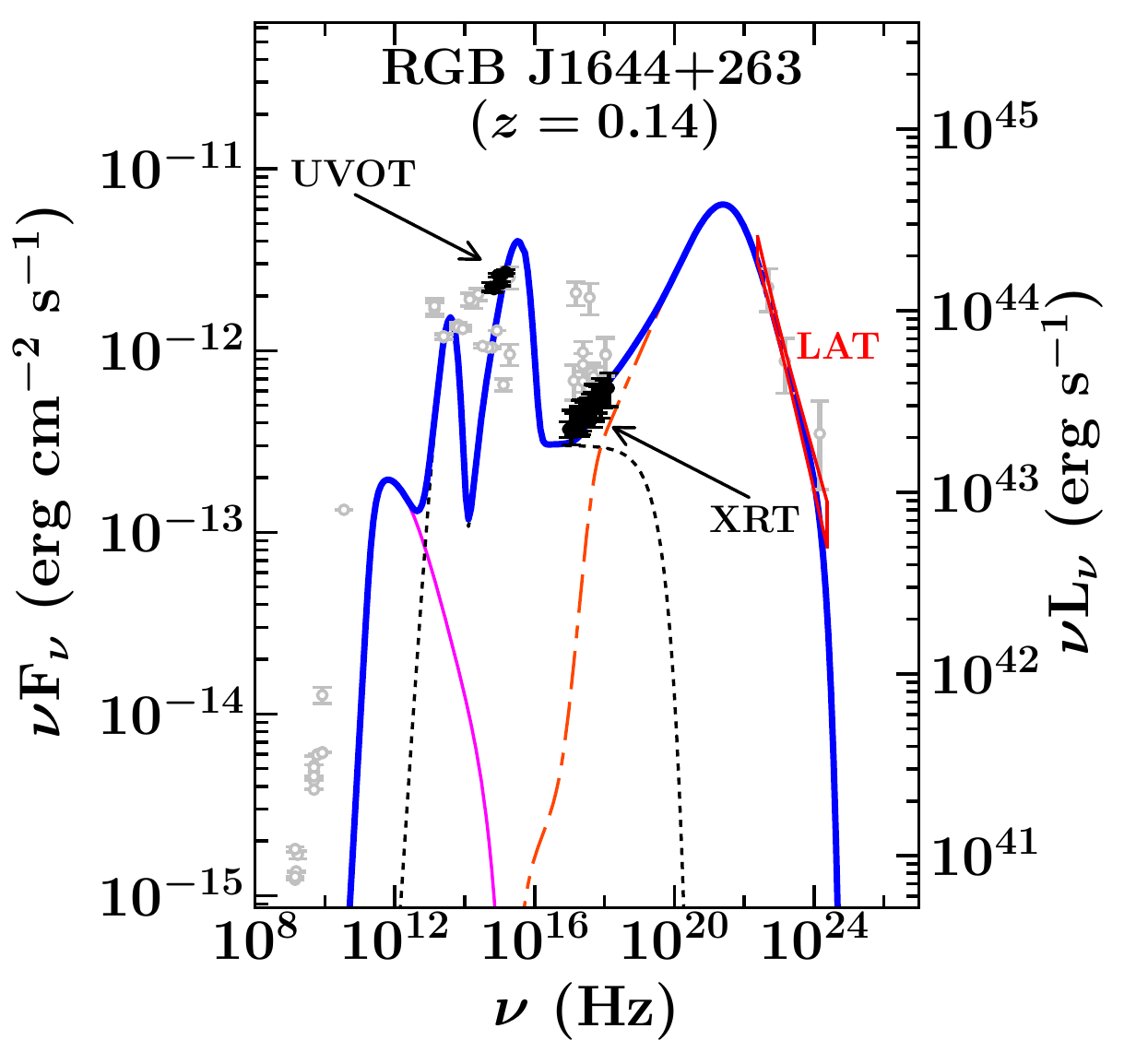}
}
\hbox{
\includegraphics[scale=0.48]{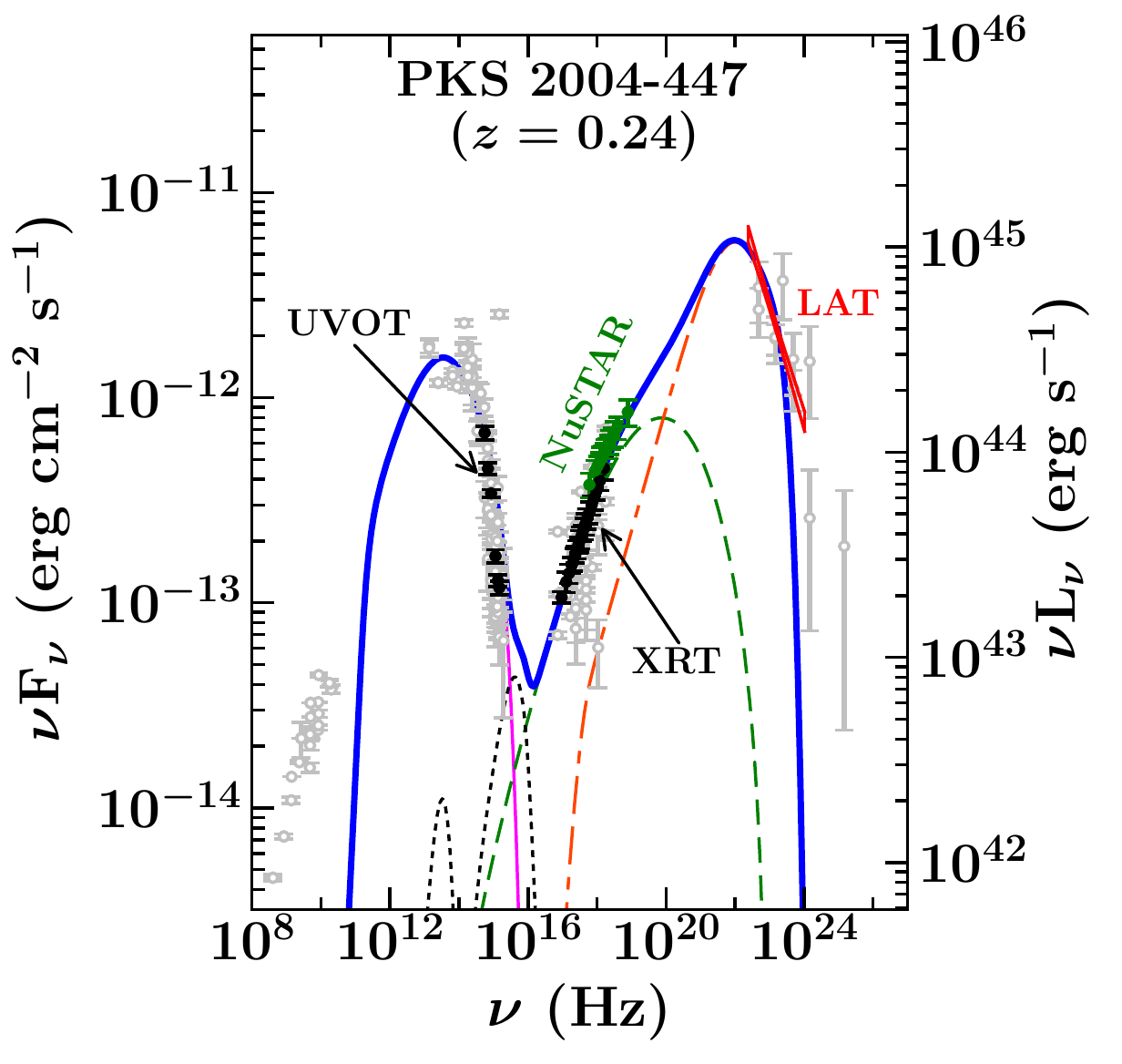}
\includegraphics[scale=0.48]{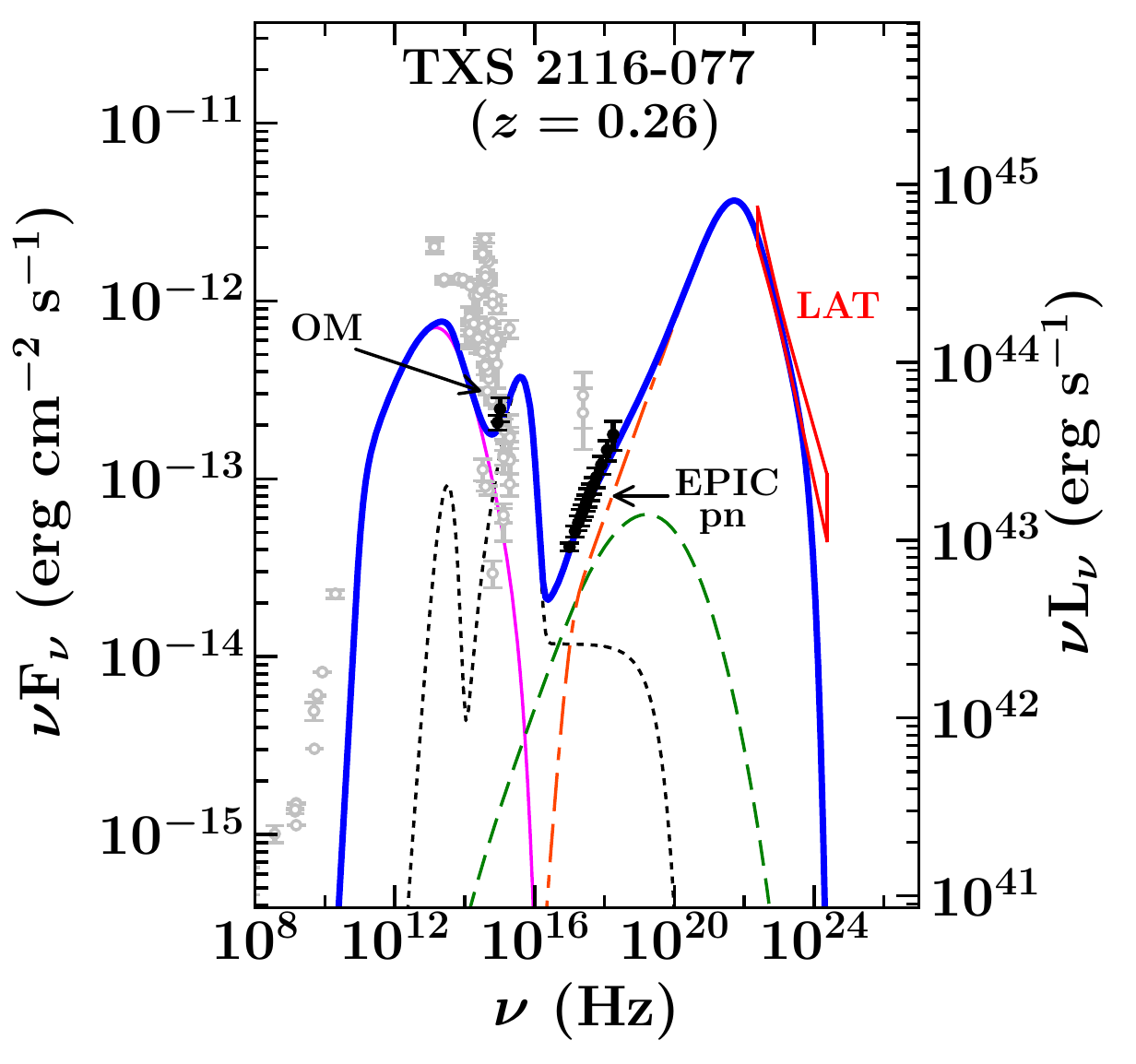}
\includegraphics[scale=0.48]{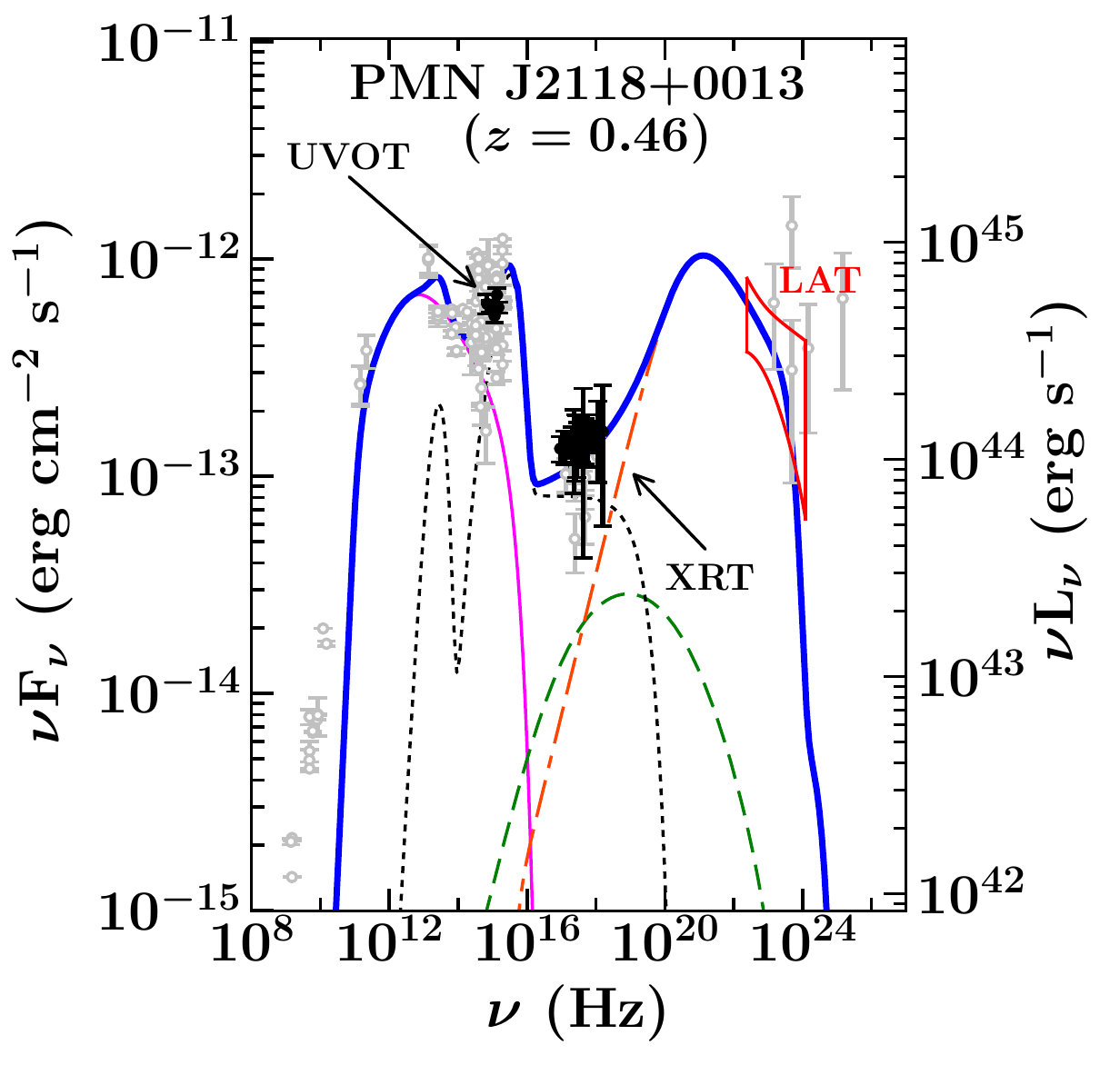}
}
\caption{Same as Figure \ref{fig:app_SED1}.\label{fig:app_SED2}}
\end{figure*}

\end{document}